\documentclass[useAMS,usenatbib]{mn2e}

\usepackage{graphicx}

\usepackage{amsmath}
\title[Amplifications of B-fields in afterglows of GRBs]
{Ambient magnetic field amplification in shock fronts of relativistic jets: an application to GRB afterglows}
\author[Rocha da Silva et al.]{G. Rocha da Silva$^{1}$; 
D. Falceta-Gon\c{c}alves$^{2,3}$\thanks{E-mail:dfalceta@usp.br}; G. Kowal$^{3}$; E. M. de Gouveia Dal Pino$^{1}$
\\$^{1}$Instituto de Astronomia, Geof\'isica e Ci\^encias Atmosf\'ericas, Universidade de S\~ao Paulo, Rua do Mat\~ao 1226,  CEP: 05508-090, S\~ao Paulo, Brazil
\\$^{2}$SUPA, School of Physics \& Astronomy, University of St Andrews, North Haugh, St Andrews, Fife KY16 9SS, UK 
\\$^{3}$Escola de Artes, Ci\^encias e Humanidades, Universidade de S\~ao Paulo, Rua Arlindo Bettio 1000, CEP 03828-000, S\~ao Paulo, Brazil}

\begin{document}

\date{}

\pagerange{\pageref{firstpage}--\pageref{lastpage}} \pubyear{2014}

\maketitle

\label{firstpage}

\begin{abstract} 
 
Strong downstream magnetic fields of order of $\sim 1$G, with large correlation lengths, are believed
to cause the large synchrotron emission at the afterglow phase of gamma
ray bursts (GRBs). Despite of the recent
theoretical efforts, models have failed to fully explain the amplification
of the magnetic field, particularly in a matter
dominated scenario. We revisit the problem
by considering the synchrotron emission to occur
at the expanding shock front of a weakly magnetized relativistic jet
over a  magnetized surrounding medium. Analytical estimates and a number of high
resolution 2D relativistic magneto-hydrodynamical (RMHD) simulations are provided.
Jet opening angles of $\theta = 0^{\circ} - 20^{\circ}$, and ambient to jet density ratios of $10^{-4} - 10^2$ were considered.
We found that most of the
amplification is due to compression of the ambient magnetic field at the contact
discontinuity between the reverse and forward shocks at the jet head, with
substantial pile-up of the magnetic field lines as the jet propagates sweeping the ambient field lines. The
pile-up is maximum for $\theta \rightarrow 0$, decreasing with $\theta$, but
larger than in the spherical blast problem. Values obtained for certain models 
are able to explain the observed intensities. The maximum correlation lengths found
for such strong fields is of $l_{\rm corr} \leq 10^{14}$cm,
$2 - 6$ orders of magnitude larger than the found in previous works.

\end{abstract}

      
\section{Introduction}

Gamma ray bursts (GRBs) liberate a significant fraction of the
rest-mass energy of their source ($>10^{51}$ erg) over intervals ranging from
a fraction of a second to minutes.  The standard fireball picture
\citep{paczynski1986, shemiandpiran1990, reesmeszaros1992, sarietal1998}
explains the otherwise puzzling ability of such sources to vary on short
timescales by arguing that the bursts are produced via a relativistic outflow
with a bulk Lorentz factor $\Gamma >100$.  At first, the relativistic flow is
dissipated internally (via internal shocks or via another form of internal
dissipation, like magnetic instabilities) that produce the prompt $\gamma$-rays.
Later the interaction of the flow with the circum-burst matter produces an
external shock and this blast wave produces the subsequent afterglow at lower
frequencies.  Observational clues concerning GRB progenitors indicate that
supernova explosions of massive stars could be the predominant sources of long
GRBs (i.e. those whose gamma emission lasts more than 2 secs and the
standard model for these objects is the Collapsar model \citep{woosley1993,paczynski1998,macfadyenwoosley1999}.
The main possible sources of short GRBs are mergers of
neutron stars (NSs) with other NSs, or with black holes \citep{eichler1989},
although other driving sources such as phase transition of a NS to a quark
star have also been proposed \citep{lugones2002}.

The field of GRBs has rapidly advanced in recent years, especially following the
launches of NASA missions {\it Swift} and {\it Fermi}, both in the past decade.  Due to their
elusive nature, observing GRBs in all wavelengths at all epochs (including
during and after the GRB) is still challenging with the current GRB detectors
and follow up telescopes.  As a result, for every new temporal or
spectral window unveiled a rich trove of new phenomenology is uncovered
\citep{zhang2011}.  The new observations have
raised new questions.

The composition of the relativistic jets that arise in GRBs is uncertain due to
the lack of direct observations.  The most important unknown parameter is the
ratio ($\sigma$) between the Poynting flux and the matter flux (here
both baryons and leptons are considered).  In the standard fireball internal
shock (IS) scenario, magnetic fields are assumed not to play dynamically
any major role, i.e. $\sigma \ll 1$.  An alternative view is that the GRB
outflow is powered by magnetic extraction from the rotational energy of a massive star
or an accreting black hole and therefore, carries a dynamically important magnetic field component, i.e. $\sigma \gg 1$.
 The GRB radiation in this case would be powered by dissipation of the magnetic field
energy in the ejecta \citep[e.g.][]{usov1992, thompson1994, meszarosrees1997,
piran1999, piran2005, lyutikovblandford2003, zhangyan2011}.  Even in a matter
dominated outflow where the magnetic field does not influence the dynamics,
magnetic fields play a crucial role at the radiation emission region. Magnetic
fields dominate the process of particle acceleration in the
collisionless shocks and also play an important role on the afterglow
synchrotron emission.

Another important aspect related to magnetically dominated
(large $\sigma$) jets is that observations  require that they
become matter dominated at some point beyond the central engine, with the
conversion of the energy  transported outward in the form of Poynting flux into kinetic energy flux. The
mechanism by which this occurs is not known yet. It could be related to gradual
acceleration of the flow \citep{heyvaerts1989,chiueh1991,bogovalov1996,lyubarski2009}, or to impulsive acceleration
\citep{granot2011,granot2012}, or even to non-ideal MHD effects such as
magnetic reconnection \citep{lyutikovblandford2003, giannios2006, lyubarski2010,
zhangyan2011, mckinney2012, levinson_begelman2013},
or magnetic  kink instabilities \citep{giannios2006, levinson_begelman2013}.
This has been known  as the $\sigma$ problem and more recent analytical and numerical studies
suggest that  this conversion may occur even before the jet breaks out from the stellar  envelope
\citep{levinson_begelman2013, bromberg2014, beniamini2014}.

The ejecta can be parametrized by $\epsilon_{B}$ and
$\epsilon_{e}$ which give the ratios of magnetic and kinetic energies with respect to the
total energy density of the ejecta, respectively.  Typical values derived from
the synchrotron emission assuming approximately energy equipartition between the
relativistic electrons and the magnetic field range from $\epsilon_{B}=10^{-5}$
to $10^{-2}$ \citep{waxman97a, galama99nature, yostetal2003,lizhao2011,santana14}.  In
general, both parameters are assumed to remain constant  in the afterglow
region.

In the extreme case mentioned above that the magnetically dominated flow dissipates
most of its magnetic energy before the breakout of the stellar envelope \citep{bromberg2014, beniamini2014},
no significant magnetic field from the source will be carried out by
the  external shock that produces the afterglow emission. This implies that only the ambient
magnetic fields swept and compressed by the ejecta will be available to accelerate the relativistic
particles responsible for the synchrotron radiation.

On the other hand, even assuming   that the ejecta drags most of the magnetic field from the source,
\citet{medvedevloeb99} considered the magnetic field of a
strongly magnetized compact object with $B \sim 10^{16}$G and found that it
cannot account for the magnetic fields observed in the
afterglow.  The average field intensity over the emitting region
scales as $\bar{B} \propto r^{-2}$. Therefore, one expects $B \sim 10^{-4}$~G and
$\epsilon_{B} \sim 10^{-7}$ at the afterglow emission, about $10^{16}$~cm away
from the source.

Other mechanisms were proposed in the literature in order to explain the origin
of the magnetic field in the afterglows of GRBs in a matter dominated scenario.
Most of them based on the growth of non-linear instabilities, such as the Weibel
instability \citep{medvedevloeb99, nishikawa05, hededal2004}.
This instability has its origin in the shock of two different
populations of collisionless plasma particles. The diffusion of part of the populations into each other
generates an anisotropy in the momentum distribution.  The magnetic field
amplification arises in order to isotropize the momentum distribution
\citep{medvedevloeb99}.  Small fluctuations of the magnetic field deflect the
particles by the Lorentz force leading to the generation of currents and the
magnetic field increases.  The deflections become stronger as the magnetic field
increases generating a runaway process.  In such instability,
however, the amplified magnetic field is randomly oriented at very short
correlation lengths ($> \delta$), where $\delta$ is the plasma skin depth
$\delta=c/\omega_{p}$ ($\omega_{p}$ is the plasma frequency), in spite of the
observed correlation lengths $l_{\rm corr} \sim 10^{10}$ $\delta$
\citep{waxman2006}.
Particle-in-cell (PIC) simulations have been performed in order to study this
problem.  For instance, \citet{kazimura98apj} found that about 5~\% of the flow
kinetic energy is converted into magnetic energy.  Also, as pointed above,
\citet{nishikawa03, nishikawa05} showed that the Weibel instability
amplifies non-uniform small scale magnetic fields only.  This could give origin to a
jitter spectra instead of a Synchroton radiation.  \citet{frederik2004} and
\citet{hededal2004} showed that the magnetic field amplitudes necessary to
accelerate particles could be provided by this instability even in the case of a
very weak upstream magnetic field, but these fields would still be small scale ones.
It is quite clear that such small scale process is unable to provide the
large scale and strong magnetic fields as needed to explain the afterglow
emission.

As stressed before, in a matter dominated scenario, we are left with the
ambient magnetic fields. The magnetic energy
density increases due to the shock compression of the interstellar
medium (ISM), which can be derived analytically from the
one-dimensional relativistic Rankine-Hugoniot (RH).
For an adiabatic shock, the RH relations predict
amplification factors of $\sim \Gamma$, being $\Gamma$ is the Lorentz factor
\citep{kennelecoroniti84, applecamenzind88, summerlin2012}.
Typical magnetic fields in the ISM of a few
$\mu$G imply  $\epsilon_{B} \sim 10^{-11}$ \citep{medvedevloeb99}.
However, when considering the confinement of the magnetized
expanding flow between the forward bow shock and the
reverse shock the amplification of the magnetic
fields is more efficient, as also qualitatively evidenced
in former numerical studies of non-relativistic and
relativistic jets \citep[e.g.][]{leismann2005}.
A systematic study of the magnetic field evolution and amplification in
such systems, sweeping a vast parametric space, is still missing though.

In this work, we revisit the problem of the
magnetic field amplification behind the shocks of GRBs. 
The paradigm considered here for the afterglow emission is that described 
e.g., in Granot \& Kronigl (2001):

{\it In the case of GRB afterglows, the most common interpretation is in terms of shocks that form at the interface between the relativistically outflowing material and the surrounding medium (with the bulk of the observed emission arising in the ``forward" shock that propagates into the ambient medium...). The radiation is inferred to be nonthermal, with the dominant emission mechanisms most commonly invoked being synchrotron and inverse Compton.}

In this context, we study numerically the time evolution
of the magnetization in the shocks generated by a relativistic
jet.  We adopt the matter dominated outflow scenario
and explore the amplification of ambient magnetic fields at the shocks by
means of two-dimensional (2D) relativistic magnetohydrodynamics (RMHD) numerical
simulations.  Our goal is to study whether the resulting compressed fields
behind the shocks are sufficient to explain the observed afterglow emission
without requiring a magnetically dominated flow scenario.

We study different possible scenarios. Specifically, we consider the
expansion of conical jets with different opening angles, from
$ \theta = 0$ (cylindrical case) up to $20^{\circ}$, before and after they break out from the stellar envelope
expanding over the interstellar gas with either smaller or  larger densities than the later.

The paper is organized as follows. In Section 2 we review the basic jump conditions in RMHD shocks. In Section 3 we describe the numerical setup and the
RMHD equations to be solved numerically in two-dimensions (2D). In section 4 we describe the numerical results from the simulations and show the magnetic field amplification due to shock compression and pile-up behind the shocks at the jet head. In Section 5 we study the coherence length of the
magnetic field using structure functions and compare them with other proposed
mechanisms of magnetic field amplification. Finally, in Section 6, we discuss our results and the implications for GRB jets and  draw our conclusions.

\section{Relativistic Shocks}

In the most simplified analytical model, a shock is considered a single
discontinuity separating the upstream and downstream media.  If considered at
the reference frame of the shock front, the steady state form of the fluid
equations (i.e. $\partial_t = 0$) provides the Rankine-Hugoniot jump conditions
for the downstream.  The relativistic jump conditions for a magnetized
case (with the shock velocity normal to the magnetic field direction) are
well described in \citet{kennelecoroniti84} \citep[see also][]{hoffmann50,mallick11}.
Here we use the same notation to
describe a shock with velocity perpendicular to the magnetic field.

\begin{equation}
 n_{1} u_{1} = n_{2} u_{2} ,
 \label{RH1}
\end{equation}

\begin{equation}
 E=\frac{u_{1} B_{1}}{\Gamma_{1}}=\frac{u_{2} B_{2}}{\Gamma_{2}} ,
 \label{RH2}
\end{equation}

\begin{equation}
  \Gamma_{1}\mu_{1}+\frac{EB_{1}}{4\pi n_{1}
u_{1}}=\Gamma_{2}\mu_{2}+\frac{EB_{2}}{4\pi n_{2} u_{2}} ,
  \label{RH3}
\end{equation}

\begin{equation}
  \mu_{1}u_{1}+\frac{P_{1}}{n_{1}u_{1}}+\frac{B_{1}^2}{8 \pi n_{1}
u_{1}}=\mu_{2}u_{2}+\frac{P_{2}}{n_{2}u_{2}}+\frac{B_{2}^2}{8 \pi n_{2} u_{2}} ,
  \label{RH4}
\end{equation}

\noindent
where $P$ is the thermal pressure, $n$ is the number density, $\Gamma$ the
Lorentz factor, $u$ is the velocity normal to the shock plane, $E$ is the electric
field in the shock frame, and $B$ the
magnetic field in the region 1 (upstream, unshocked region) and the region
2 (downstream, shocked region).  The
factor $\mu$ is the specific enthalpy, which for a relativistic gas with a
polytropic index $\gamma$ is defined by:

\begin{equation}
 \mu=1 +\frac{\gamma}{\gamma-1}\left(\frac{P}{n m c^2}\right).
\end{equation}

\noindent
In the case of a relativistic adiabatic shock $\gamma \rightarrow 4/3$.

From the set of equations (1-5) above the magnetic amplification ratio
$B_{2}/B_{1}$ is obtained:

\begin{equation}
Y \equiv
\frac{B_{2}}{B_{1}}=\frac{N_{2}}{N_{1}}=\frac{\Gamma_{2}u_{1}}{\Gamma_{1}u_{2}}.
\end{equation}

\noindent
Notice that the measured number density N relates to the proper
density through the relation $N=n \Gamma$ \citep{gallant1992}.

According to the conservation equations above, the amplification of the magnetic
field occurs due to the strong shock compression.  The basic
assumption of a fluid frozen into the magnetic fields results in an
equal jump condition for both $\rho$ and $B$.  Therefore, for strong shocks,
part of the kinetic energy is converted to magnetic energy.

\begin{figure}
 \center
 \includegraphics[width=0.45\textwidth]{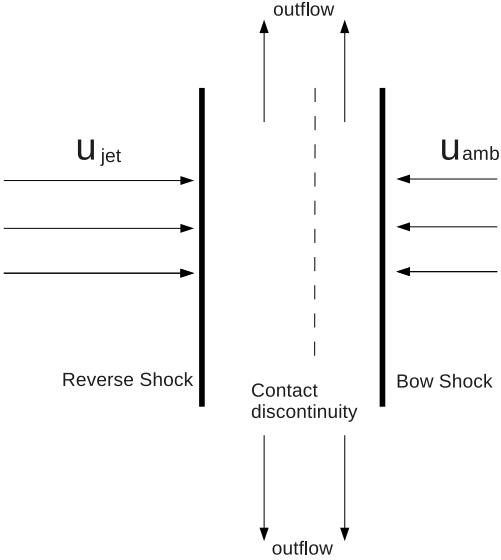}
 \caption{Idealized and simplified picture of the shock fronts generated by a relativistic jet 
 expanding over the ambient medium at rest. Differently to what happens in an isotropically expanding blast wave, 
 the downstream ambient material is able to flow along the contact discontinuity. This results in a lower 
 downstream pressure and in a thinner shock front, compared to the isotropic case.}
 \label{fig:shock}
\end{figure}

This scenario is more complex if the shocked region is bounded by two shocks.
Actually, this is the case of the high speed jet propagating over the
ambient medium.  At the reference frame of a supersonic shock there are two
incoming flows, the relativistic jet from one side and the
ambient gas from the other with a contact discontinuity between them, where the
kinetic linear momenta are equal.  If compression leads to significant lateral 
expansion an outflow is expected to emerge in the direction perpendicular to
the inflows (see Fig.\ref{fig:shock}), and this problem cannot be solved in one dimension.

Following Fig.\ref{fig:shock}, the upstream jet gas is decelerated
at the shock discontinuity on the left (shock 1) and its downstream shocked
material is pushed outwards in the lateral direction.  The ambient
material is shocked at the discontinuity on the right (shock
2), enters the shock region, and leaves outward, as well. The
equilibrium of momentum between both downstream flows
occurs at the contact discontinuity, and turbulent mixing of the fluids at
this surface may occur.

Earlier three-dimensional (3D) numerical studies of hydrodynamical non-relativistic jets
\citep{chernin1994} have demonstrated that this mixing depends mainly on the jet
Mach number and the density ratio between the jet and the ambient gas.  For small
values of both parameters (Mach numbers $< 6$ and density ratios $< 3$) turbulent mixing and entrainment
become  important - a condition typically  fulfilled, e.g., by certain classes
of AGN jets \cite[see e.g.][and references therein]{dalpino93, ragacabrit1993,
stonenorman93, dalpino2005} and for further hydrodynamical studies
\citep{folini00, folini06} \cite[check also][for similar studies in
non-relativistic MHD flows]{cerqueira97, diego2012}.

These authors also showed that the momentum transfer and width of the shocked
region is strongly affected by the thermal radiative
cooling of the shocked material.  Strong cooling decreases the turbulent mixing
(as part of the internal energy of the shocked material is radiated away) and
also shrinks the shock region, as the downstream internal energy is small
compared to the upstream kinetic one.  A consequence of non-uniform cooling and
thin shock regions is the growth of the non-linear thin layer instability
\citep{vishniac94} and the Rayleigh-Taylor instability which can break the bow
shock region into a healthy clumpy structure \citep{blondin90, dalpino93, dalpino94,
stonenorman93}.

In the case of magnetized shock-bounded slabs, the upstream gas drags field
lines into the shocked region.  Depending on the orientation of the
upstream fields, the downstream magnetic field lines are not carried
away with the outflow. 2D MHD numerical simulations of non-relativistic converging flows reveal
that part of the magnetic field component perpendicular to the shock velocity  $B_\perp$
is not advected, instead, it piles-up and remains parallel to the contact discontinuity surface \citep{diego2012,diego2014}.
Then, the downstream shocked plasma flows along the
amplified field lines outwards to fill the cocoon surrounding the jet beam.  Since the jet is continuously pushing the
ambient gas forward there is a constant inflow of ambient magnetic field
lines into the bow shock region, causing the {\it piling-up
effect}.

\begin{figure}
 \center
 \includegraphics[width=0.45\textwidth]{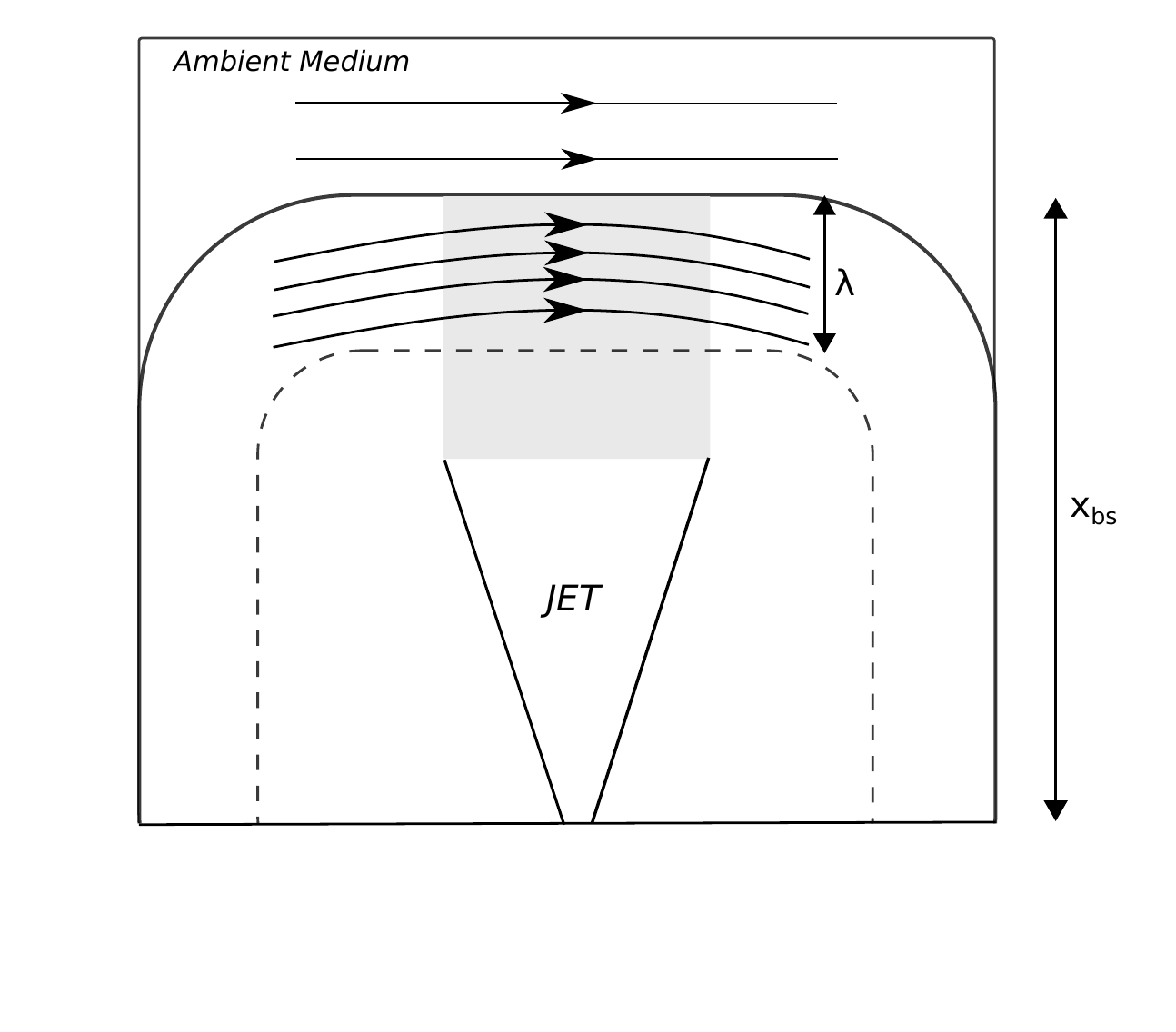}
 \includegraphics[width=0.45\textwidth]{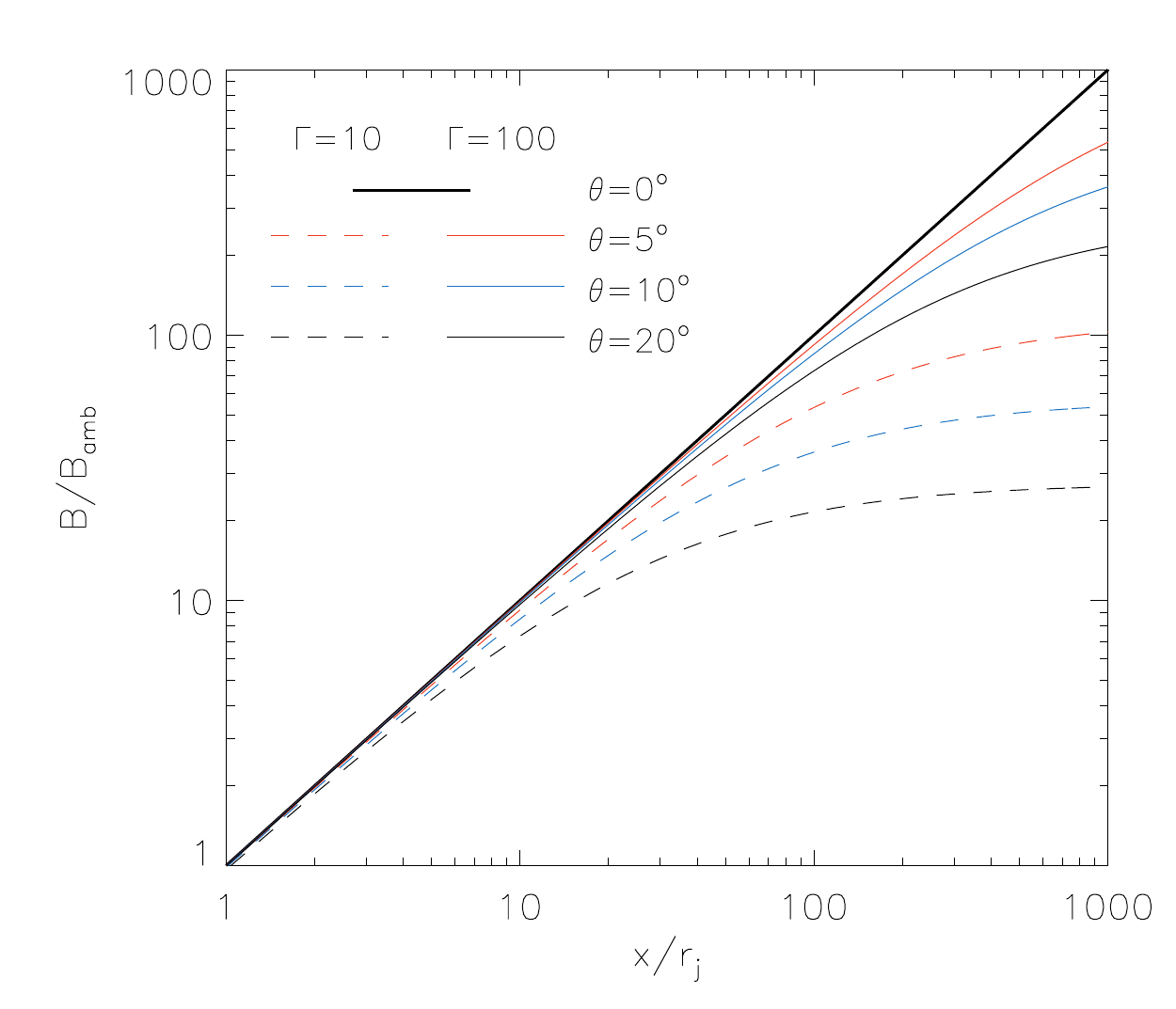}
 \caption{Up: Schematic representation of the pile-up effect of the magnetic field lines as the jet propagates in an ambient medium with uniform magnetic field.  Bottom: Pile-up effect on the magnetic field intensity at the shock region for collimated and wide jets as a function of the distance, as given by  Eqs.\ref{eq:pileup} and \ref{eq:rjcon}, respectively, for different jet opening angles, and for $\Gamma =10$ (dashed) and
 $\Gamma =100$ (solid).}
 \label{fig:pileupdraw}
\end{figure}

If we consider the pile-up effect of the ambient magnetic field only, magnetic flux conservation implies  a piled-up
magnetic field intensity in the shock frame given by:

\begin{equation}
B_{\rm x} \simeq B_{\rm amb} \left( \frac{x_{bs}(t)}{\lambda} \right)^\alpha ,
\label{eq:pileup}
\end{equation}

\noindent
where $B_{\rm x}$ is the magnetic field that is squeezed behind the shock structure after the bow shock at the jet head has propagated  a distance  $x_{bs}(t)$ and $\lambda$ represents the width of the shock region. Here, $\alpha \rightarrow 1$ if the field is
parallel to the contact discontinuity and $\alpha \rightarrow 0$ if the field
lines are mostly perpendicular to the discontinuity. Fig.\ref{fig:pileupdraw} sketches the pile-up effect. The arrows represent an initially uniform magnetic field in the ambient medium and as the jet propagates it sweeps the magnetic field lines which are compressed within the double shock structure, i.e., between the forward bow shock and the reverse jet shock.

An analytical estimate of $\lambda$ is not trivial though, mostly because of the asymmetric
morphology of the shock region. The shock thickness for spherical relativistic blast waves has
been derived as $\lambda \sim R/\Gamma$ \citep{bland76}, being $R$ the shock wave radius. Since $R=R(t)$, the
thickness $\lambda$ is also a function of time. This expansion of $\lambda$ with time may be understood from
the conservation of matter and energy. The shock dynamics is that of a one-dimensional radial Riemann
problem, but with an uniformly expanding shocked volume as the shell expands.
The accumulation, as the blast wave moves, results in local increase of enthalpy that leads
to an expansion of the shock thickness.

This scenario is different for the jet case though, which is
not well-described by an one-dimensional Riemann problem. Here the shocked gas flows away from the axis of symmetry. If a steady state is achieved and if the jet is collimated into a quasi-cylindrical shape, i.e. $\theta \rightarrow 0$, there
is no net enhancement of local enthalpy and $\lambda$ is constant with time.
In this case, by considering mass conservation at the dashed area of Fig.\ref{fig:pileupdraw} one obtains, for
the  $\theta \rightarrow 0$ (cylindrical) case:

\begin{equation}
\lambda_{\rm cyl} \simeq \frac{r_{\rm jet}}{2}\frac{n_{\rm j,1}u_{\rm j,1}+n_{\rm A,1}u_{\rm sh}}{n_{\rm j,1}Y_{\rm j}u_{\rm j,2}+n_{\rm A,1}Y_{\rm A}u_{\rm A,2}}
\label{eq:rjcil}
\end{equation}

where $r_{\rm jet}$ represents the radius of the jet at the working surface, $Y$ the jump in density
between downstream and upstream flows, indices $j$ and $A$ account for jet and ambient values, respectively, and $u_{\rm sh}$
represents the speed of the shock region in the observers reference frame:
\footnote{which is obtained from momentum flux conservation assuming that the ambient
pressure is negligible, as the gas is cold, and the jet pressure is much smaller
than the jet shock ram pressure}

\begin{equation}
u_{\rm sh} \sim u_{\rm j,1}\frac{\left(n_{\rm j,1} \Gamma_1^2 n_{\rm A,1}^{-1} \right)^{1/2}}{1+\left(n_{\rm j,1} \Gamma_1^2 n_{\rm A,1}^{-1} \right)^{1/2}}.
\end{equation}

Since in the case of a well-collimated (cylindrical) jet $r_{\rm jet}$ is constant as the shock front moves
further away of the central source, for $\Gamma_1 \gg 1$, we obtain:

\begin{equation}
\lambda_{\rm cyl} \sim \frac{\sqrt{2}}{2} \eta r_{\rm jet}
\end{equation}

\noindent
where $\eta = n_{\rm A,1}/n_{\rm j,1}$.

As discussed later on in the paper, there are observational evidences - as well as results from
numerical simulations - indicating that core-collapse GRB jets may be, in reality, not well collimated after
the breakout of the stellar envelope. Observationally, power-law break decay during the afterglow emission
has been well modelled by means of conical jets, with opening angles as large as $20^{\circ}$, being
$\theta_{\rm j} < 10^{\circ}$ in a vast majority of objects \citep[see][]{sari99,bloom2003,frail01,zeh06,
tchekhovskoy2009, bromberg2011,mizuta2013}.

In the case of a conical jet the rate at which gas is loaded into the
shock region varies with time. This because the Mach disk, i.e. the area of the jet working over the shocked gas,
increases as the jet propagates forward, away from the central source. To obtain a modified
analytical approximation for this case, we separate the fluxes of gas into the shock region in two,
one being exactly the same as considered in Eq.\ref{eq:rjcil}, and the other
being the net increase due to the increased radius of the jet, i.e. $\Phi_{\rm tot}=\Phi_{r_{\rm jet,0}}+\Phi_{\Delta r_{\rm jet}}$.
Let us consider a simple case in which the opening angle $\theta$ is constant. Since $\Delta r_j \propto x(t)  tan \theta$,
Eq.\ref{eq:rjcil}, with now $r_{\rm jet}=r_{\rm jet}(t)=r_{\rm jet,0}+\Delta r_j$, results in:

\begin{equation}
\lambda_{\rm con}(t)  \sim \lambda_{\rm cyl} \left(1+\frac{x(t) \tan\theta}{r_{\rm jet,0} \Gamma_1}\right)
\label{eq:rjcon}
\end{equation}

Therefore, the pile-up must occur at shorter distances in the case of  conical jets, with a departure
from the linear growth of $B$ and eventual saturation of the magnetic field amplification,consistent with causality constraints.

Using  Eqs.\ref{eq:pileup} and \ref{eq:rjcon}, we  computed the pile-up effect which is shown in Fig.\ref{fig:pileupdraw}
 (bottom) as a function of the distance to the central source, for different jet parameters. We clearly see in the Figure that for large values of $\Gamma$, the difference between collimated ($\theta \rightarrow 0$) and wide jets decreases substantially.

As stated before, despite the extensive numerical multidimensional study
that can be found in the literature of magnetized relativistic jet flows, a
systematic study of the amplification of ambient magnetic fields by relativistic jets,
particularly in the context of GRBs, is still missing.  In the following
sections we will explore this issue and test the scenarios above  considering
relativistic MHD numerical simulations of both collimated and wide jets propagating
over a weakly magnetized ambient.

\section{Governing Equations and Numerical Setup}

The evolution of our system is governed by the special relativistic
magnetohydrodynamic equations (SRMHD) which can be written in the general
conservative form
\begin{equation}
\partial_t\mbox{\bf U} + \nabla \cdot\mbox{\bf F}(\mbox{\bf U}) = 0,
\label{eq:genform}
\end{equation}
where ${\bf U}$ is the vector of conserved variables
\begin{align}
{\bf U}  = & \left( D, {\bf S}, {\bf B}, E \right)^T \nonumber \\
         = & \Bigl( \Gamma \rho, \left( \xi + B^2 \right) {\bf v} - \left( {\bf v} \cdot {\bf B} \right) {\bf B}, {\bf B}, \nonumber \\
           & \xi + \frac{1}{2} \left( B^2 + v^2 B^2- ( {\bf v} \cdot {\bf B})^2 \right) - p - D \Bigr)^T,
\label{eq:cons}
\end{align}
and ${\bf F}$ is the tensor of fluxes
\begin{align}
{\bf F}  = & \Bigl( D {\bf v}, \nonumber \\
           & \left( \xi + B^2 \right) {\bf v} {\bf v} - \frac{{\bf B} {\bf B}}{\Gamma^2} - \left( {\bf v} \cdot {\bf B} \right) \left( {\bf B} {\bf v} + {\bf v} {\bf B} \right) + {\bf I} p_{\rm tot}, \nonumber \\
         &{\bf v} {\bf B} - {\bf B} {\bf v}, E {\bf v} + p_{\rm tot} {\bf v} - \left( {\bf v} \cdot {\bf B} \right) {\bf B} \Bigr)^{T},
\label{eq:flux}
\end{align}
where $D$ is the rest mass density, $\mathbf{S}$ is the momentum density, $E$ is
the energy density, $\rho$ is the mass density, ${\bf v}$ is the fluid velocity,
${\bf B}$ is the magnetic field, $p_{tot} = p + p_{\rm mag}$ is the total
pressure, $p$ is the gas pressure, $p_{\rm mag} = \frac{1}{2} \left( (B /
\Gamma)^2 + ( \mathbf{v} \cdot \mathbf{B} )^2 \right)$ is the magnetic pressure,
$\Gamma = \left[ 1 - (v / c)^2 \right]^{-1}$ is the Lorentz factor, and for the
case of an ideal equation of state with a constant polytropic index $\gamma$,
the measure of enthalpy $\xi$ is given by
\begin{equation}
\xi = \Gamma^2 \left( \rho + \frac{\gamma}{\gamma-1} p \right).
\label{eq:enth}
\end{equation}

The above set of equations was solved using the GODUNOV code (check {\it http://amuncode.org} for the public available source code) which implements the Godunov-framework of the
hyperbolic equation numerical solution \citep{godunov59} extended by methods
suitable to solve the SRMHD equations.  The code has been extensively tested
and applied to several astrophysical problems \citep[e.g.][]{kowal10, fal10a,
fal10b, fal10c, santos-lima10, kowal11a, kowal11b, kowal12, santos-lima12, santos-lima13, poidevin13}
In the work presented here we used the
$5^{th}$ order monotonicity-preserving (MP) reconstruction \citep{suresh97,
he11} of the Riemann states, the approximate HLLC Riemann solver
\citep{mignone06} in order to calculate the numerical approximation of the
fluxes $\mathbf{F}$.  The solution advances in time using the $3^{rd}$ order
four-stage explicit optimal Strong Stability Preserving Runge-Kutta SSPRK(4,3)
method \citep{ruuth06}.  In order to keep the divergence of magnetic field
minimum, we use the hyperbolic divergence cleaning approach by \cite{dedner02}.

A non-straightforward element of the solution of the relativistic MHD
equations is the determination of the primitive variables $\mathbf{Q} = (\rho,
\mathbf{v}, \mathbf{B}, p)$ from their conservative representation $\mathbf{U}$
(see Eq.~\ref{eq:cons}).  While in the non-relativistic case the conversion
requires only simple algebraic manipulations, here we are forced to use
iterative methods.  A number of such methods has been compared in \cite{noble06}
with the conclusion that their 1Dw scheme is the most accurate and robust one
and therefore, it is also employed in our calculations.

\subsection{Initial setup}

Significant progress has been achieved in the past years regarding relativistic
jet simulations both in the framework of extragalactic jets  \citep{marti1997, aloy1999, hughes2002}
and of  GRB jets \citep{komissarov1999, macfadyen2001, zhang_woosley2003, leismann2005, tchekhovskoy2010,tchekhovskoy2008, morsony07,mizuta2009,lazzati2009,decolle2012,mizuta2013,bromberg2014},
most of which were performed in the low $\sigma$ regime
and, due to computational limitations, in two-dimensions\footnote{The 2-dimensional approach in this work means that the system is 
considered in a planar symmetry, not axial, since the external magnetic field has to be kept uniform. Such approach is limited, 
obviously, and underestimates the dynamics of the flows perpendicular to the plane, i.e. ortogonal to the magnetic field. A more 
detailed discussion about this is presented later in the manuscript, but a more comprehensive picture will be provided in a future 
work where a full 3-dimensional modelling is presented.}, but none focussed on the
investigation of the interaction of the shocks of the ejecta with the ambient magnetic field
after the breakout  of the collapsing stellar envelope. 

There is some debate in the literature regarding the GRB jet opening angle at the breakout \citep[see e.g.][and references therein]{lazzati05,morsony07,
tchekhovskoy2008,mizuta2013,bromberg2014}. When inside the stellar envelope, collimation of a Poynting flux-driven jet may occur due to
net currents driven locally, as well as by the surrounding  pressure. The energy dissipation
at the jet  shock head increases the total pressure of a hot cocoon that develops around the jet which helps collimating the jet.
Once the jet breaks out of the stellar envelope,
it may become wider due to reduced pressure \citep[][]{bromberg2011, mizuta2013}.
There is in fact evidence in favour of conical jets.
\citet{tchekhovskoy2009}, for instance, finds from his simulations  Lorentz factors $\Gamma \sim 100 - 5000$ and opening
angles $\theta_{\rm j} \sim 0.1 - 10^{\circ}$, reproducing inferred properties of GRB jets.
On the other hand, the confinement inside the
envelope of a Poynting flux dominated jet due to both magnetic and cocoon pressure  can be so large that  the jet can emerge from the
stellar envelope with a radius of the order of the source light cylinder radius ($R_L$) and
likely remain confined well after the breakout providing a consistent  jet scenario for both the
prompt gamma emission and the formation of a photosphere \citep{levinson_begelman2013}.

Observationally, the opening angle is inferred by fitting the break in the power-law
decay of the afterglow emission with the fluxes expected from an emitting plasma subject to
relativistic beaming \citep{rhoads99}. The fit model depends on several simplifications, such as the
density distribution of the surrounding medium (e.g. for winds or ISM), and radial dependencies within the
jet. Under these conditions, the vast majority of GRBs data results in $\theta_{\rm j} < 10^{\circ}$, with fiducial
estimates at $\theta_{\rm j} \sim 4^{\circ}$ \citep{sari99,bloom2003,frail01,zeh06}.

While this question of the jet opening angle  is still debatable \citep[e.g.][]{lazzati05, morsony07,tchekhovskoy2009,bromberg2014}, 
in this work we explore different possible values for this parameter.
At the inlet of the computational domain we start with a jet that
has just emerged from the collapsing stellar envelope into the ambient medium.
For the sake of simplicity the jet in our simulations is not launched from first principles, but injected as boundary condition. The opening angle is therefore a free parameter and in our simulations three conditions have been tested for it, namely $\theta_{\rm j}=0^{\circ}$,$10^{\circ}$ and $20^{\circ}$. Also,  $\theta_{\rm j}$ is set as constant as the jet emerges from the collapsing stellar envelope into the ambient medium.

The initial setup of the jet beam is built with a region of continuous injection
of material into the computational domain of radius $R_{\rm j}$, which defines the jet radius,
set at the left vertical boundary of the box domain.
The bottom horizontal and
the left vertical boundaries of the box are assumed to
be reflective, while the other are open boundaries allowing the material to leave the domain.

The ambient gas density $\rho$ and pressure $p$ are assumed initially uniform in the
whole domain. Since we are interested in the
study of the amplification of the  magnetic field in the shock region of a matter
dominated flow, we set a weak uniform ambient
magnetic field initially perpendicular to the propagation of the jet, corresponding to
$P_{mag} / P_{th}= 10^{-5}$.

Another important parameter for the dynamical evolution of jets,
though not critical for the purpose of this work as we discuss further below in the
paper, is the ratio between the surrounding ambient
and the jet densities ($\eta = \rho_{\rm amb}/\rho_{\rm
jet}$).
Traditionally,  relativistic jet propagation models for microquasars,  AGNs and GRBs
 assume an underdense relativistic flow, i.e., with
a density smaller than that of the ambient medium. One of the justifications for this assumption is the general absence of thermal emission in the  shocks of these jets.
In the collapsar scenario, if the GRB jet is magnetically driven,
the launch occurs probably with $\eta > 1$
\citep[see e.g. simulations by][]{lopez13}, as predicted by the semianalytical models and
numerical simulations referenced in the previous sections.
After breaking out of the envelope, far from the stellar material, the jet may
change its mass regime and the jet density may become larger than that of the ambient, i.e.
$\eta \ll 1$. This is discussed, for instance, by \citet{lazzati05} and \citet{morsony07}.
Simulations of the jet-envelope interaction indicate a transition between the regimes of
$\eta \simeq 10^{4}-10^{5} $ (at the central region of the stellar envelope) and
$\eta < 10^{-6}$ after the jet breaks the outer boundary of the star.

Since the dynamical evolution of the jet over an uniform ambient medium may differ considerably 
depending on the parameter $\eta$ chosen, for the sake of completeness, we studied the dynamical 
evolution of jets in both regimes, i.e. $\eta < 1.0$ and $\eta > 1.0$, sweeping a parametric 
range  $10^{-4} \leq \eta \leq 10^2$. In this sense we can study the different morphologies 
and magnetic field amplification for the phases of the jet interacting with the surrounding 
media right after the breakout of the stellar envelope and further out.

The space of parameters investigated in this work is presented in Table~\ref{tab}.
In all cases the
jet is initially relativistic (with Lorentz factors $\Gamma$ = 2, 10, or 100)
and supersonic, i.e. the initial Mach number, defined as $M \equiv v_{\rm jet}/c_s$ is set as 10 
for all models, where the sound speed is given by 
$c_s=\sqrt{ \gamma \left(\gamma -1\right) P / \left[\left(\gamma-1\right)\rho + \gamma P \right] }$. 

The dimensions of the simulated computational box is $(L_x,L_y) =
(48,12)$ in code units.  The adopted code unit for the distance is $5R_j$.  The
code unit for time is defined as $5 R_j/c$, where $c$ is equal 1 in our
simulations. The simulations were performed with a resolution of $4096
\times 1024$ cells.

%%%%%%%%%%%%%%%%%%%%%%%%%%%%%%%%%%%%%%%%%%%%%%%%%%%%%%%%%%%%%%%%%%%%%%%%%%%%%%%%%%%%%%%%
%		              Table - parameters                                        %
%%%%%%%%%%%%%%%%%%%%%%%%%%%%%%%%%%%%%%%%%%%%%%%%%%%%%%%%%%%%%%%%%%%%%%%%%%%%%%%%%%%%%%%%
\begin{table}
\centering
\caption{Parameters used in each simulation run. We explore the dependence with
the density ratio, polytropic index, Lorentz factor and opening angle. The angle equal
$0^{\circ}$ refers to the case where the jet is injected with a cylindrical geometry.}
\begin{tabular}{cccccc}
\hline
$\gamma$ &	Mach	& $\Gamma$ &	$\rho_{\rm amb}/\rho_{\rm
jet}$ & $\theta_{\rm j}$ & $Model$ \\
\hline
\hline
1.1  &	10	&2   &	$10^{2}$  & $0^{\circ}$ & NA1cyl\\
1.1  &  10	&10  &	$10^{2}$  & $0^{\circ}$ & NA2cyl\\
1.1  &	10	&100 &  $10^{2}$  & $0^{\circ}$ & NA3cyl\\
1.33 &  10	&2   &  $10^{2}$  & $0^{\circ}$ & AD1cyl\\
1.33 &	10	&10  &  $10^{2}$  & $0^{\circ}$ & AD2cyl\\
1.33 &	10	&100 &  $10^{2}$  & $0^{\circ}$ & AD3cyl\\
1.33 &  10      &10  &  $10$      & $0^{\circ}$ & AD4cyl\\
1.33 &  10      &10  &  $1$       & $0^{\circ}$ & AD5cyl\\
1.33 &  10      &10  &  $10^{-4}$ & $0^{\circ}$ & AD6cyl\\
1.33 &  10      &10  &  $10^{-2}$ & $0^{\circ}$ & AD7cyl\\
1.33 &  10      &10  &  $10^{-3}$ & $0^{\circ}$ & AD8cyl\\
1.33 &  10      &100 &  $10^{-4}$ & $0^{\circ}$ & AD9cyl\\
\hline
1.1   &	10	&10  &  $10^{2}$  & $10^{\circ}$  & NA1con\\
1.1   &	10	&100 &  $10^{2}$  & $10^{\circ}$  & NA2con\\
1.33  &	10	&10  &  $10^{2}$  & $10^{\circ}$  & AD1con\\
1.33  &	10	&100 &  $10^{2}$  & $10^{\circ}$  & AD2con\\
1.33 &  10      &10  &  $10^{-4}$ & $10^{\circ}$  & AD3con\\
1.33 &  10      &100 &  $10^{-4}$ & $10^{\circ}$  & AD4con\\
1.33 &  10      &10  &  $10^{-4}$ & $20^{\circ}$  & AD5con\\
1.33 &  10      &100 &  $10^{-4}$ & $20^{\circ}$  & AD6con\\
\hline
\hline
\end{tabular}
\label{tab}
\end{table}
%%%%%%%%%%%%%%%%%%%%%%%%%%%%%%%%%%%%%%%%%%%%%%%%%%%%%%%%%

The thermal radiative cooling of the hot shocked ambient plasma may result in thin and unstable shock regions.
However, radiative losses of GRB jets are dominantly non-thermal, mainly synchrotron and inverse-Compton processes.
 The actual role  of these processes in the cooling of the shocked plasma at the afterglow phase is not clear yet \citep{granot2001}.
 For this reason, we run most of our models under an adiabatic regime ($\gamma = 4/3$) and,
in order to mimic the  action of the thermal radiative cooling  at the bow shock region upon the magnetic field
amplification, we run the same models with a reduced effective polytropic index of $\gamma = 1.1$ (these models are referred as
{\it NA} in Table \ref{tab}).

Notice that we have used an uniform value of $\gamma$ for the whole computational domain.
Recent works have focused on the stability and
thermodynamical aspects that can influence the jet dynamics \citep{bodo2013}.
\citet{mignoneandmckinney2007}  explored the effects
of  varying smoothly the gas enthalpy in the propagation of a relativistic
jet (with an polytropic index $\gamma=4/3$) into a
non-relativistic medium (with $\gamma=5/3$).
\footnote{Long before Taub (1948) had shown that in order to preserve the consistency with the 
relativistic kinetic theory, the specific enthalpy $\mu$ has to satisfy the inequality:
$(\mu-\Theta)(\mu-4\Theta) \geq 1$,
where $\Theta = p/ \rho$ and $\mu$ is the enthalpy of the relativistic gas and their
proposed equation of state satisfies the Taub inequality.}
Their two-dimensional simulations revealed a slower evolution of the jet and changes in the shape of the cocoon,
but the main conclusion was that the overall structure of the relativistic jet with the modified enthalpy equation is similar to the case
with uniform $\gamma=4/3$.

The simulations were run until the propagating jet reached the right vertical boundary of the computational domain,
except for the model with Lorentz factor $\Gamma = 2$, for which the jet power was too little to drill the ambient gas
through to the right boundary. The outcomes of the simulations are described in the following section.

\section{Numerical Results}

In this section we present the results from the simulations and comparisons between the different models.

\subsection{Jet/Ambient morphologies}

\subsubsection{$\eta > 1.0$ (light jets)}

Let us first discuss the morphologies of the shocked
material surrounding  light jets. These models can be
specially suitable for right after the breakout of the jet into the environment of the core-collapse GRB.

\begin{figure*}
 \center
{
 \includegraphics[width=0.45\textwidth]{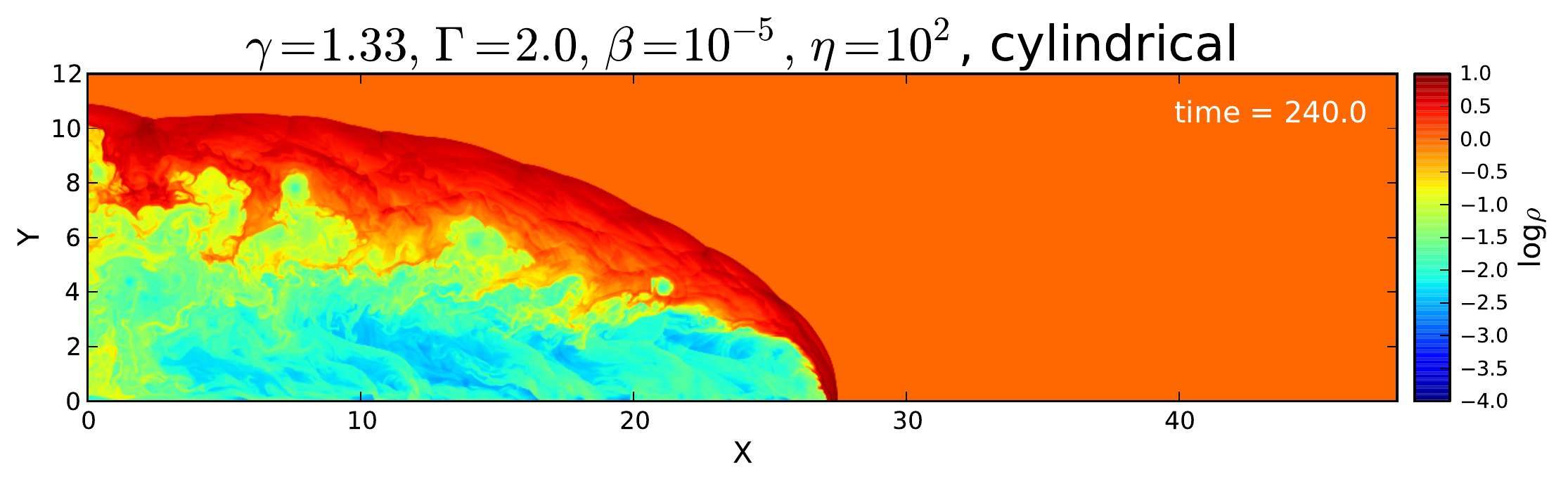}
 \includegraphics[width=0.45\textwidth]{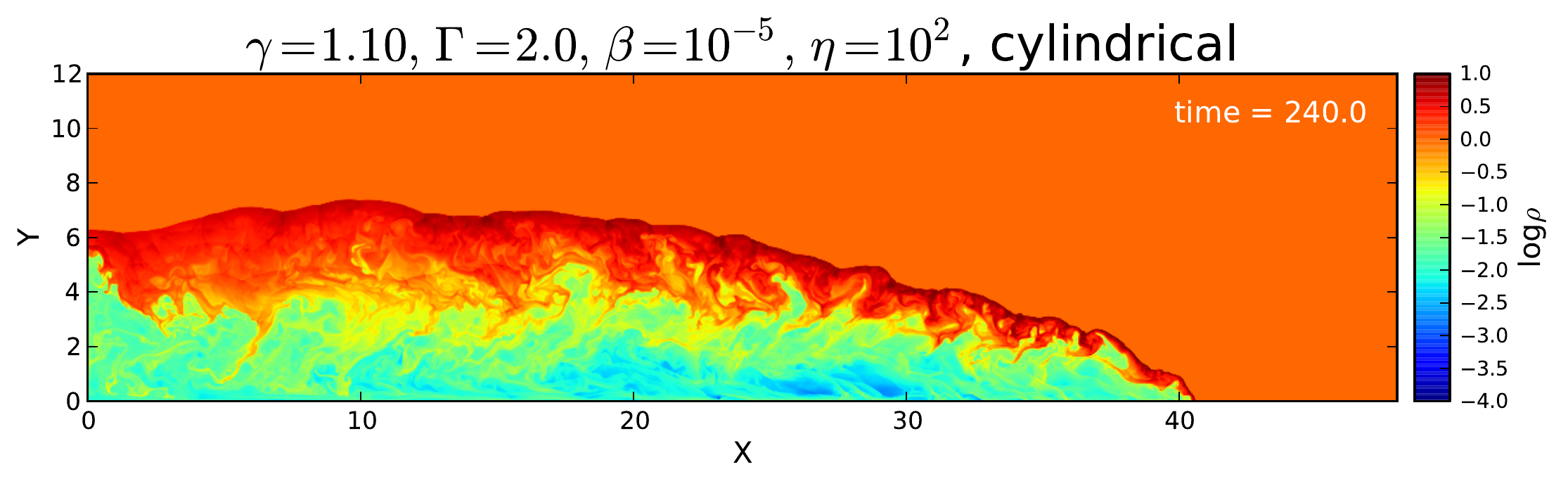}
 \includegraphics[width=0.45\textwidth]{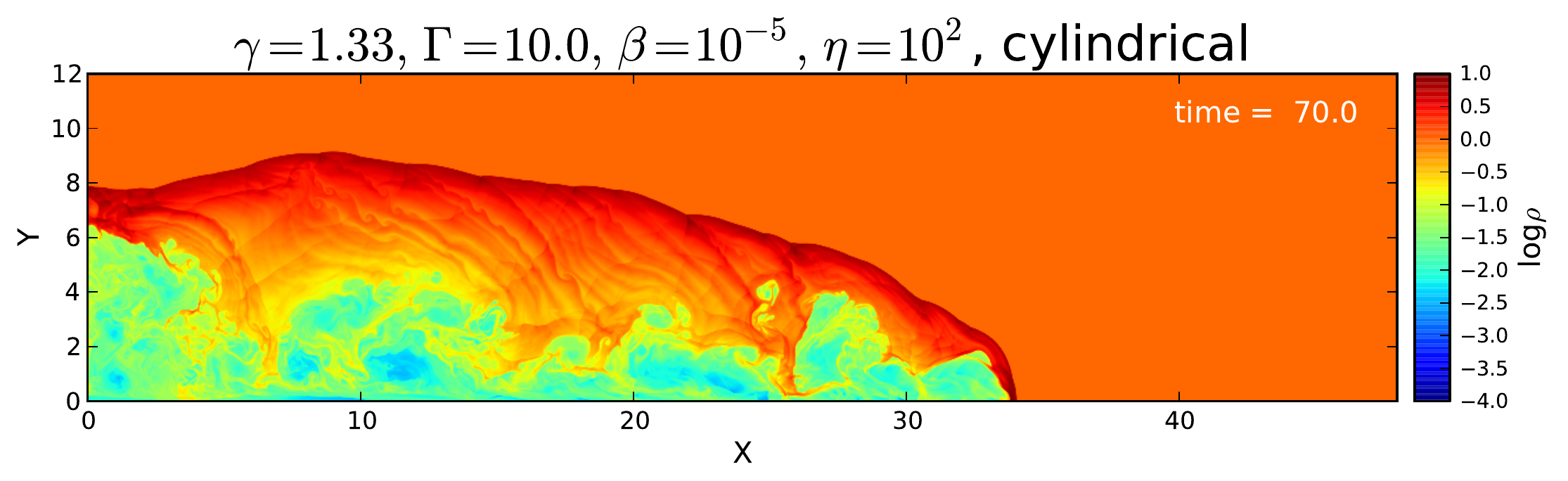}
 \includegraphics[width=0.45\textwidth]{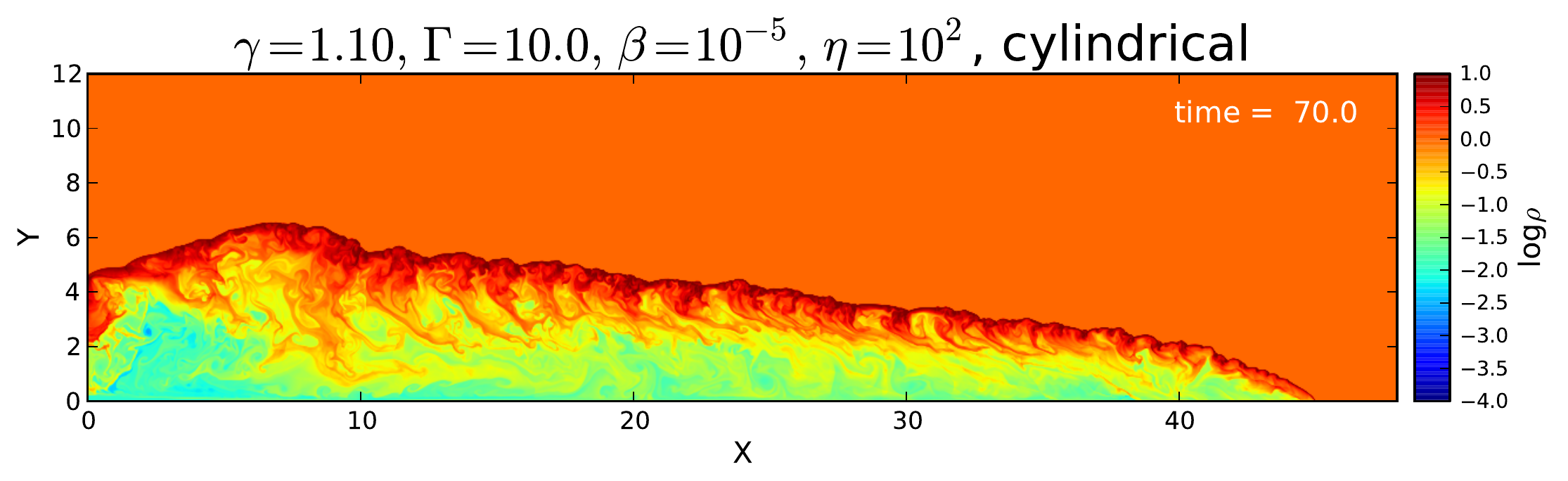}
 \includegraphics[width=0.45\textwidth]{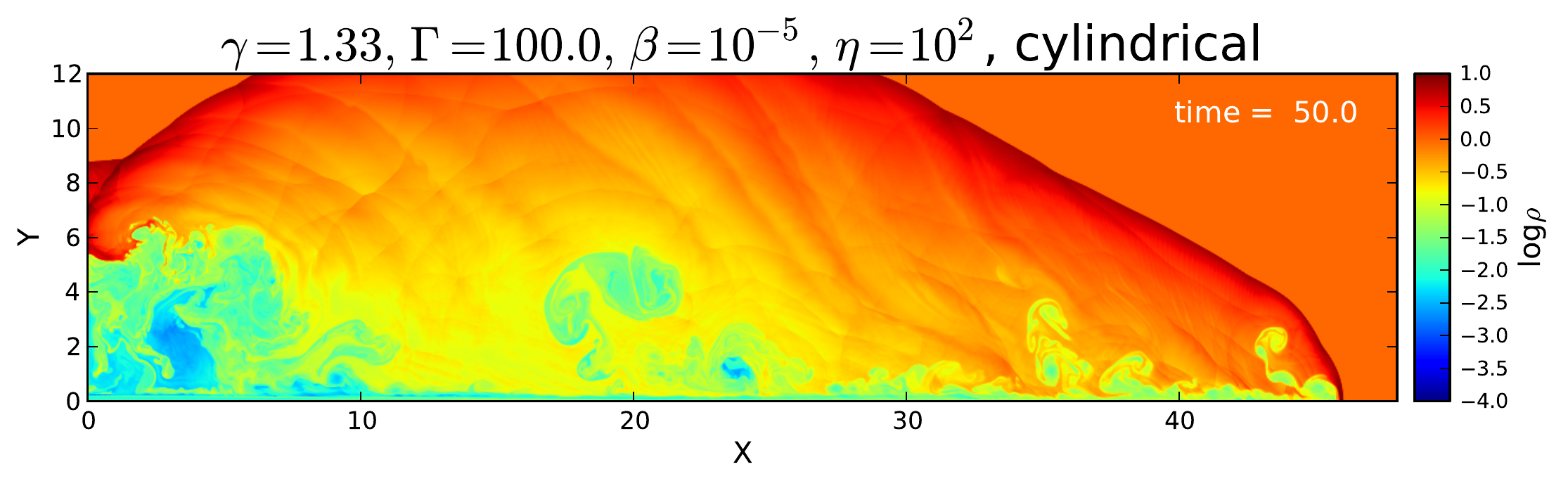}
 \includegraphics[width=0.45\textwidth]{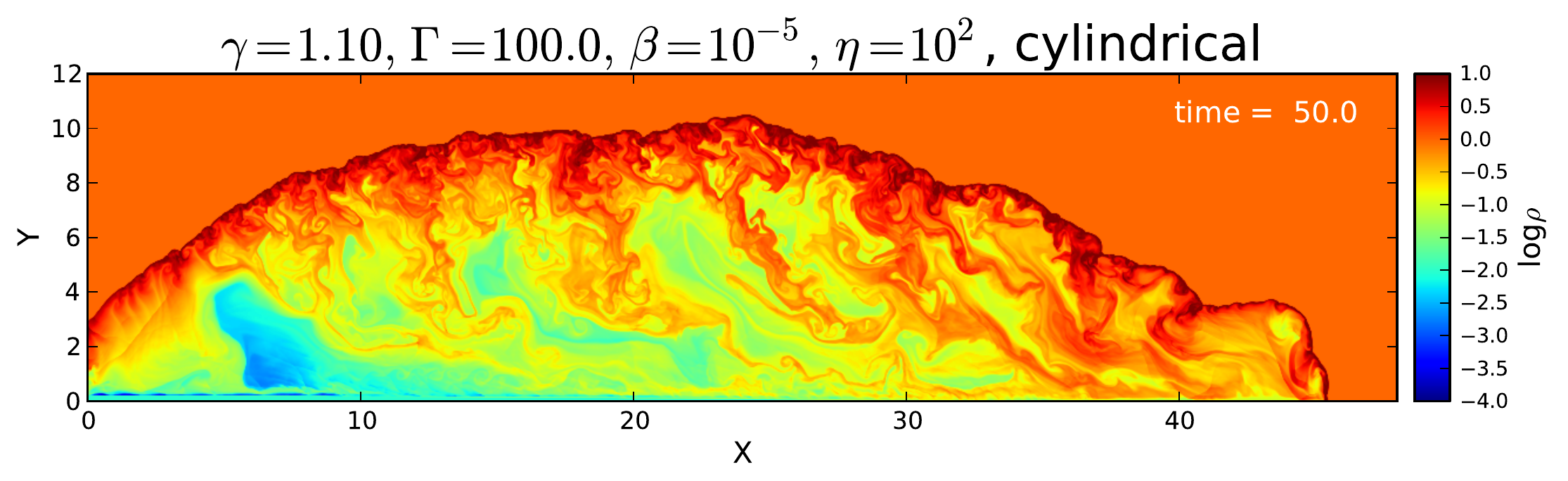}
 \includegraphics[width=0.45\textwidth]{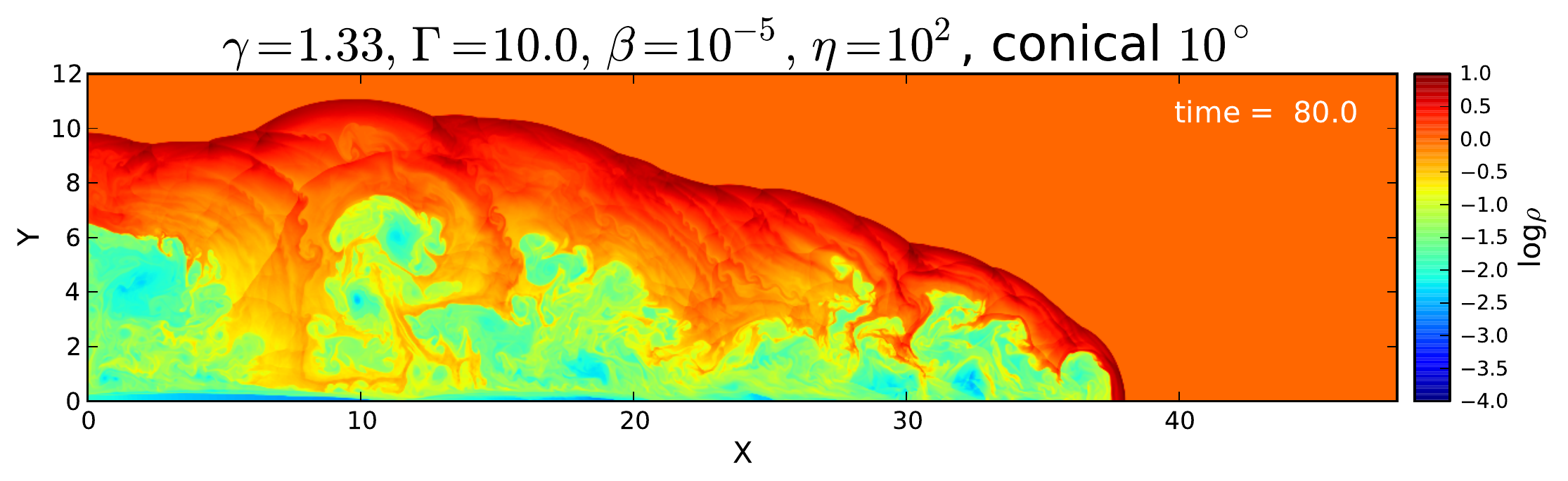}
 \includegraphics[width=0.45\textwidth]{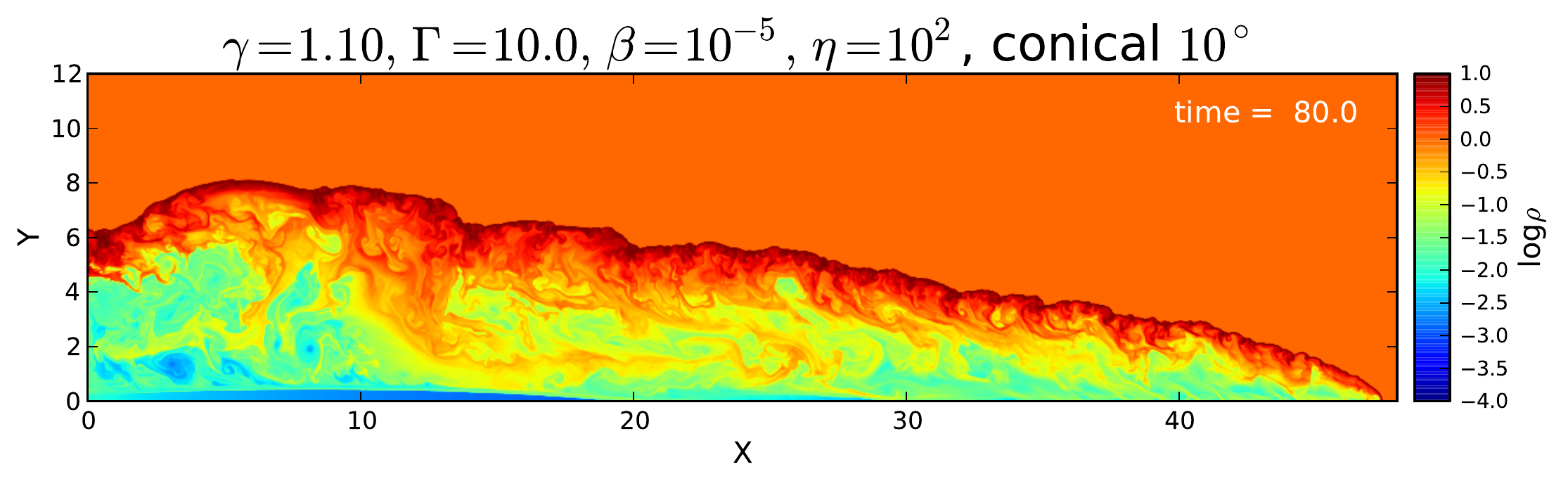}
 \includegraphics[width=0.45\textwidth]{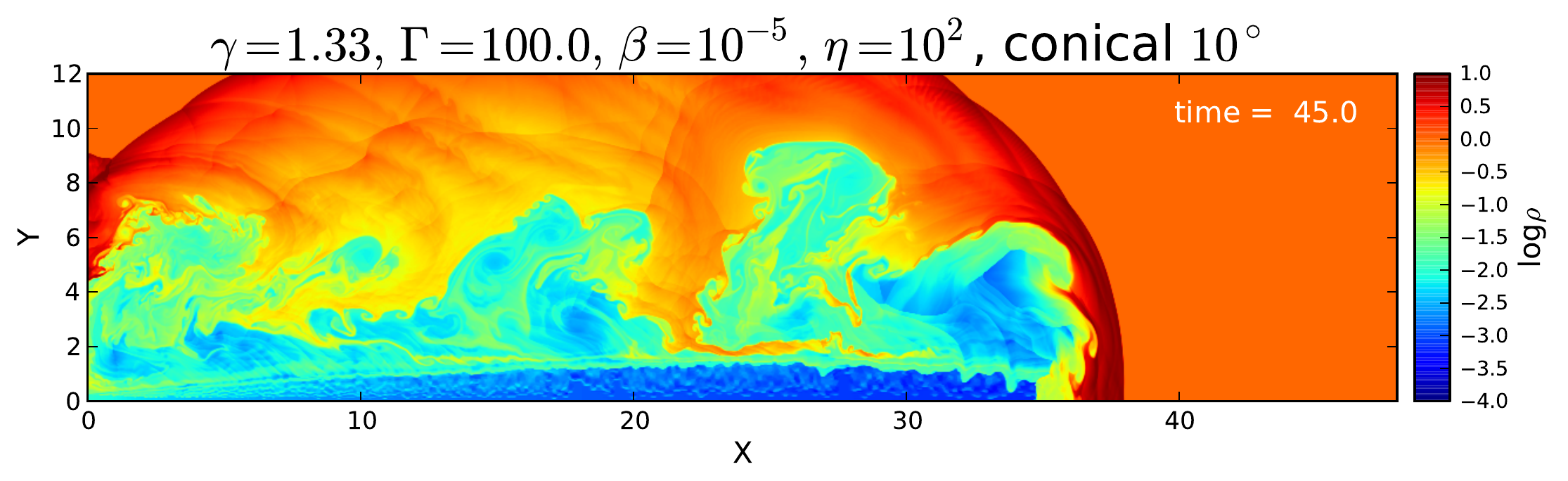}
 \includegraphics[width=0.45\textwidth]{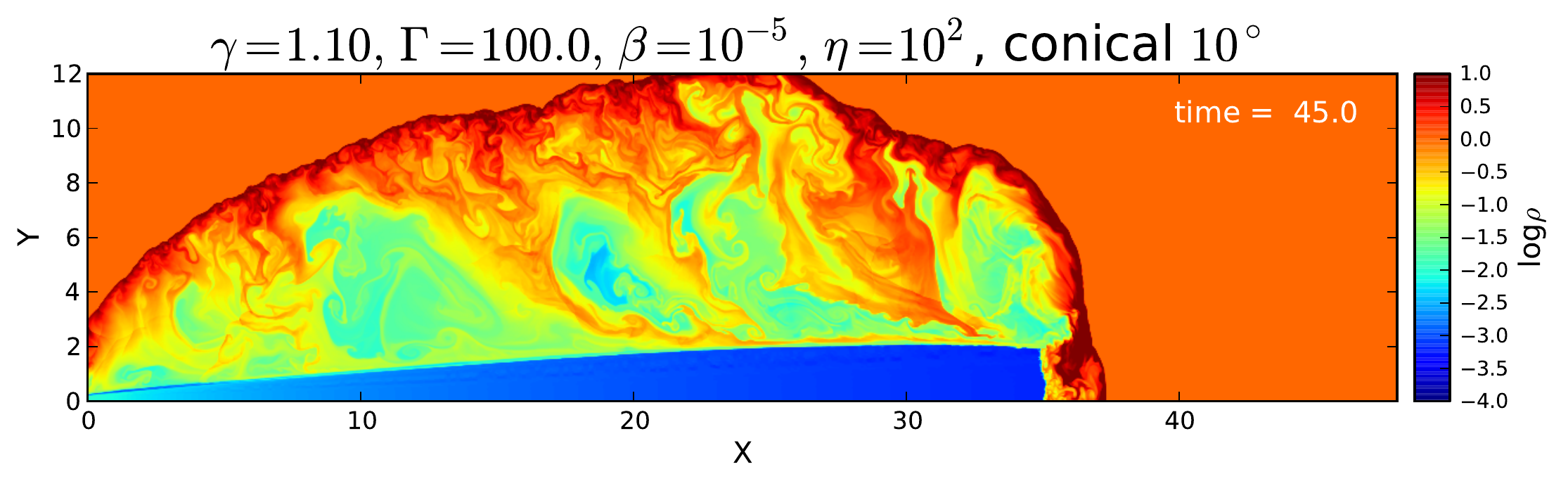}
 \caption{Distribution of logarithm of density for different light jet models with $\eta = 10^2$. Left column maps
 represent adiabatic ($\gamma = 4/3$) simulations and the non-adiabatic ($\gamma_{\rm eff} = 1.1$) models are shown in the right column. Models were run with $\Gamma = 2$, $\Gamma = 10$
 and $\Gamma = 100$ and the geometry was tested both for collimated ( $\theta \rightarrow 0$) and wide jets. Numbers represent the time of the given snapshot in the simulation in code units.}
\label{fig:density}
}
\end{figure*}

In Fig.\ref{fig:density} we present snapshots of the density distribution for the adiabatic
($\gamma = 4/3$) (left column) and  non-adiabatic ($\gamma_{\rm eff} = 1.1$) (right column) models of Table
\ref{tab} with different Lorentz factors and jet opening angles.

All models evidence the formation of a bow shock structure as the jet head sweeps the ambient gas.
The ambient shocked material is deposited into a cocoon that surrounds the beam. Although not obvious
in the snapshots shown in Fig.\ref{fig:density}, a double shock structure soon develops. Besides the
forward bow shock, a reverse internal shock decelerates the jet beam and shocked jet material is also
deposited into the internal part of the cocoon. The low density portion of the cocoon is of  jet shocked
material, while the denser one is composed of shocked ambient gas.

The interaction of the hot shocked gas of the cocoon with the beam material drives  Kelvin-Helmholtz (KH)
instabilities \citep[e.g.][]{birkinshaw96}  which in turn induce both the formation of internal shocks
pinching the beam and strong turbulent mixing and entrainment, as  detected in former numerical studies
of  non-relativistic jets \citep{dalpino93,chernin1994}.

Also, as expected from earlier studies of thermal radiative cooling jets
\citep[e.g.,][]{blondin90,dalpino93}, the  effects above are much stronger in the adiabatic jets (in the left
panels of Fig. \ref{fig:density}) since in these cases the internal energy of the  shocked material in the
cocoon is much larger than in their non-adiabatic counterparts (in the right side of the Fig.\ref{fig:density}).
In the latter cases, the enthalpy of the gas in the cocoon is much smaller due to the adoption of a $\gamma=1.1$
index to mimic thermal radiative cooling in the shocked  dense ambient gas.

We should remark  that in a more realistic calculation the effective value for $\gamma$ would be dependent on
the local properties of the plasma, and the cooling function. The adoption of a single value of $\gamma=1.1$
for the whole system in the case of the right side models of Fig. 2 is, therefore, a simplification and the
comparison with the adiabatic models should be taken with caution. These non-adiabatic models actually represent
extreme examples.

Models with higher Lorentz factor obviously reach the boundary of the spatial domain earlier and therefore,
look less evolved. The higher propagation velocity results a smaller loading of shocked jet material and larger
spreading of the shocked ambient gas into the cocoon, which makes the driving of shear KH instabilities and turbulent
entrainment less prominent than in lower Lorentz factor jets.

The  models with smaller Lorentz factor  ($\Gamma = 2$), specially the adiabatic one ($\gamma=4/3$), present
a cocoon with a larger portion of  low density shocked jet gas and smaller portion of high shocked ambient gas. This
is because the jet beam has not enough power to drill through the dense ambient gas and also retains much more shocked
jet material.

In the non-adiabatic models, the bow shock layer is thinner. The high velocity of the upstream jet flow interacting
with a thin layer gives rise to the Vishniac instability \citep{vishniac94} which breaks the layer and enhances the
growth of turbulence, particularly in the outer parts of the cocoon. It is also clear in these models the impact of
the turbulence on the diffusion and mixing of the gas in the shocked plasma.

The morphology and general properties of the density distributions, as described above, do not differ much for
conical jets. This is expected since the cocoon
pressure readily becomes important in the case of these low density jets, i.e. with $\eta<1$, resulting in similar dynamics once
the jet is collimated by the cocoon.

The maximum density at the shock region is also dependent on  the  parameters $\gamma$ and $\Gamma$.
Fig. \ref{fig:density} indicates that larger Lorentz factors result in larger shocked densities which is
consistent with the relativistic RH jump conditions (see eqs. \ref{RH1} to \ref{RH4}). Also, the effective
thermal radiative cooling introduced in the non-adiabatic jets by  decreasing $\gamma$ to 1.1 is expected,
according to the jump conditions, to increase the density of the compressed shocked material at the same
time that it decreases its internal pressure. The pressure reduction of the downstream gas also  decreases
the velocity at which it is pushed outwards. All these effects are  detected in Fig.\ref{fig:density} and are
compatible with earlier studies of non-relativistic radiative cooling jets \citep{blondin90,dalpino93,stonenorman93}.
We show in Fig.\ref{fig:profiles} the Lorentz factor, density and magnetic field amplification factor at the jet
axis, along y=0, for models with $\eta= 10^{-2}$.

\subsubsection{$\eta < 1.0$ (heavy jets)}

After  the jet breaks the stellar envelope it may expand over an underdense interstellar medium. In this case,  the jet density may eventually become orders of magnitude larger than that of the ambient gas.
Numerical issues constrain the density contrast of the simulations, which has been fixed here to a minimum possible value $\eta = 10^{-4}$.
The larger density of the jet makes it easier to expand over the ambient medium, thus reducing and delaying certain shock effects, such as
the development of a large pressure cocoon.

\begin{figure*}
 \center
{
 \includegraphics[width=0.45\textwidth]{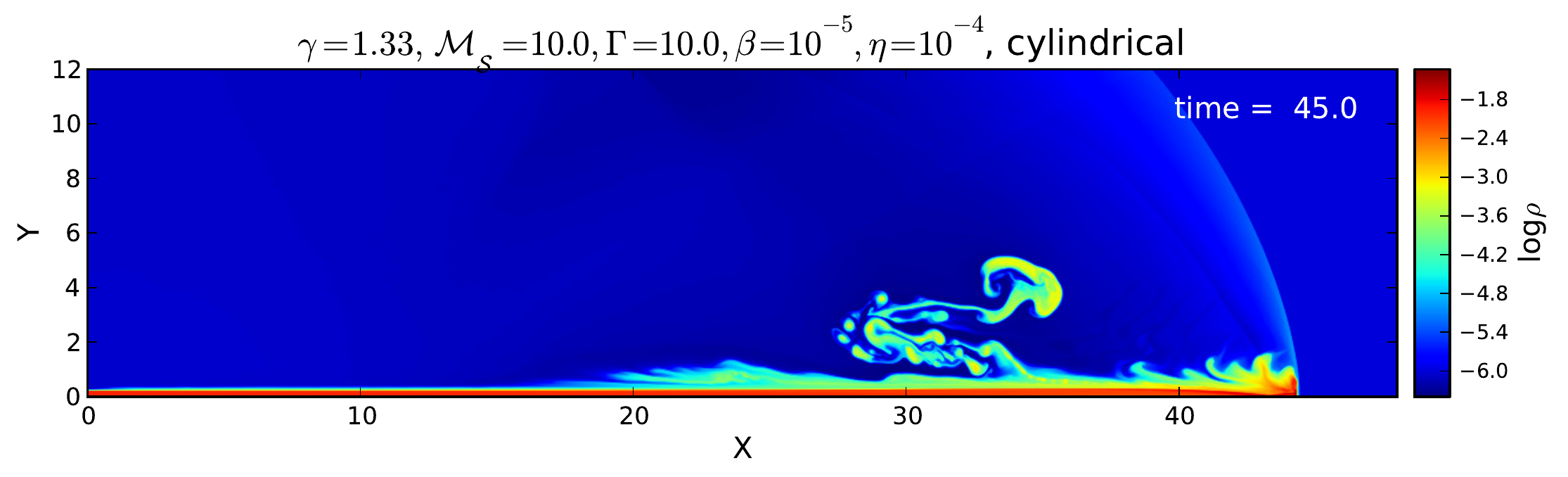}
 \includegraphics[width=0.45\textwidth]{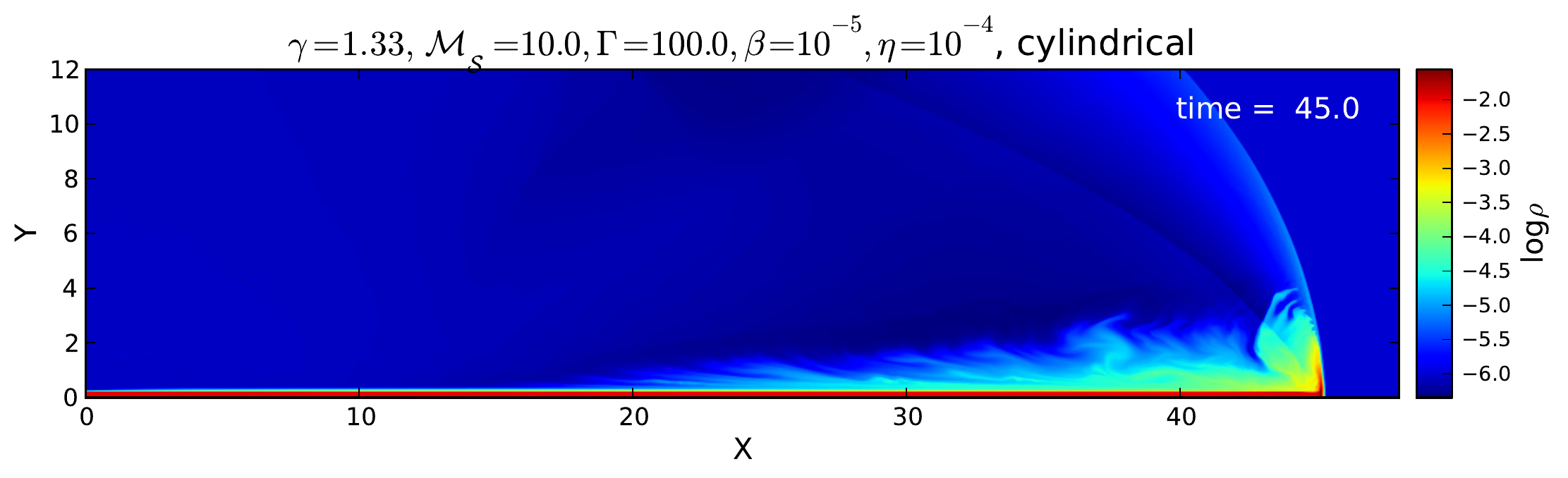}
 \includegraphics[width=0.45\textwidth]{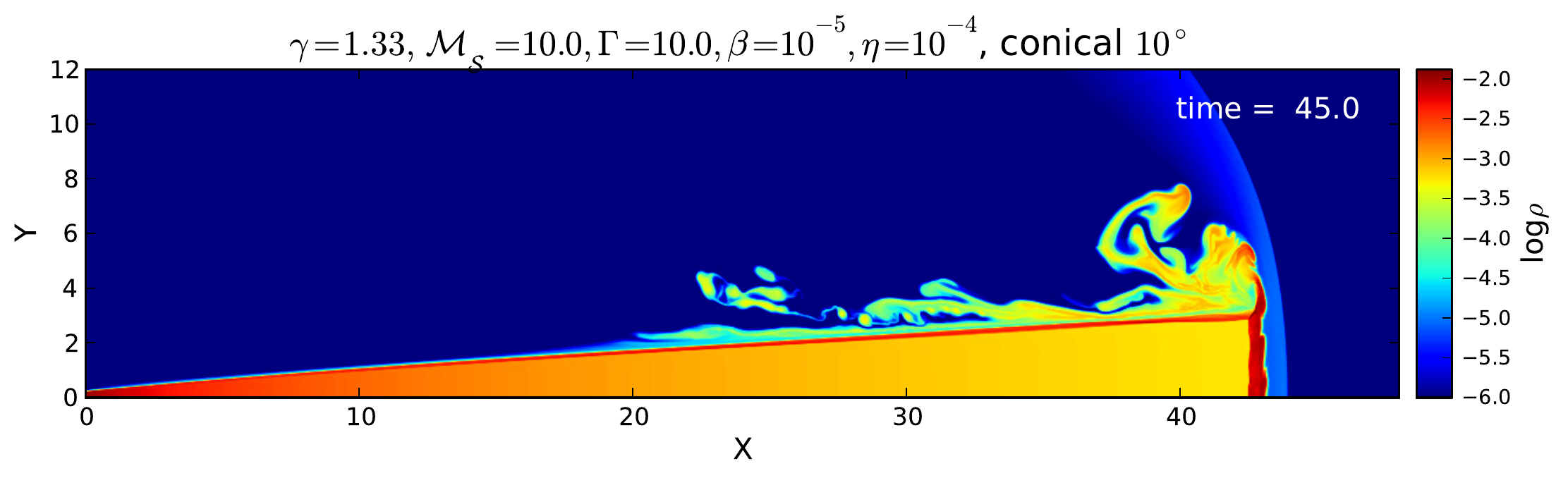}
 \includegraphics[width=0.45\textwidth]{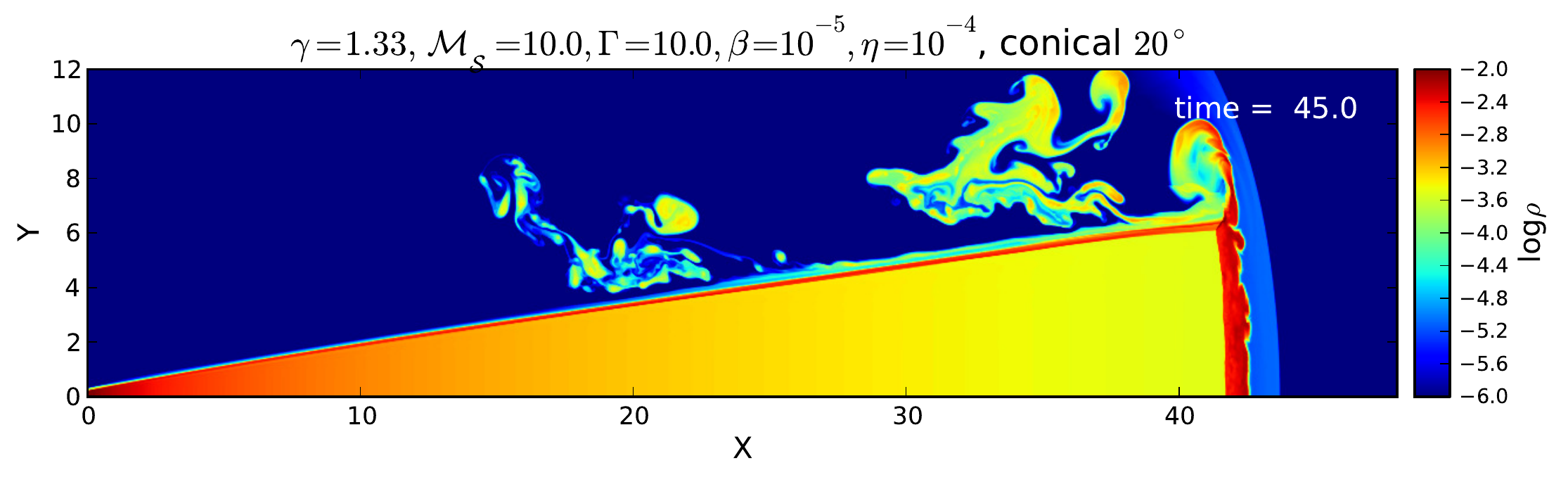}
 \includegraphics[width=0.45\textwidth]{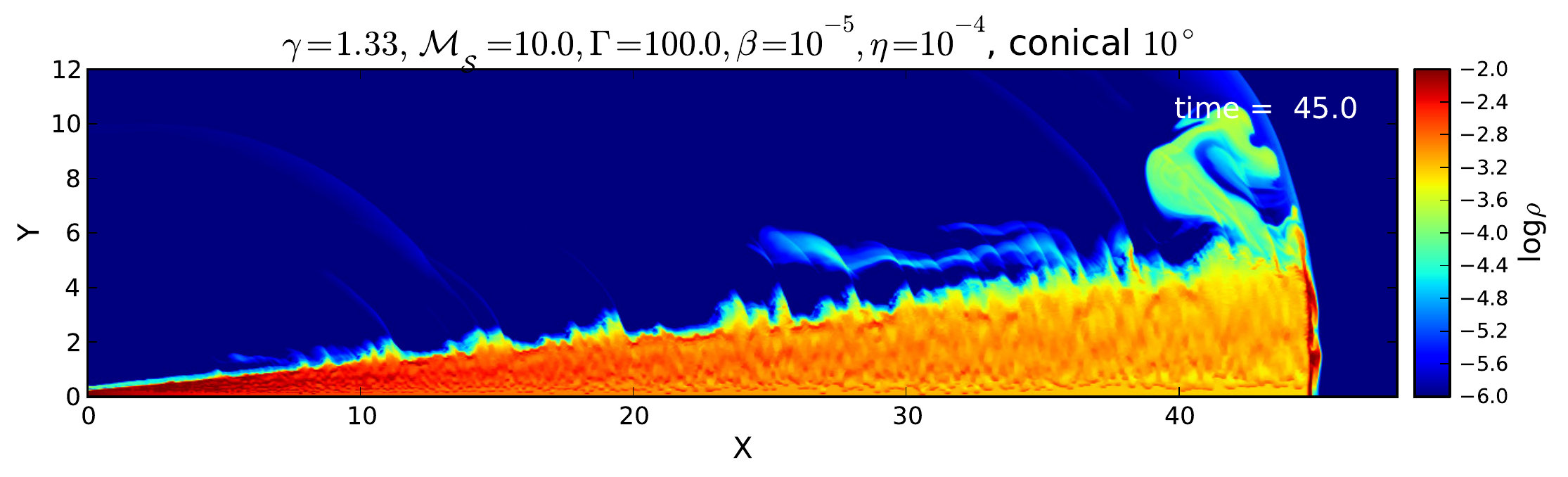}
 \includegraphics[width=0.45\textwidth]{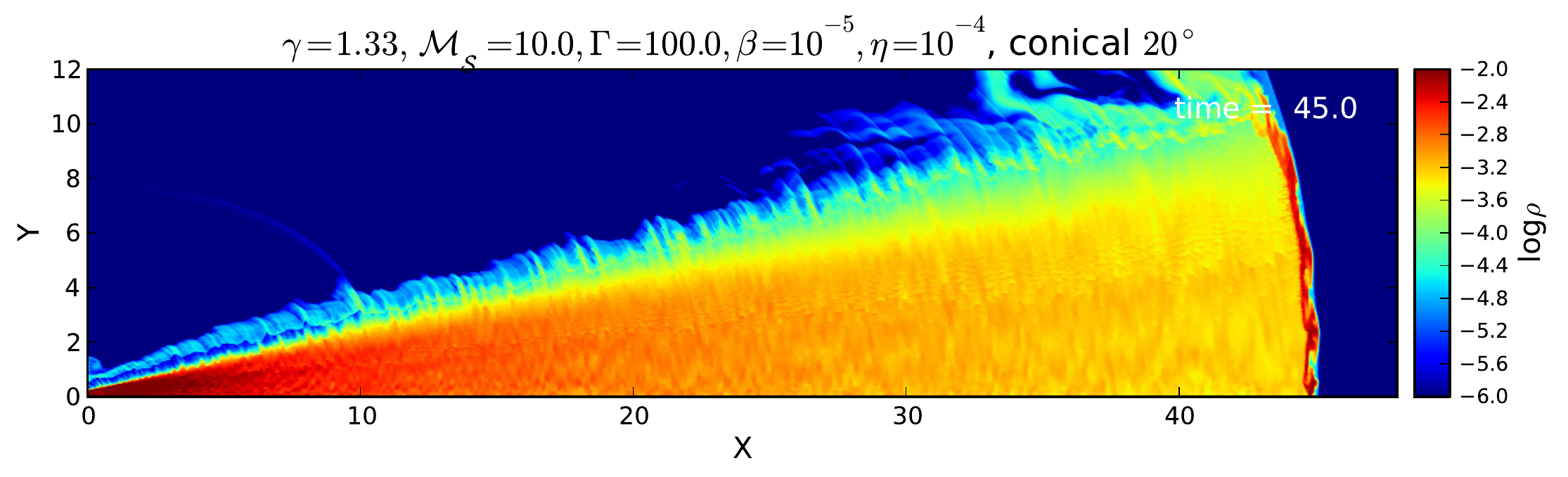}
 \includegraphics[width=0.45\textwidth]{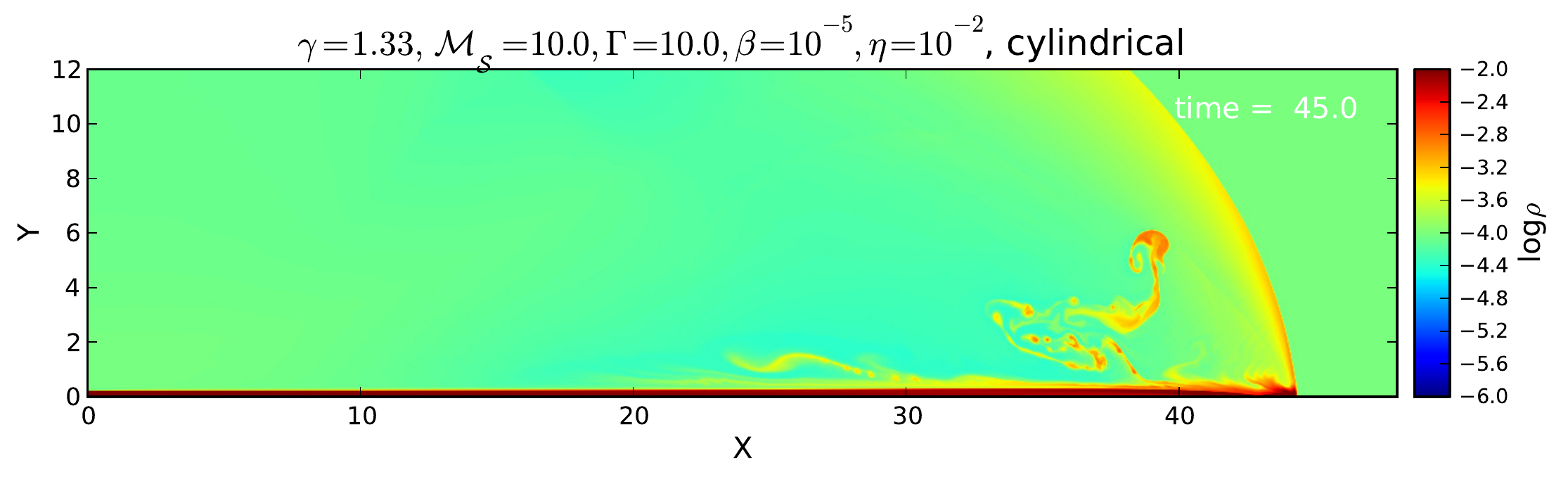}
 \includegraphics[width=0.45\textwidth]{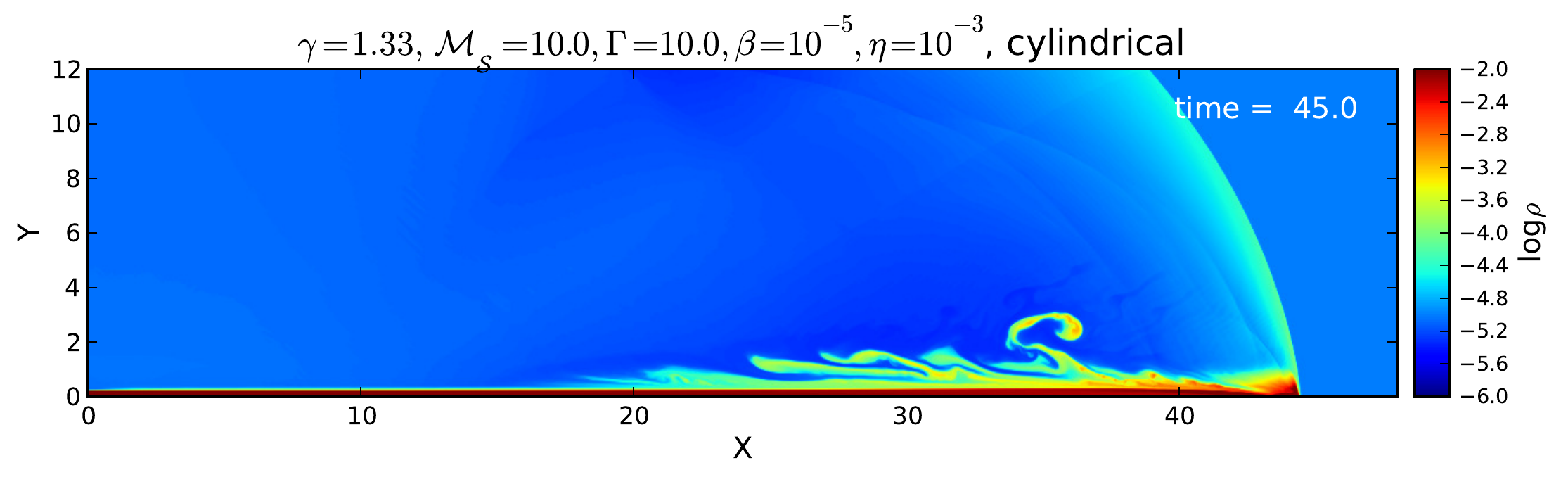}
 \caption{Distribution of logarithm of density for  different adiabatic, heavy jet models (with $\eta <1$). Models were run with $\Gamma = 10$ and $\Gamma = 100$ and the geometry was tested both for collimated ( $\theta \rightarrow 0$) and wide jets, with the opening angle varying between $0^{\circ}$, $10^{\circ}$ and $20^{\circ}$. Numbers in white represent the time of the given snapshot in the simulation in code units.}
\label{fig:eta}
}
\end{figure*}

We can see in  Fig.\ref{fig:eta} that the morphology of the jet changes substantially with the increase
in the jet density. For very dense jets the ambient pressure is negligible and have little impact on the propagation of
the jet. The shock region shows less turbulence compared to the low density jets.

Fig.\ref{fig:profiles2} depicts the Lorentz factor, density and magnetic field amplification factor at the jet
axis, along y=0, for models with $\eta= 10^{4}$.

\begin{figure*}
\center
{
\includegraphics[width=0.45\textwidth]{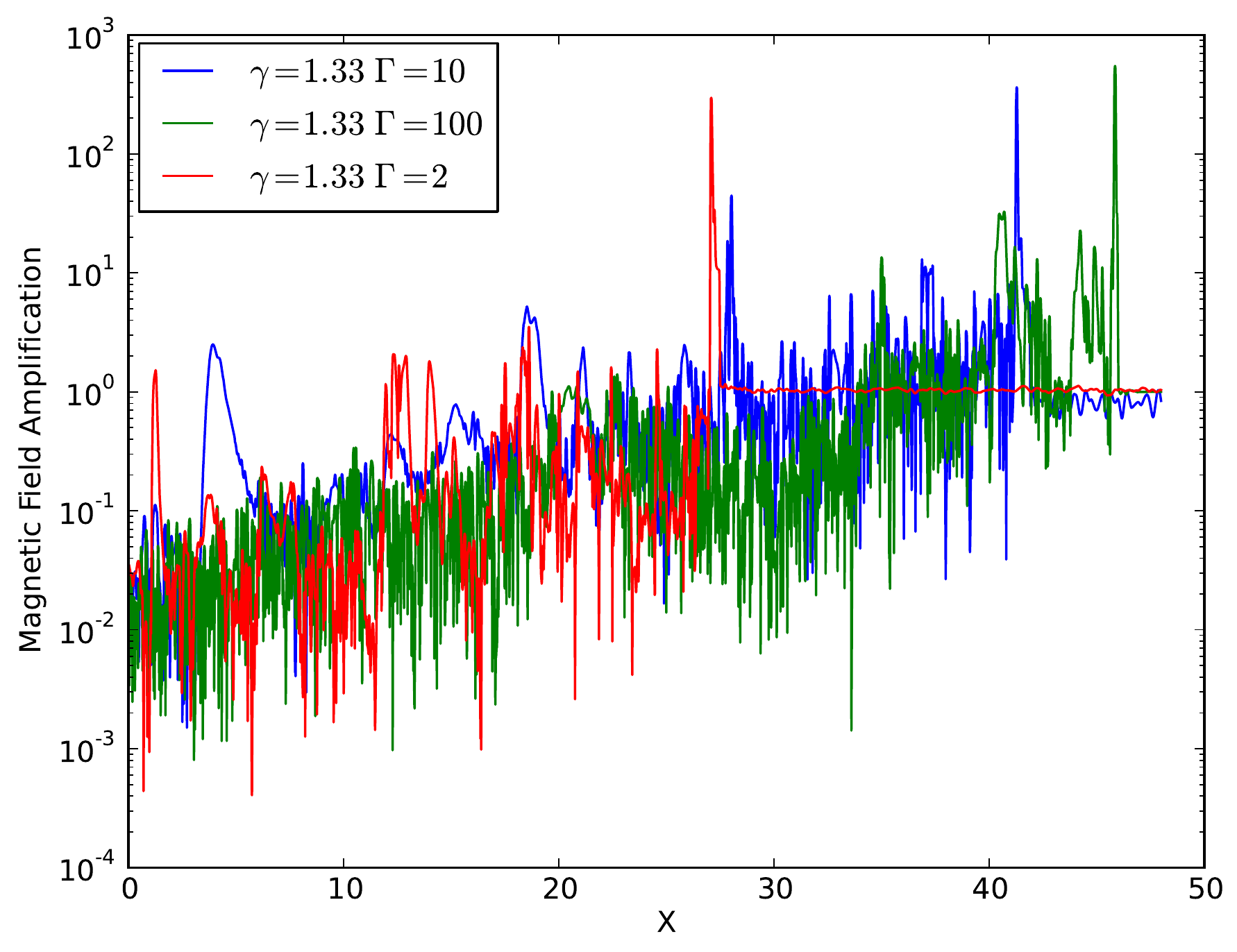}
\includegraphics[width=0.45\textwidth]{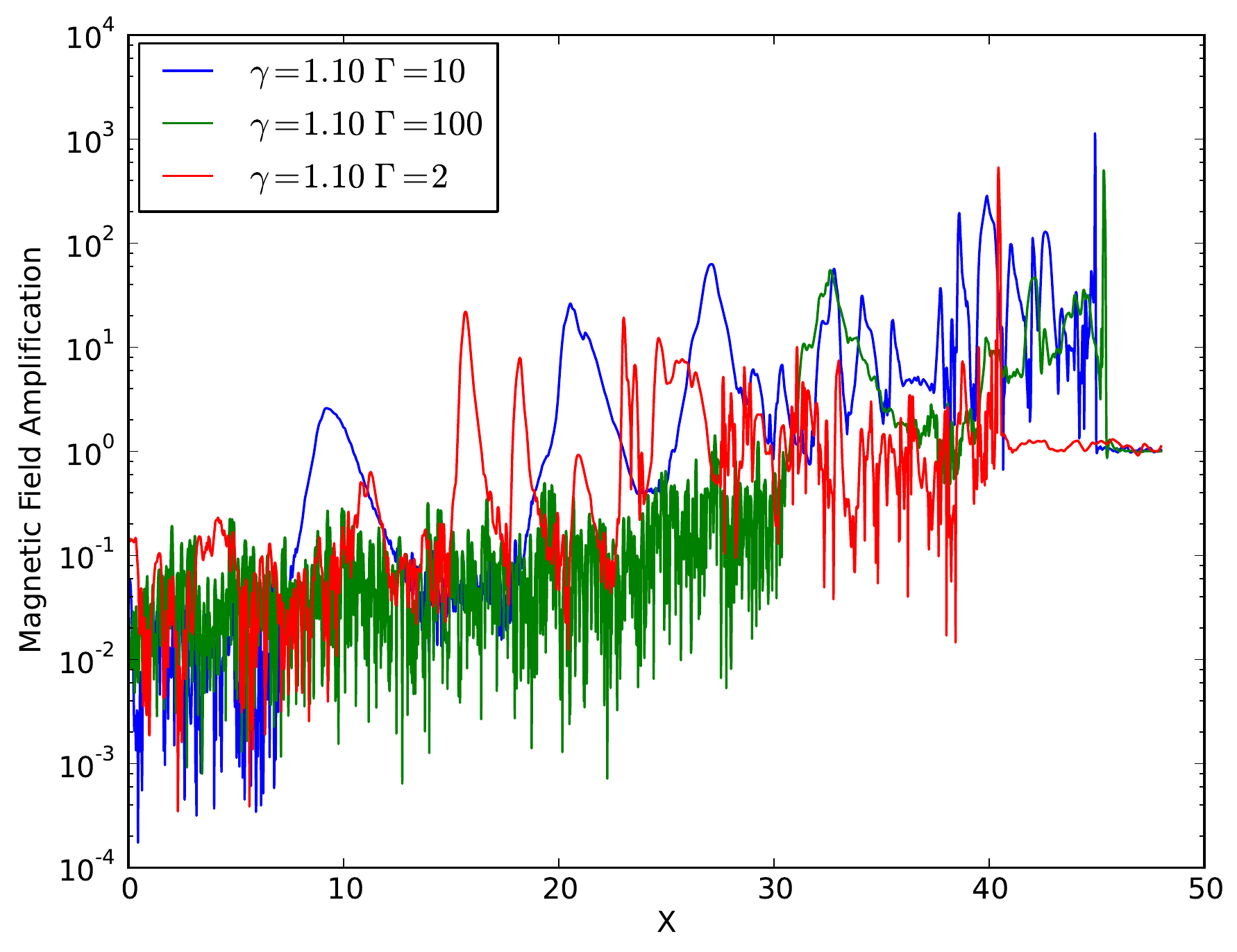}
\includegraphics[width=0.45\textwidth]{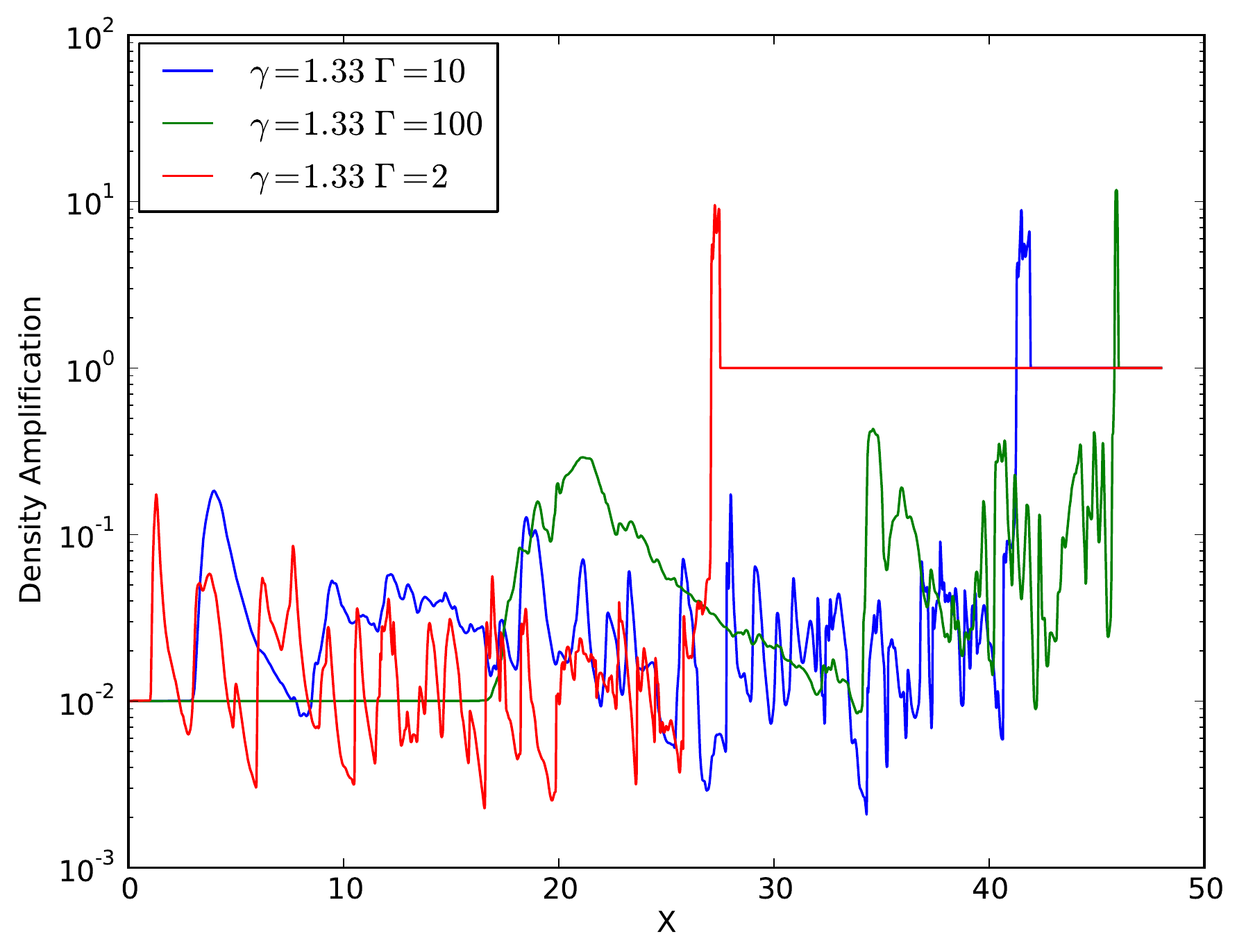}
\includegraphics[width=0.45\textwidth]{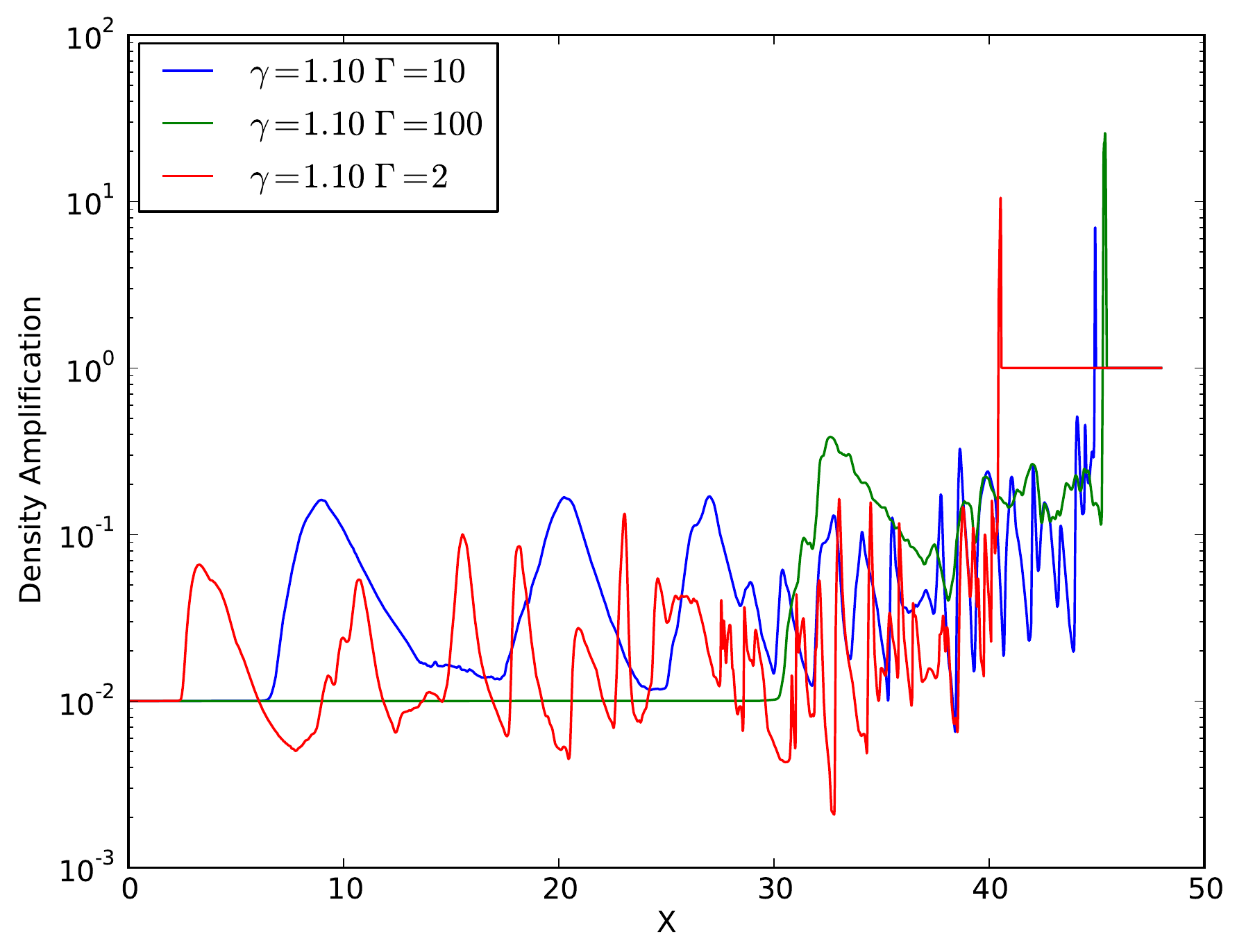}
\includegraphics[width=0.45\textwidth]{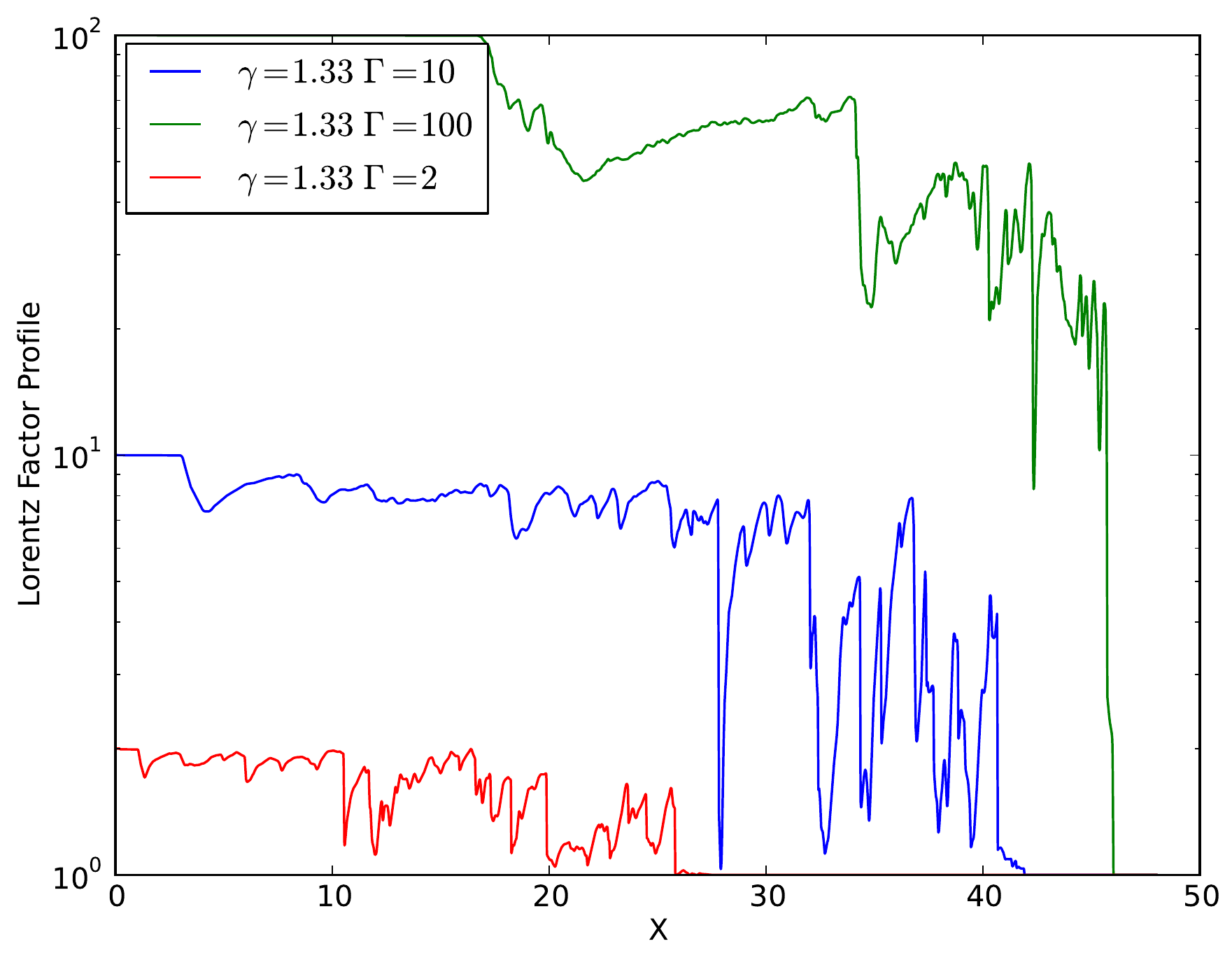}
\includegraphics[width=0.45\textwidth]{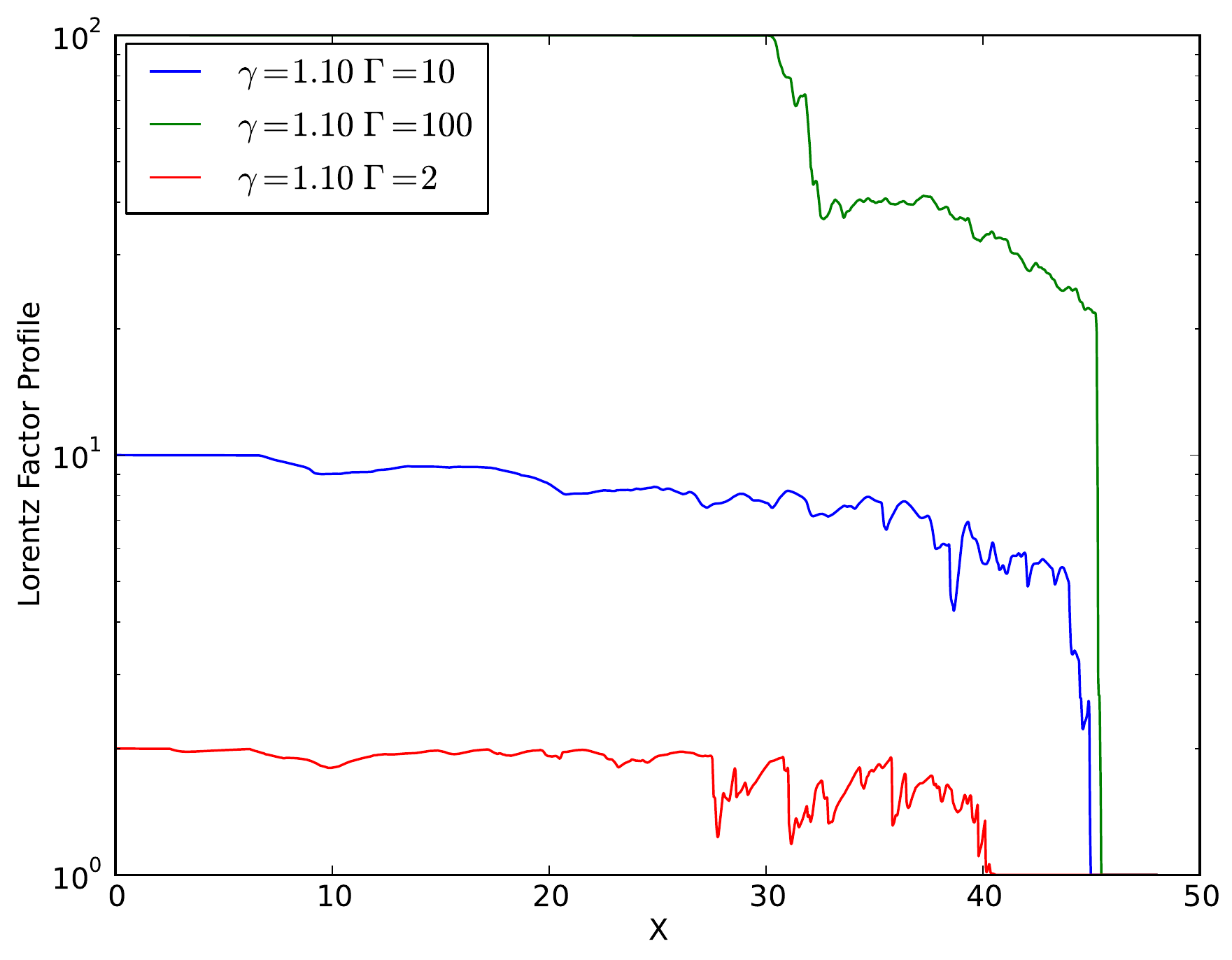}
\caption{Profiles of the magnetic field amplification factor, the density amplification
factor and the evolution of the Lorentz factor, obtained at $y=0$, for the $\gamma = 4/3$ 
and $1.1$ jet models, with $\eta=10^2$.}
\label{fig:profiles}
}
\end{figure*}

\begin{figure}
\center
{
\includegraphics[width=0.45\textwidth]{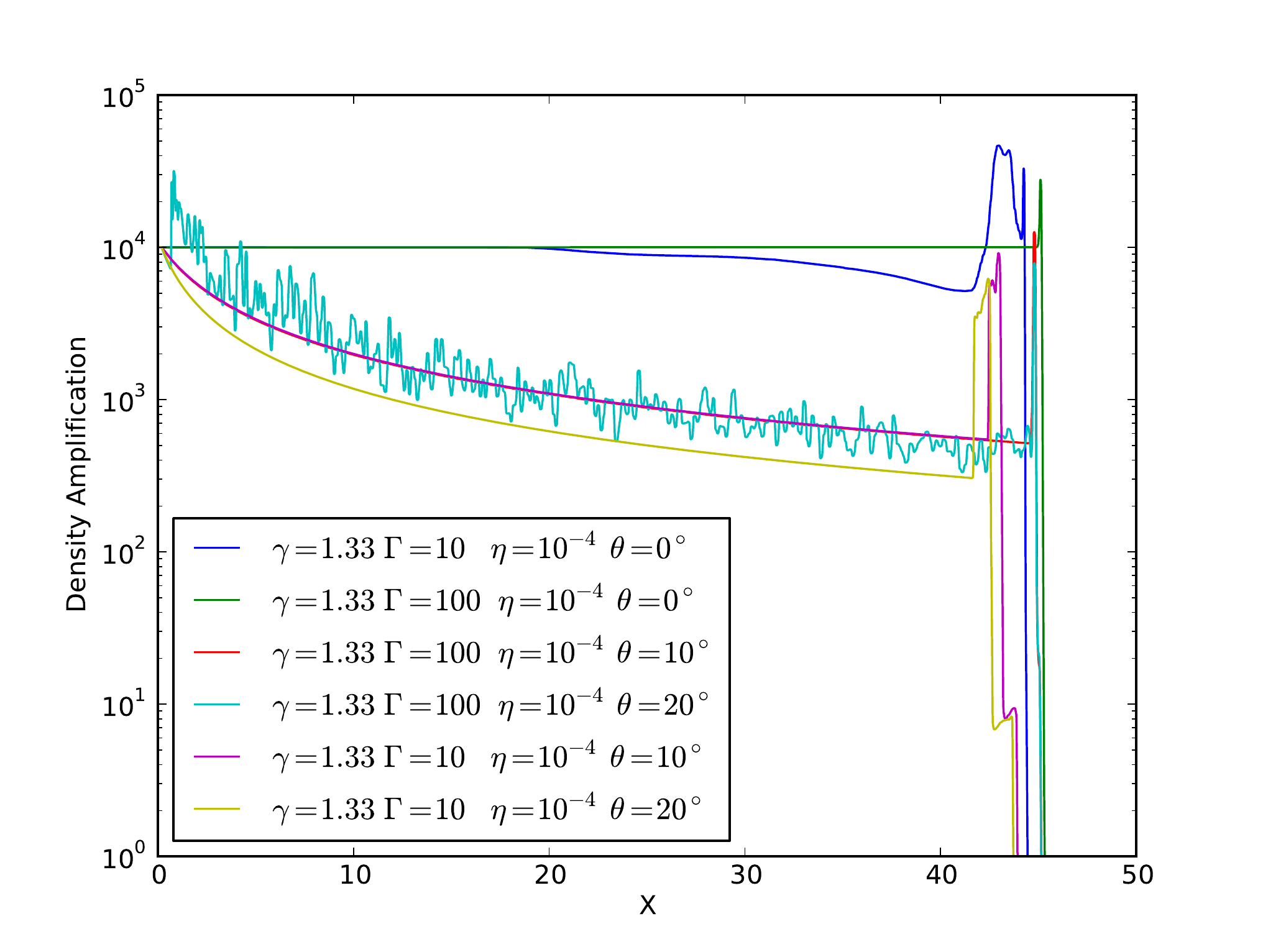}
\includegraphics[width=0.45\textwidth]{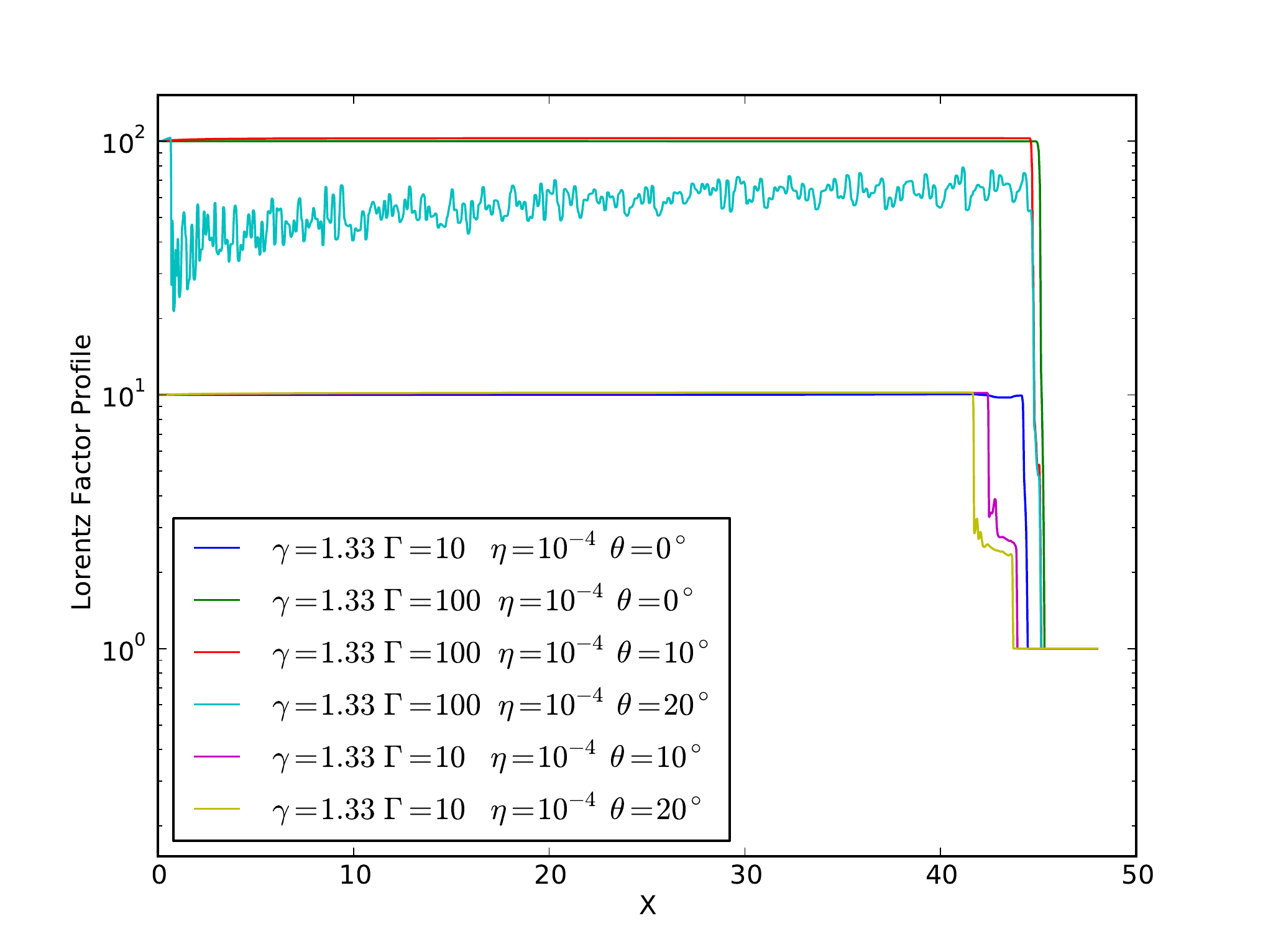}
\includegraphics[width=0.45\textwidth]{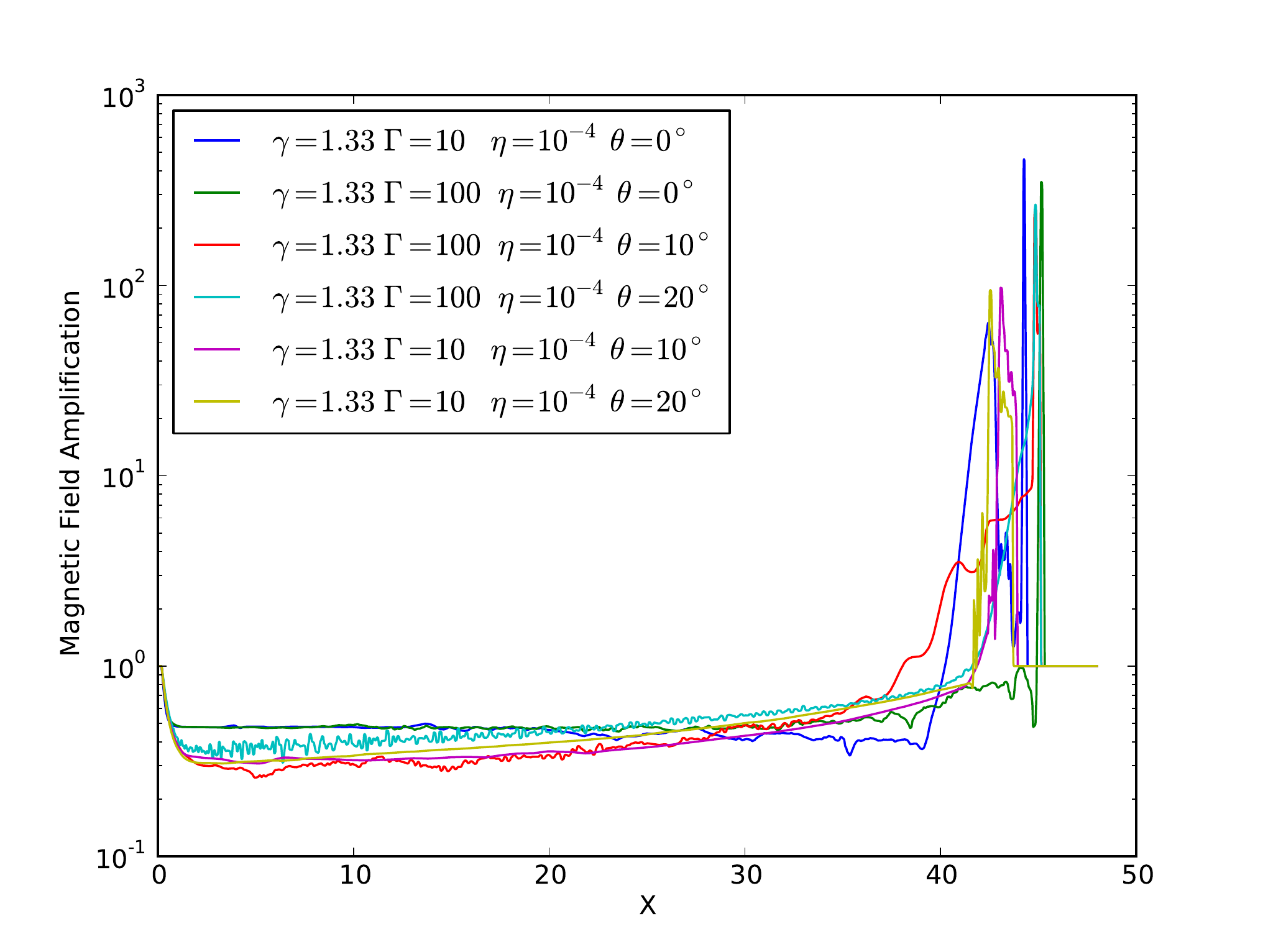}
\caption{Profiles of the magnetic field amplification factor, the density amplification
factor and the evolution of the Lorentz factor, obtained at $y=0$, for the adiabatic 
($\gamma = 4/3$) heavy jet models with $\eta=10^{-4}$.}
\label{fig:profiles2}
}
\end{figure}

The comparison of  Figs.\ref{fig:profiles} and \ref{fig:profiles2} (see also Figs. \ref{fig:density} and \ref{fig:eta}) indicate that the density amplification in the interface between the cocoon and the external medium is larger with increasing density ratio $\eta$. An important fact is that for $\eta \gg 100$ the jet is too light and has little momentum to push the ambient gas. In this case the jet decelerates  quickly and does not evolve to larger radii, as already pointed out in \citet{marti1997}.

In the next subsection  we will see in detail how these parameters affect the spatial distribution of the magnetic field.

\subsection{Magnetic Energy}

\begin{figure*}
 \center
{
 \includegraphics[width=0.45\textwidth]{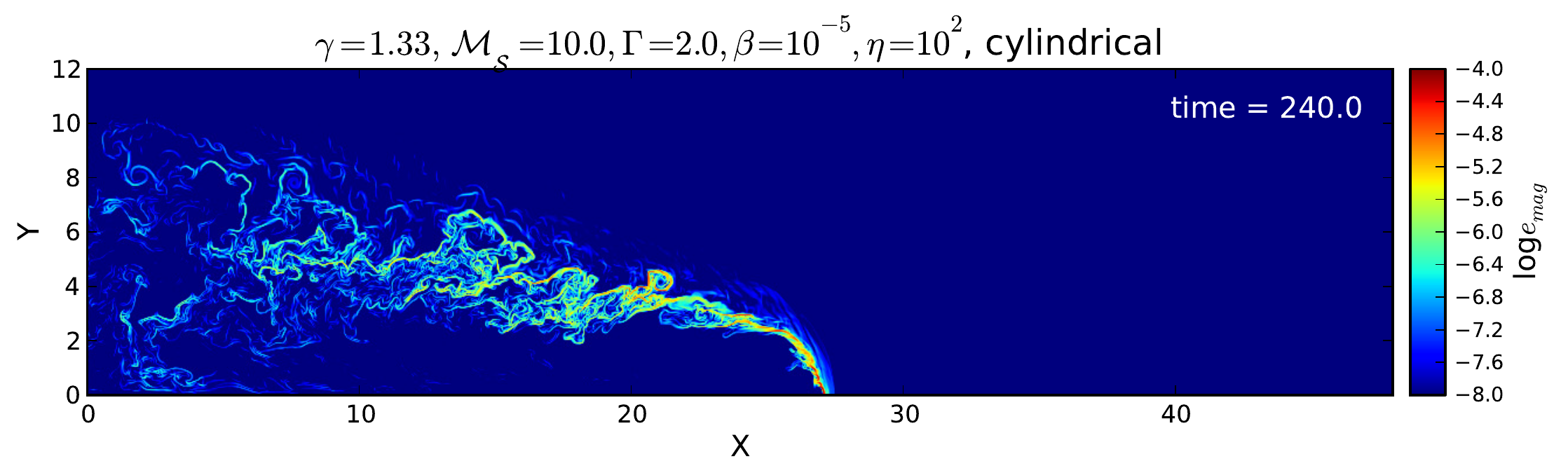}
 \includegraphics[width=0.45\textwidth]{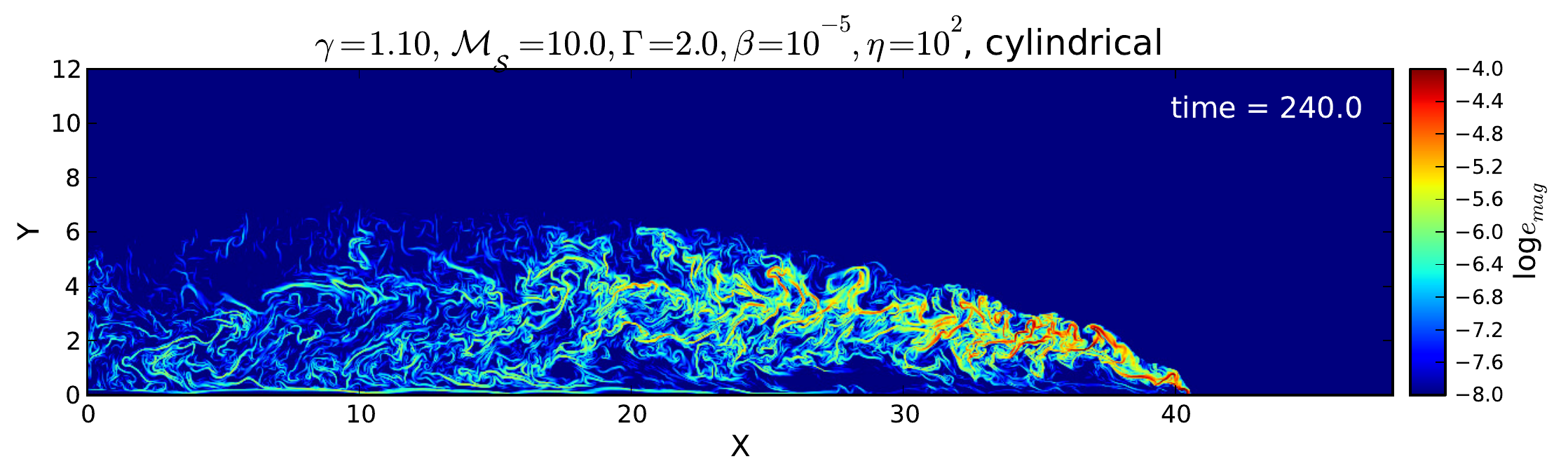}
 \includegraphics[width=0.45\textwidth]{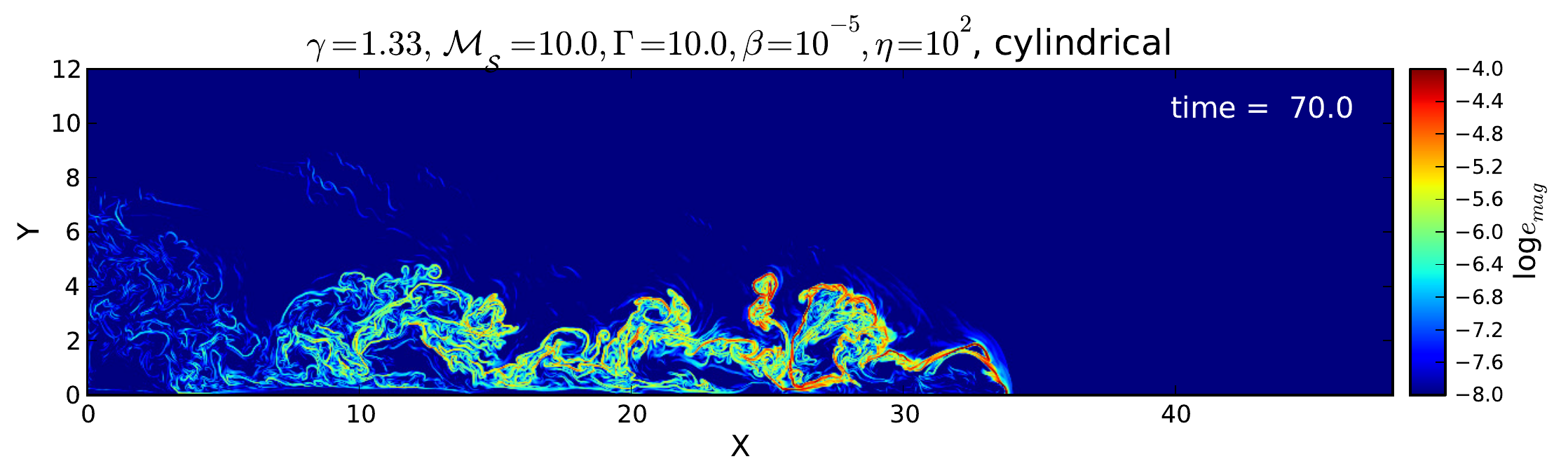}
 \includegraphics[width=0.45\textwidth]{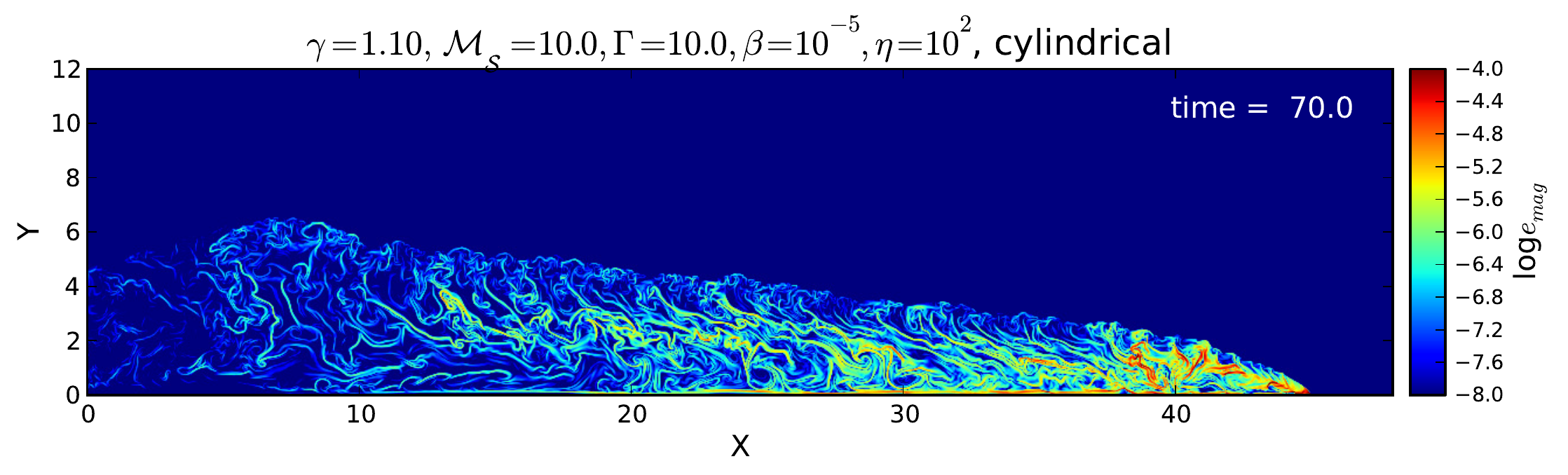}
 \includegraphics[width=0.45\textwidth]{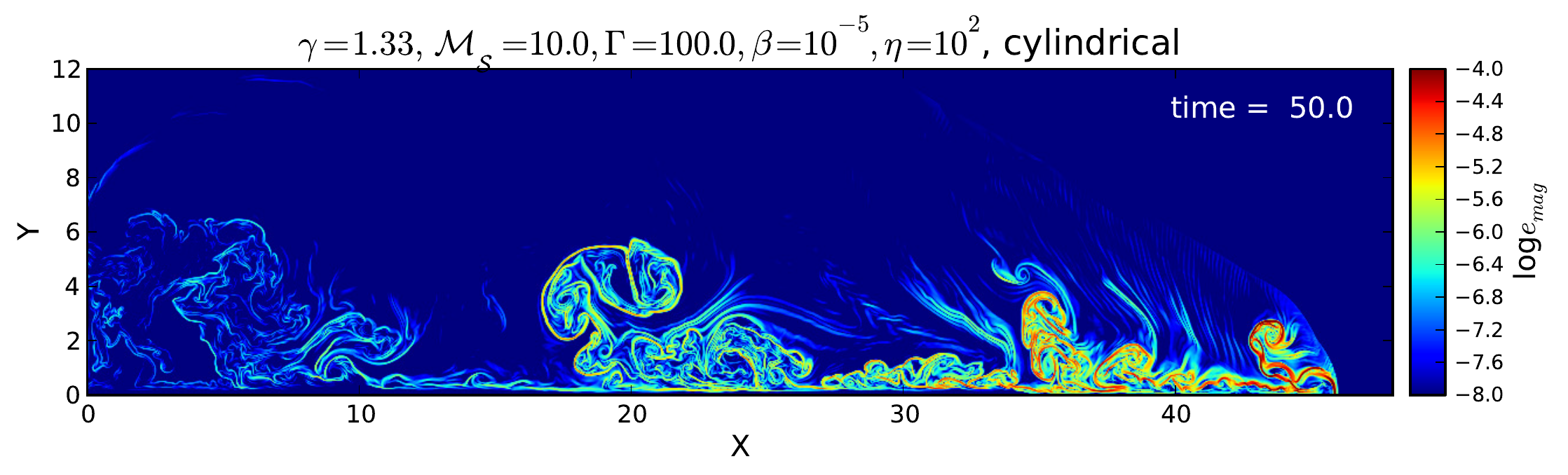}
 \includegraphics[width=0.45\textwidth]{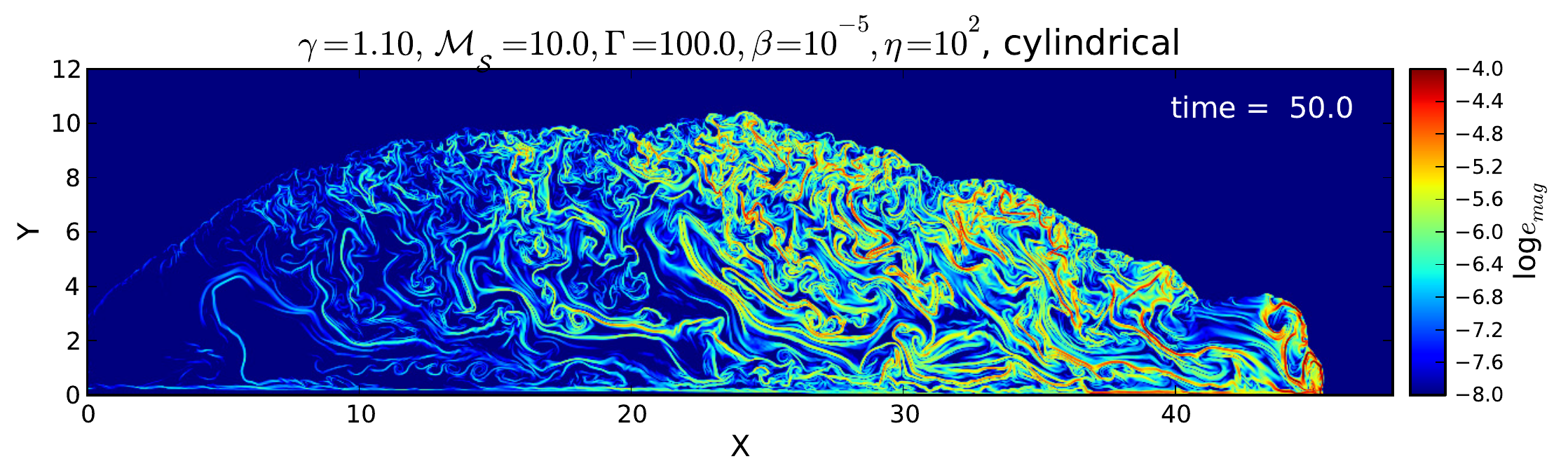}
 \includegraphics[width=0.45\textwidth]{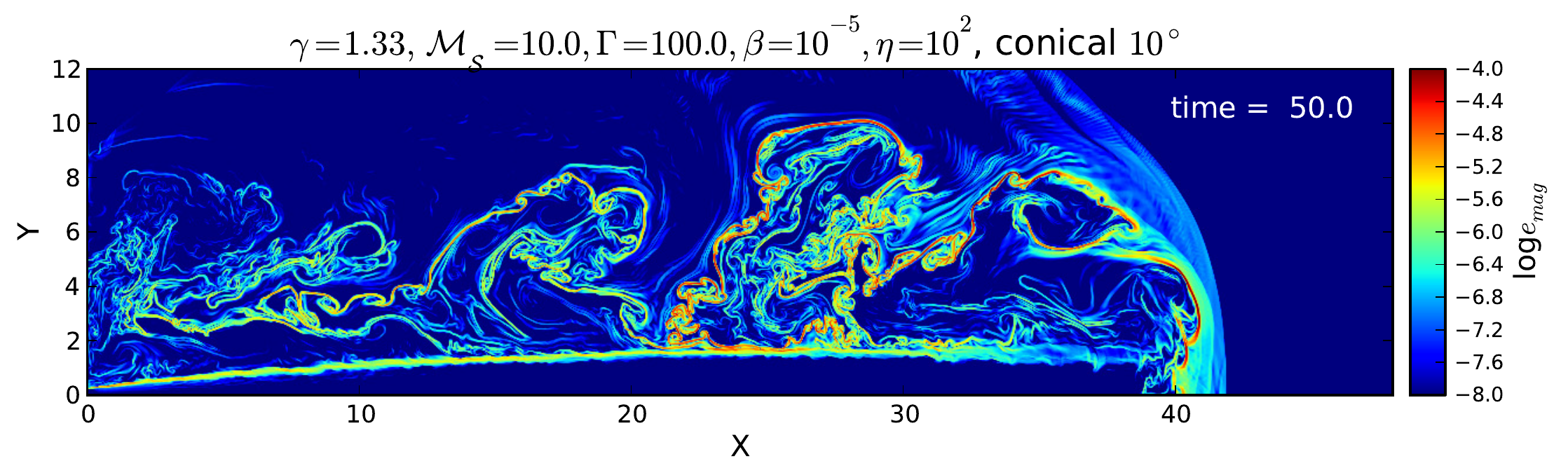}
 \includegraphics[width=0.45\textwidth]{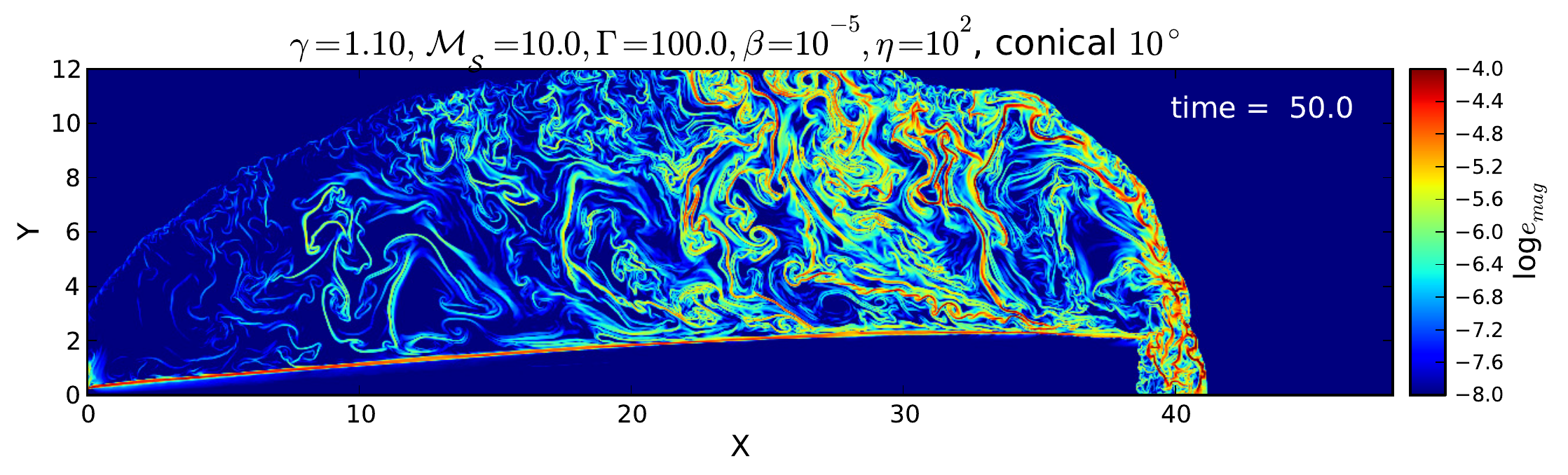}
 \includegraphics[width=0.45\textwidth]{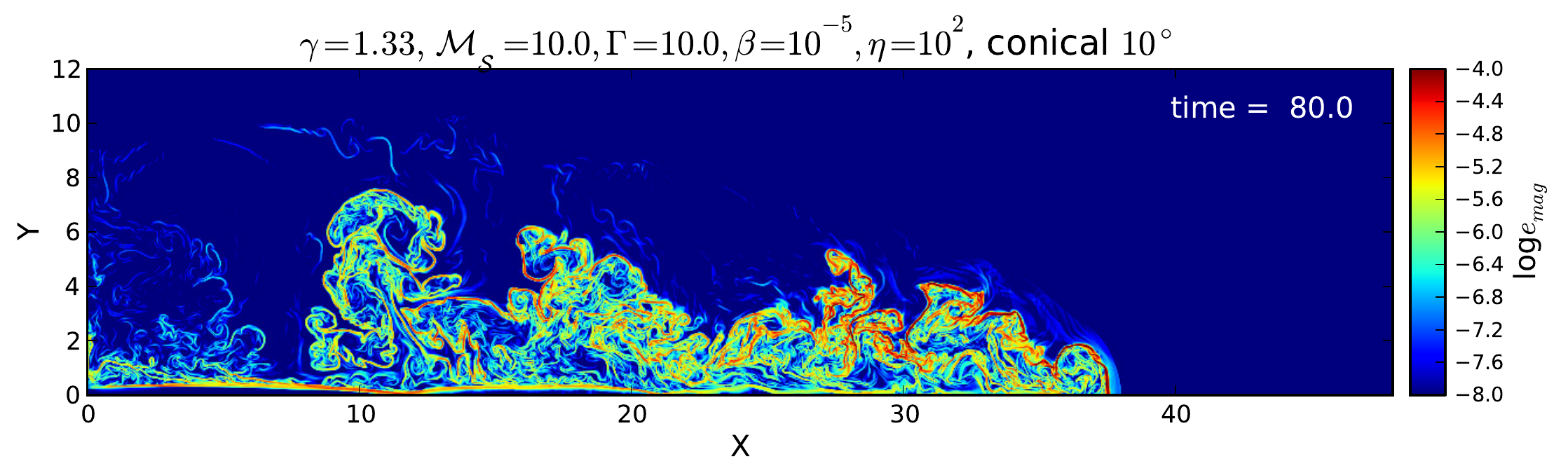}
 \includegraphics[width=0.45\textwidth]{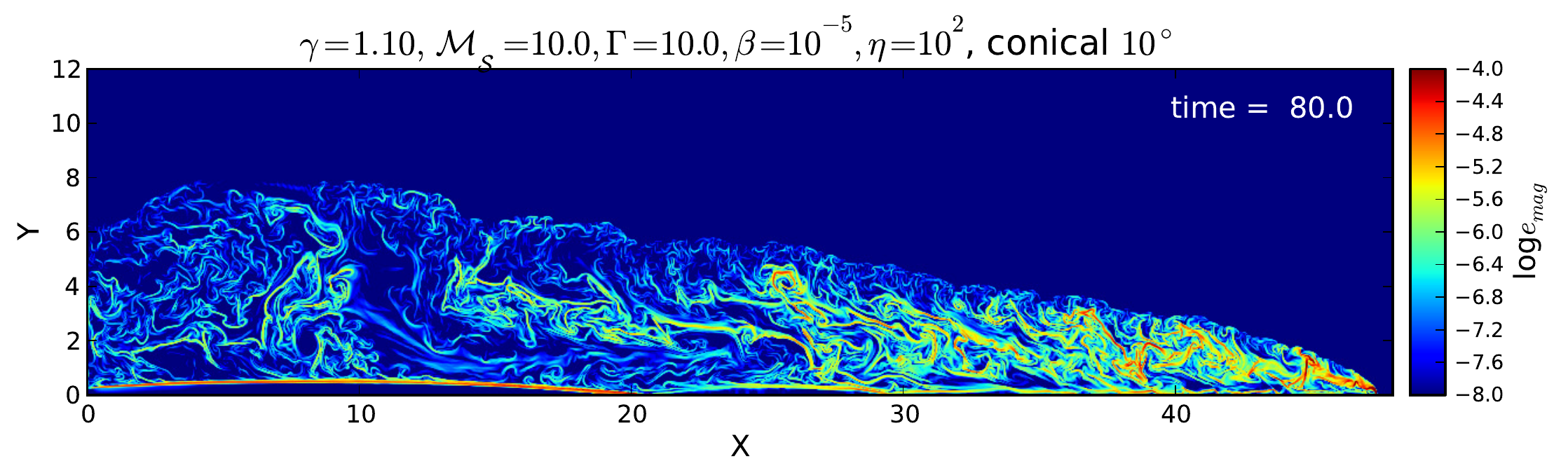}
 \caption{Same description of Fig.\ref{fig:density} but for the logarithm of 
 magnetic energy density, for the jet models with $\eta=10^2$.}
\label{fig:emag}
}
\end{figure*}

The simulated jets are initially non-magnetized, while the ambient medium is weakly magnetized, therefore
the magnetization of the downstream shocked gas is mostly due to the ambient field dragged into the shock at
the head and the cocoon. The spatial distribution of magnetic pressure for  different models with $\eta=10^2$ is shown in
Fig.\ref{fig:emag}.

Similarly to what is observed in the density distributions, there are striking differences in the magnetic
pressure distributions of the models. For the adiabatic models (left side panels) the high
intensity magnetic fields are located at the interface (or contact discontinuity) that separates the shocked jet
and ambient downstream flows, (i.e., the low and high density portions of the cocoon). These high magnetic intensity
regions basically contour the low density region as seen in Fig.\ref{fig:density}. The main reason for this is that
the ambient magnetic field lines enter into the cocoon dragged by the shocked downstream  flow. These are not able,
though, to enter into the shocked jet downstream material. At the ambient downstream, the gas is deflected and flows
along the contact discontinuity, as clearly seen in the adiabatic cases. The magnetic field lines, on the other hand,
simply accumulate at the contact discontinuity (piling-up there). The maximum intensity of $B$ occurs at the head of
the shock for all models.

\begin{figure*}
 \center
{
 \includegraphics[width=0.9\textwidth]{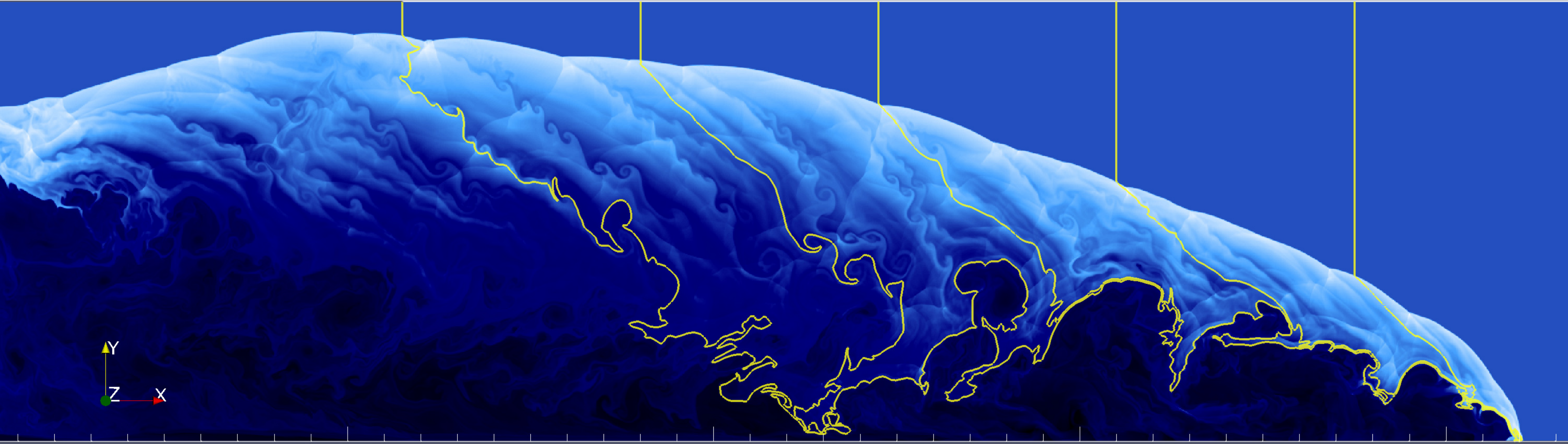}
 \caption{Logarithmic  density distribution for  model AD2 of Table 1, with $\gamma = 4/3$, $\Gamma = 10$, and 
 $\eta=10^2$ at $t=90$. The 5 lines drawn over the density plot represent magnetic field lines, each line starts at the 
 vertical coordinate $y=12.0$ (top boundary) and horizontal coordinates $x=[11.5, 18.0, 24.5, 31.0$ and $37.5]$.}
 \label{fig:pileup}
}
\end{figure*}

Fig.\ref{fig:pileup} shows the density distribution for the well-collimated (cylindrical) jet model AD2cyl, with $\gamma = 4/3$ and $\Gamma = 10$ at $t=90$,
overplotted with five selected high intensity magnetic field lines. The streamlines depicted follow the magnetic field
lines starting at the vertical coordinate $y=12.0$ (top boundary) and horizontal coordinates $x=[11.5, 18.0, 24.5, 31.0$
and $37.5]$. The ambient region presents the initial vertical field lines. At the shock regions, the lines are deflected
and stretched, as expected for a super-Alfv\'enic flow (i.e., with velocity higher than the local Alfv\'en speed).
As seen in Fig.\ref{fig:pileup} the field lines do not diffuse to the low density  region of the cocoon. On the contrary,
they accumulate at the contact discontinuity region between the ambient shocked material and the jet shocked material,
where therefore, the magnetic field intensity is larger.

The comparison of the adiabatic models in the left side of Fig.\ref{fig:emag} with the non-adiabatic ones in
the right hand side, indicates that the maximum values of the magnetic fields are slightly larger in the non-adiabatic cases.
This is compatible with the RH jump conditions for radiative cooling flows which predict a larger density of the shocked
material and therefore, a larger amplification of the magnetic field behind the shocks, than in the adiabatic counterparts.

\begin{figure*}
 \center
{
 \includegraphics[width=0.45\textwidth]{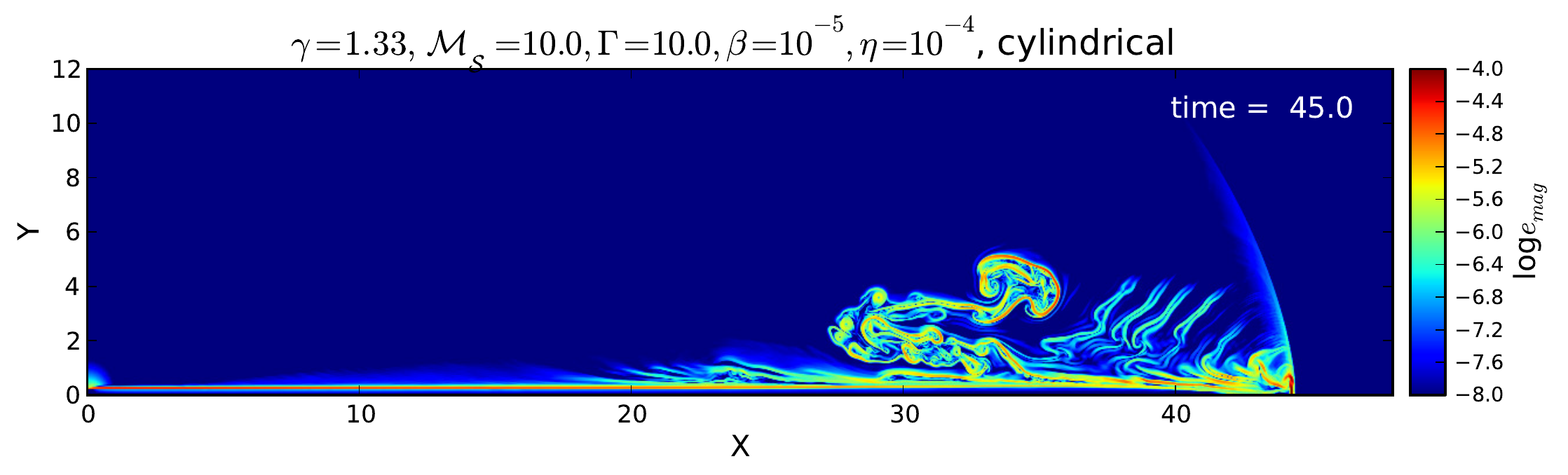}
 \includegraphics[width=0.45\textwidth]{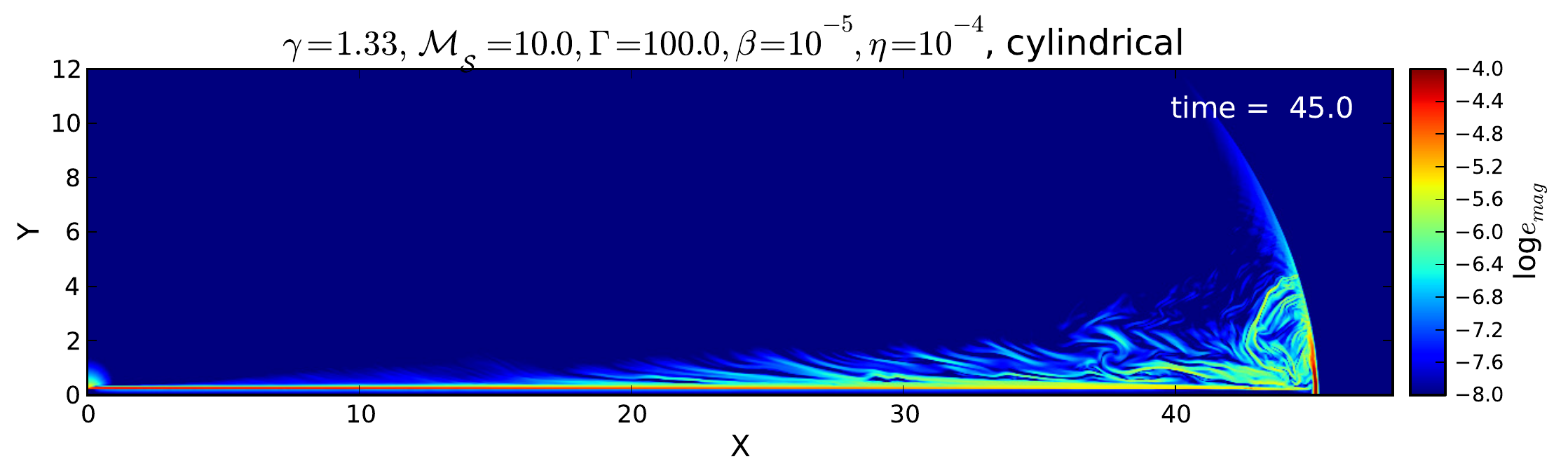}
 \includegraphics[width=0.45\textwidth]{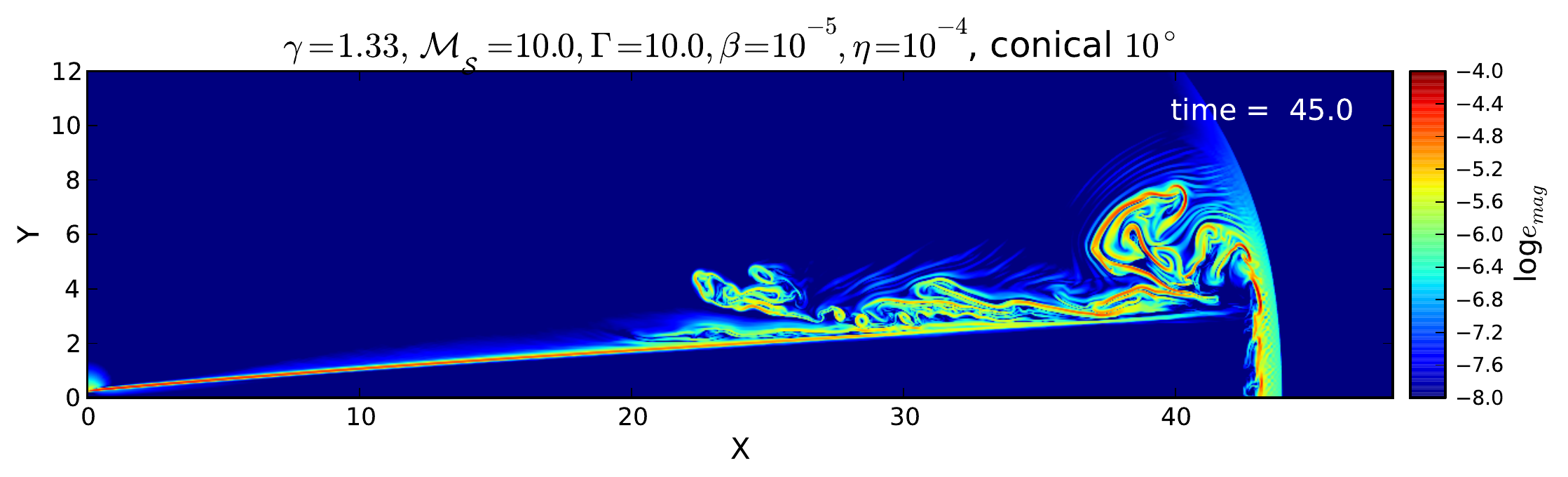}
 \includegraphics[width=0.45\textwidth]{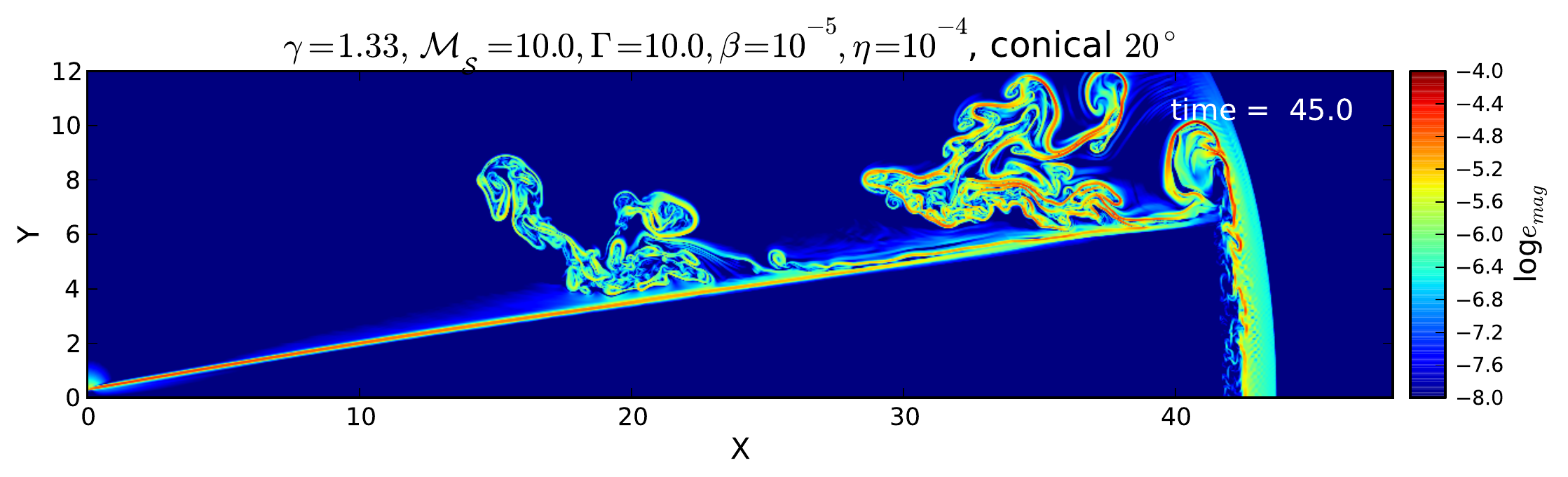}
 \includegraphics[width=0.45\textwidth]{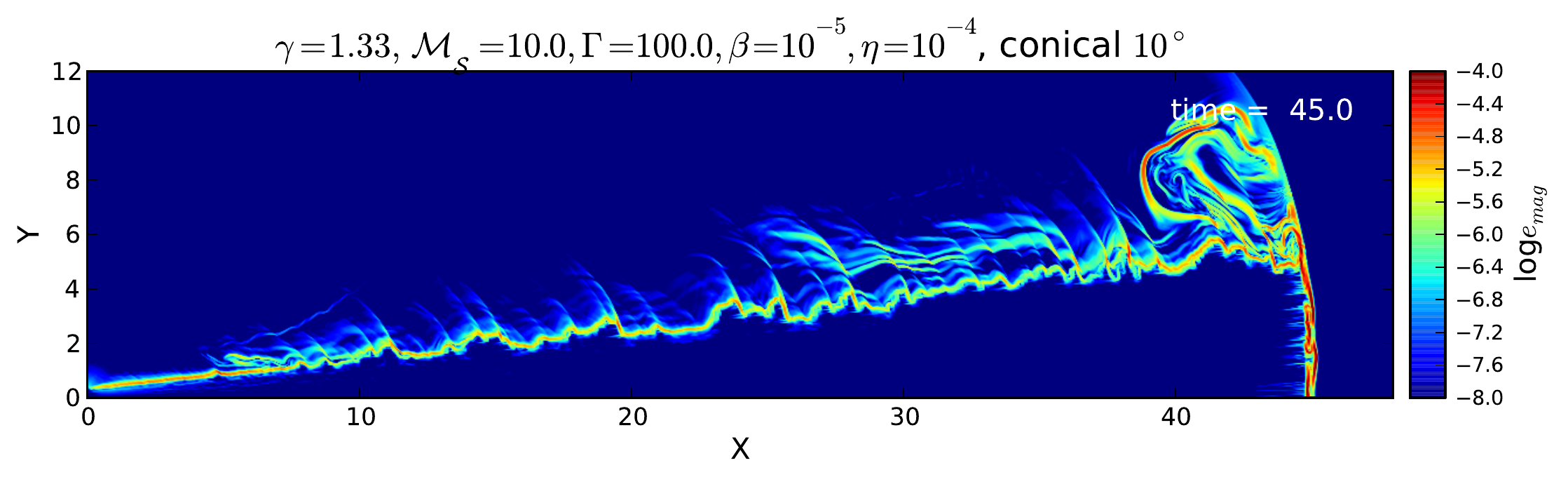}
 \includegraphics[width=0.45\textwidth]{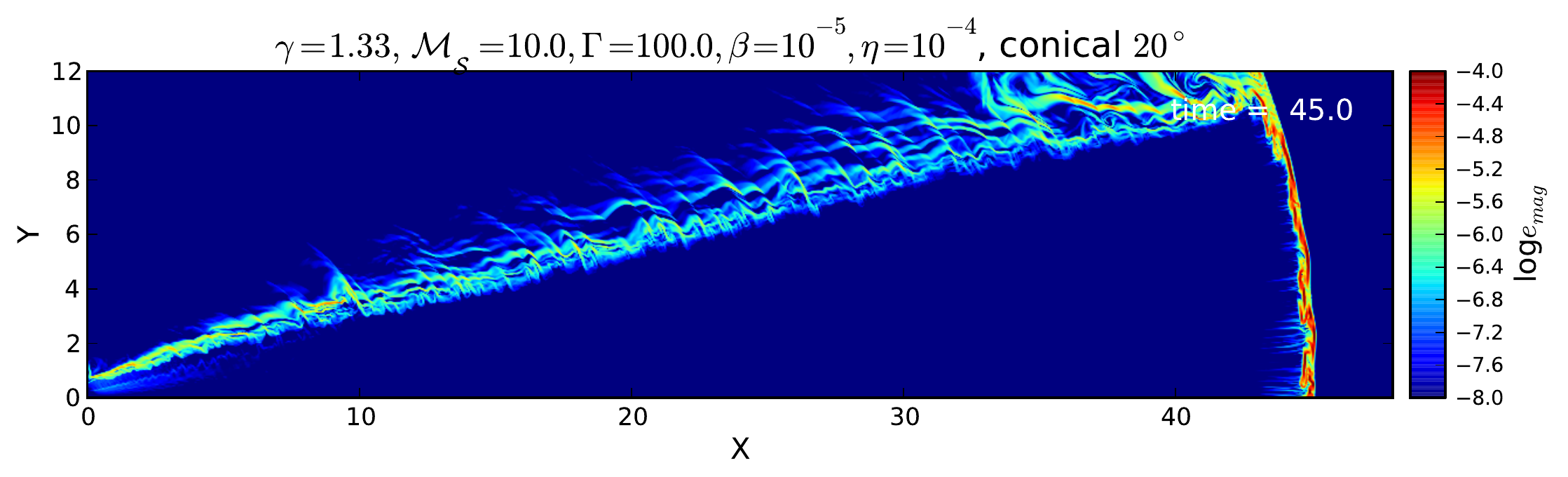}
 \includegraphics[width=0.45\textwidth]{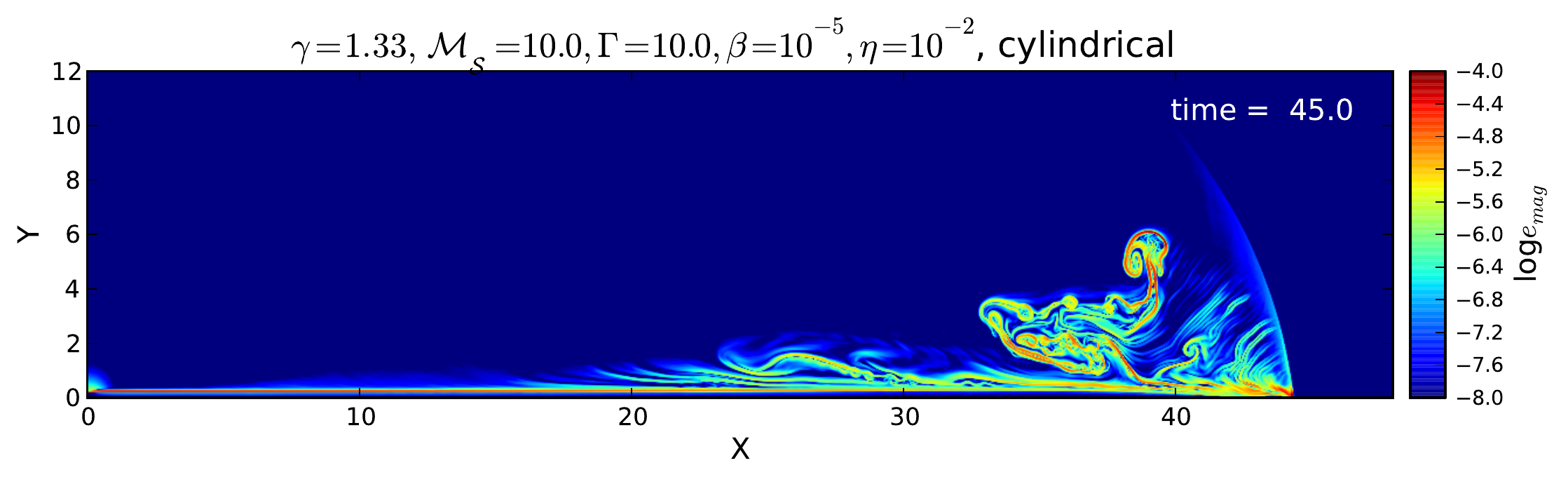}
 \includegraphics[width=0.45\textwidth]{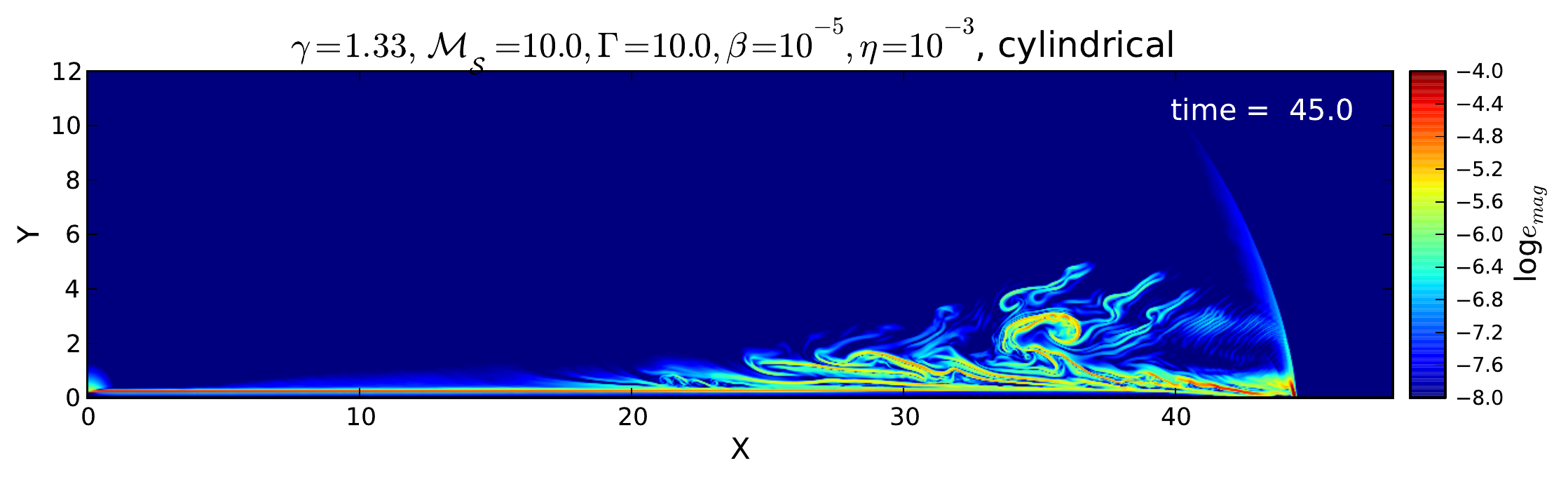}
 \caption{ame description of Fig.\ref{fig:density} but for the logarithm of 
 magnetic energy density, for the jet models with $\eta=10^{-4}$ to $\eta=10^{-2}$, 
 $\Gamma=10$ and $\Gamma=100$ and the opening angle varying between $0^{\circ}$ and $20^{\circ}$.}
\label{fig:eta2}
}
\end{figure*}

Heavy jets ($\eta <1$) on the other hand (Fig.\ref{fig:eta2}), have all similar magnetic field distributions, as already noted in the case of their density distributions. The absence of a prominent cocoon reduces the internal turbulence and its role in diffusing magnetic field lines.
Nevertheless, let  us perform a more careful analysis of the overall results.

\begin{figure}
 \center
{
 \includegraphics[width=0.45\textwidth]{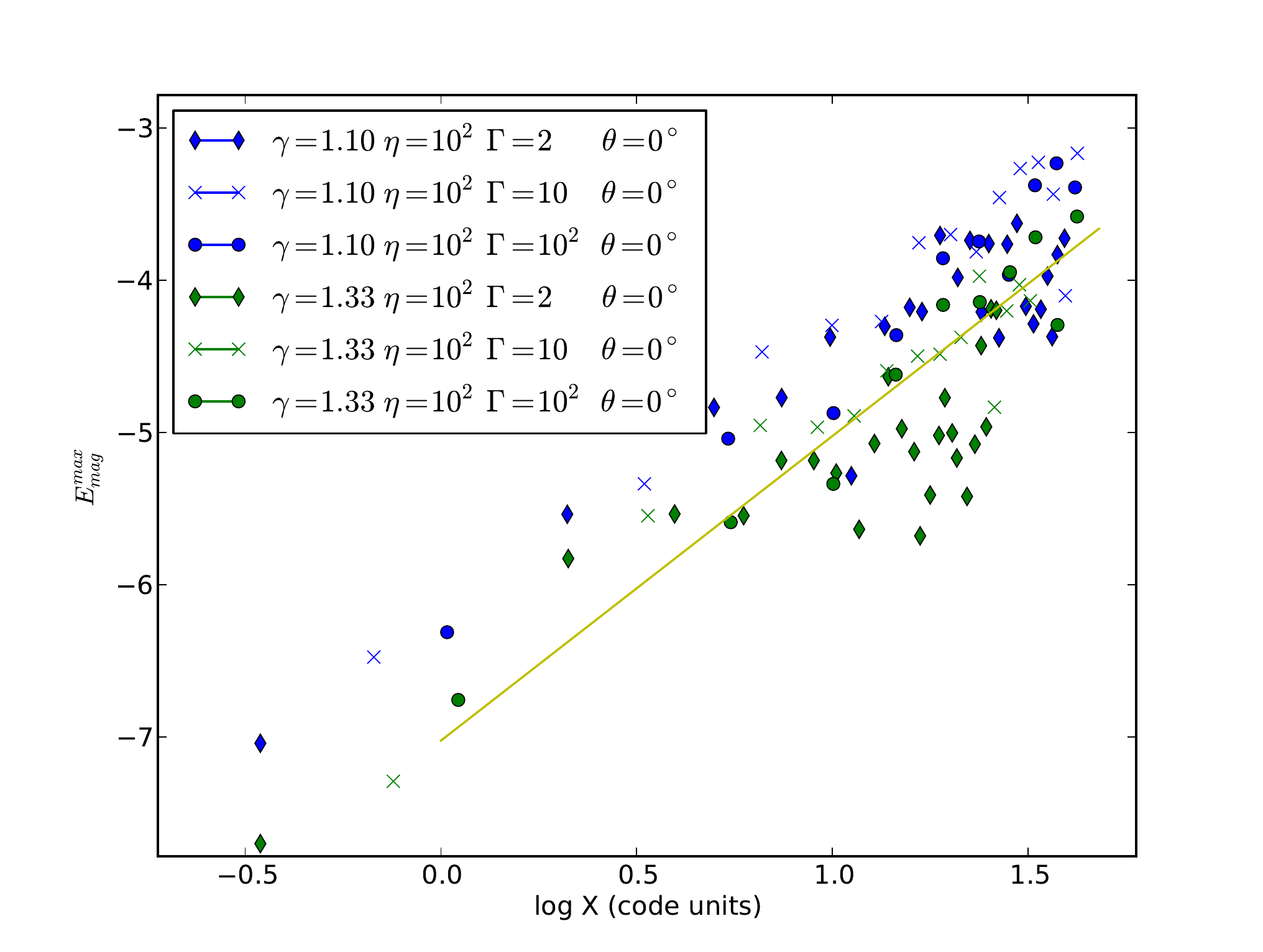}
 \includegraphics[width=0.45\textwidth]{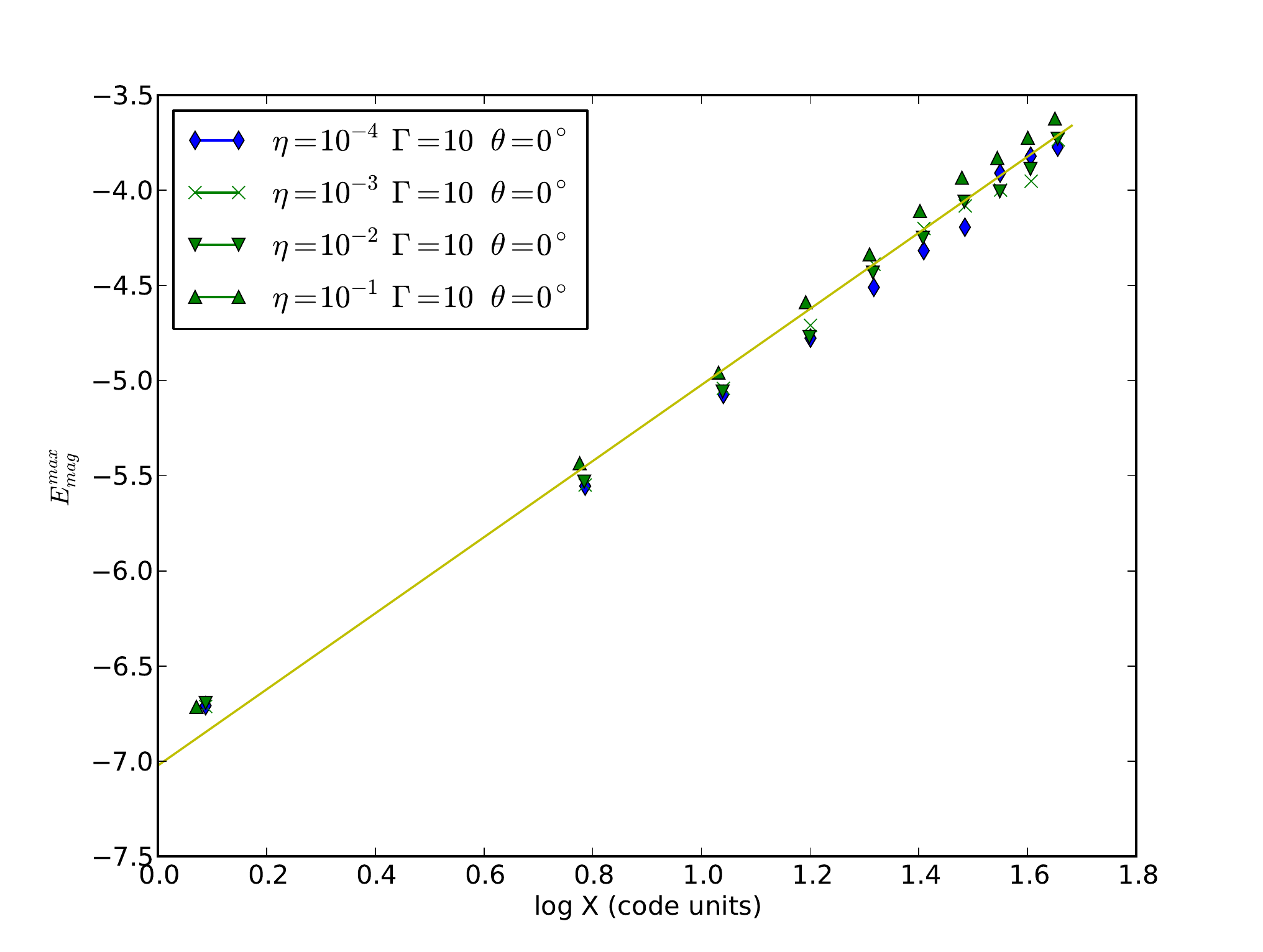}
 \includegraphics[width=0.45\textwidth]{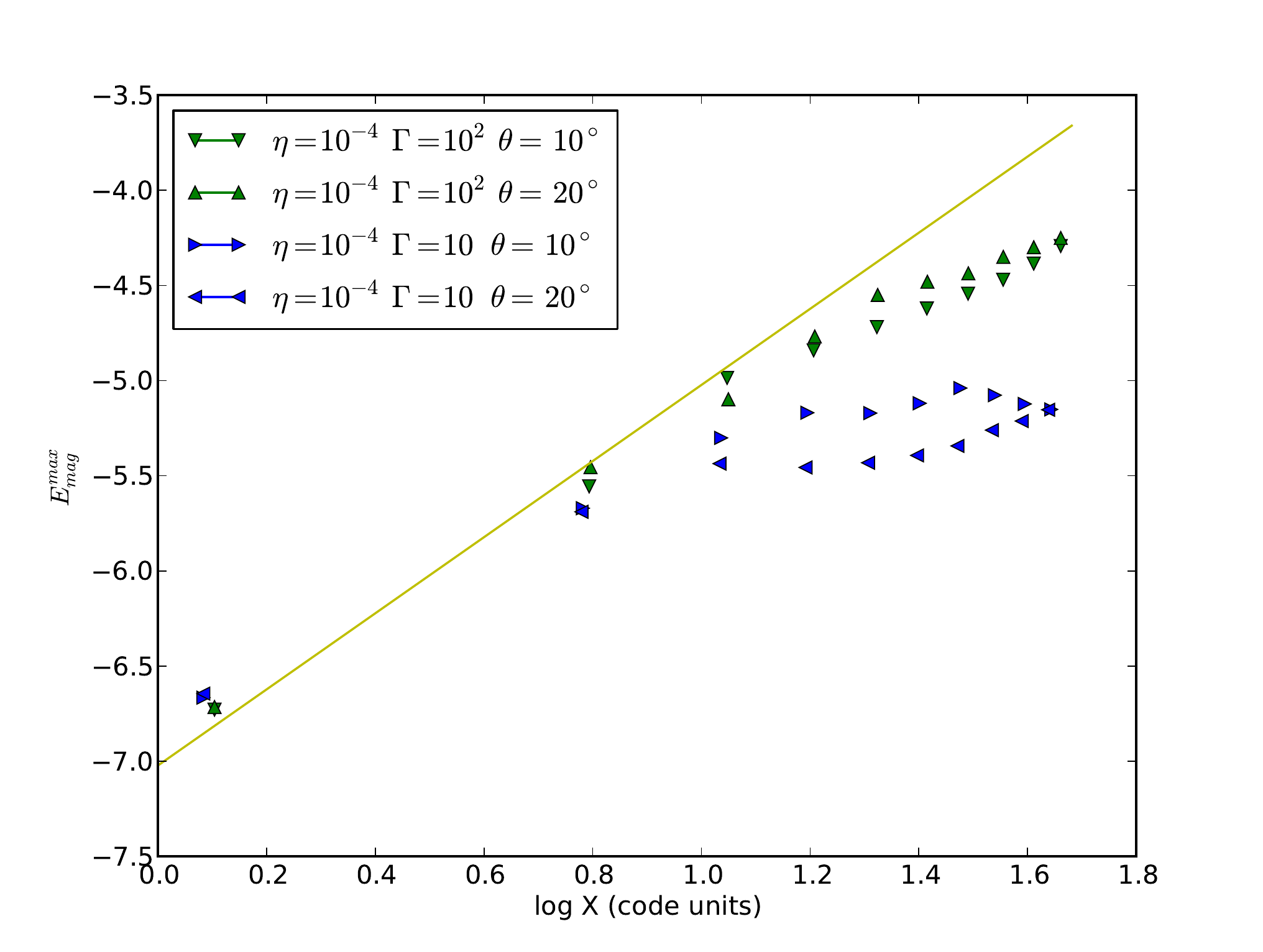}
 \caption{Maximum magnetic energy density
 (in erg $cm^{-3}$) as a function of the jet head position;
 top: cylindrical ($\theta=0$) light jets,  middle: cylindrical ($\theta=0$) heavy jets, and bottom: wide jets, with different Lorentz factors, and opening angles. The correlation $E_{\rm max}^{\rm mag} \propto B_{\rm max}^2 \propto x^2$ is very similar for all collimated jet models. The solid line with a  slope of $\zeta=2$ was drawn for reference.}
\label{fig:maxmag}
}
\end{figure}

The maximum magnetic energy density ($E_{\rm max}^{\rm mag}$) is generally located at the head of the bow shock region, as
the jet expands.
$E_{\rm max}^{\rm mag}$ as a function of the location of the bow shock head in the x-direction is shown in
Fig.\ref{fig:maxmag}. Each snapshot created from the simulations is shown as a point in the plot. The top and middle diagrams show the results for all collimated jets (with light jets being depicted in the top panel and heavy jets in the middle panel). The solid line
represents the correlation $E_{\rm mag}^{\rm max} \propto x^2$, for comparison.  Notice that the line is not a statistical
fit, but is in good agreement with all models with $\theta = 0$. Most strikingly, all models, independent on the
Lorenz factor, the polytropic index, or the density ratio $\eta$, present similar $E_{\rm mag}^{\rm max}$ at the same position of the shock
head. This result is consistent with the magnetic field pile-up effect discussed in Section 2 and with Eq.
\ref{eq:pileup} which predicts $B_{ampl} \propto x^\alpha$,
with a maximum $\alpha \simeq 1$ for a compressed magnetic field parallel to the discontinuity.

In the bottom diagram of Fig.\ref{fig:maxmag}, we show the evolution of the maximum magnetic field intensity for the conical
jets. It is clear the dependence of $E_{\rm max}^{\rm mag}$ with the opening angle $\theta$ and $\Gamma$ in consistence with
the analytical prediction  of Eq.\ref{eq:rjcon}.

The results above clearly show that the pile-up effect is maximized in the case of $\theta \rightarrow 0$, as one should expect. 
In fact, the piling-up is maximized in the forward shock region where the jet velocity is nearly normal to the magnetic field lines. 
Thus, although a conical geometry may offer a larger area for the forward shock to sweep the ambient magnetic lines, most of the 
shock front is oblique which will weaken the piling-up. Also, the net flux of plasma out of the shock region is limited (causality 
is not broken here), and once the Mach disk becomes large enough local enthalpy cannot be considered as constant any longer. 
After this transition phase the shock width scales linearly with $x$ and the pile-up effect ceases. This is well accounted in Eq.\ref{eq:rjcon}.

We  stress here that the results obtained in Fig.\ref{fig:maxmag} are  nearly insensitive to the jet-ambient
density ratio $\eta$. This result is actually not  surprising.
The accumulation of the compressed ambient magnetic field lines depends on $\eta$ through the propagation
velocity of the forward bow shock into the ambient medium (see eqs. 7 to 9), i.e., $\beta_{bs} = \beta_j (1+ L^{-1/2})$,
where $L$ measures the ratio between the jet energy density and the ambient rest mass energy density,
$L= \mu_j \Gamma_j^2/ \eta$, and $\mu_j \sim 1$ is the specific enthalpy of the jet \citep[][]{bromberg2011}.
For the typical large values of $\Gamma_j \sim 10 - 100$ of GRB jets, it turns out that in general $L \ll 1$,
even for $\eta$ varying in a broad range like the one investigated here $\eta = 10^{-4}$ to $10^2$, so that
$\beta_{bs}$ and the pile-up effect are nearly insensitive to this parameter.

\subsection{Structure function of ${\bf B}$ and its correlation length}

The amplification of $B$ as seen in these models is particularly important because, regardless of the magnetization of the jet itself, as the beam sweeps the ambient gas, the ambient magnetic field lines are dragged, amplified by compression, and piled-up into the shock region. Also, as important as the total magnetic field intensity is its correlation length.

In any model of magnetic field amplification, theoretical predictions must also provide arguments for obtaining sustainable large scale magnetic fields.

One way of determining the correlation length of the magnetic field distribution is by means of the second order structure function (SF) \citep[e.g.][]{kowal2007,diego2008}, defined as:

\begin{equation}
{\rm SF}(l)= \langle |{\bf B}({\bf r}+{\bf l})-{\bf B}({\bf r})|^{2} \rangle,
\label{eq:sf}
\end{equation}

\noindent
where ${\bf B}({\bf r})$ represents the magnetic field vector at a given position ${\bf r}$, and ${\bf l}$ the spatial increment for the structure function. The increment $l$ is a vector taken to be parallel to the local orientation of the field line. In this sense the SF measures the statistical changes on the magnetic field along the streamlines. Notice that ${\rm SF}_{l \rightarrow 0} \rightarrow 0$, while as $l$ increases the structure function also increases up to a saturation level. The scale length at which the SF saturates represents the largest scales of the magnetic fluctuations, i.e. the correlation length.

\begin{figure}
 \center
{
 \includegraphics[width=0.45\textwidth]{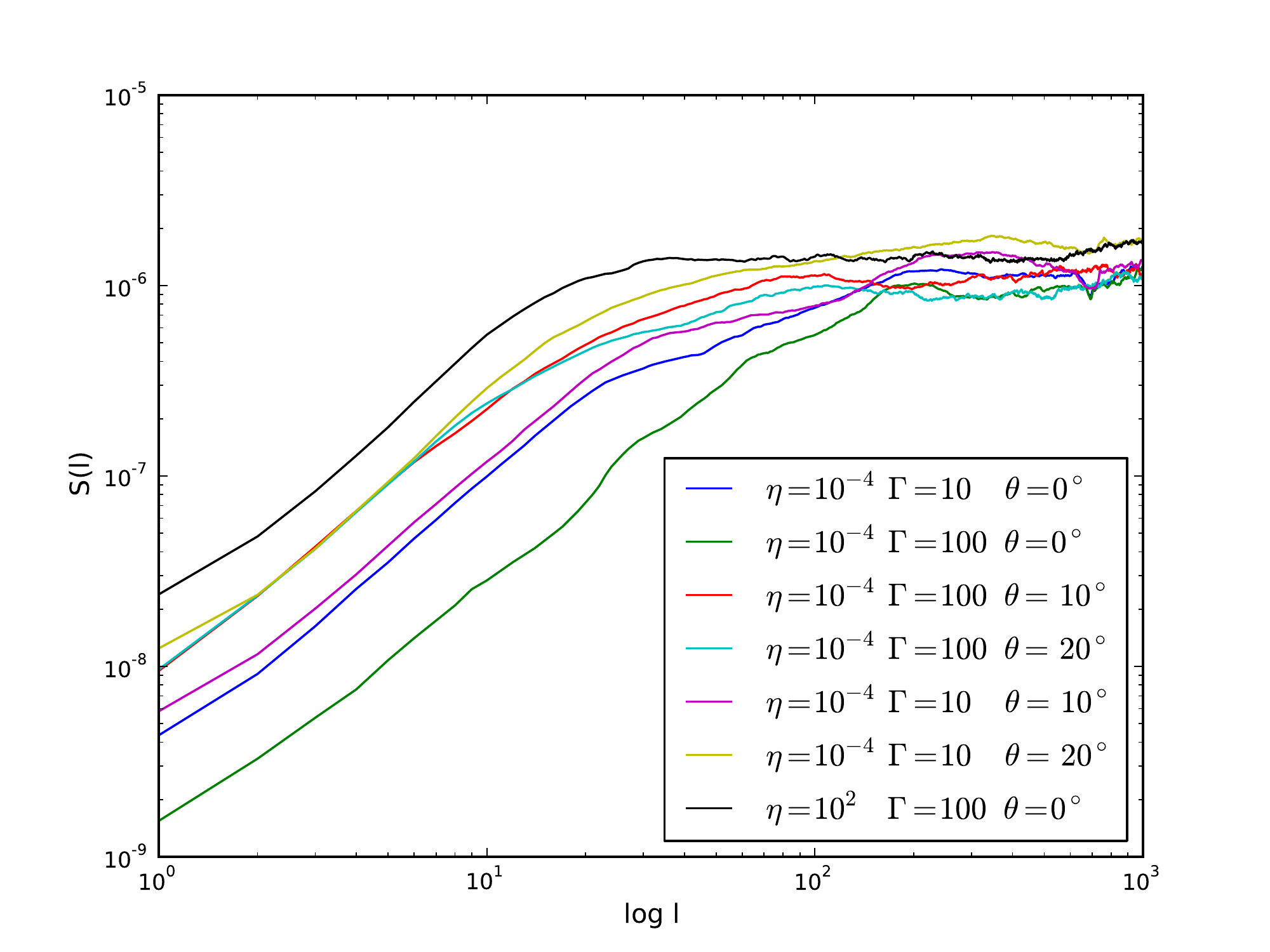}
 \caption{Structure functions (SFs) of magnetic field lines
 for models varying the parameters $\eta$,  $\Gamma$, and the opening angle. The horizontal axis is shown in number of pixels.
 Notice that the SFs are calculated along the magnetic field lines and the total pathways are therefore larger
 than the size of the box.}
\label{sf}
}
\end{figure}

We performed the structure function  calculations for the magnetic field lines anchored into the
shock head only - this because we focus on determining the correlation length of the maximum
amplified magnetic fields. In Fig.\ref{sf} we present the SFs (S) obtained for the selected
adiabatic models. The non-adiabatic models were not plotted to avoid superposition with the depicted
curves, as they present very similar profiles to those of the adiabatic counterparts.

For  light jets with $\eta=10^2$, the saturation of the SFs occurs, in all models, at
length scales of  which represent
$l_{\rm sat} \simeq 0.35 - 0.46$ in code units. These values correspond to $\sim 3 - 5$ times the shock thickness
$\lambda$ for the adiabatic and non-adiabatic models, respectively.
Heavy jets with $\eta=10^{-4}$ present larger coherence lengths. For these models the saturation of the
SFs occurs at lengthscales in the range of $l_{\rm sat} \simeq 60 - 200$ pixels, depending on the opening
angle of the jet, which represents $l_{\rm sat} \simeq 2.3$ in code units. Larger coherence lengths occur
for smaller opening angles.

\section{Discussion and Conclusions}

In this work we have explored  the possible magnetic field amplification and correlation lengths behind  the shock head region of non-magnetized, light and heavy  relativistic jets propagating into weakly magnetized environments aiming at comparisons with GRB jet afterglows scenarios in a matter dominated regime. For this we have carried out 2D relativistic MHD (RMHD) simulations considering different values of the jet bulk Lorentz factor ($\Gamma =$ 2,  10, and 100) and the density ratio between the jet and the environment ($\eta = \rho_{amb}/\rho_{j} = 10^{-4} - 10^{2}$). We have focussed on relativistic adiabatic jets (with an adiabatic index $\gamma = 4/3$), but  for comparison have also considered systems with $\gamma = 1.1$  in order to mimic the effects of a potential strong thermal radiative cooling in the shocked ambient material at the jet head. All the jets were expanded for approximately the same extension, so that the jet with the highest Lorentz factor was the less evolved one.  We have also 
tested the effects of the jet geometry, considering different opening angles from $\theta=0^{\circ}$ (cylindrical jet) to $\theta=20^{\circ}$. Our findings are summarized below.

The magnetic field is amplified by shock compression and accumulates at the contact discontinuity (pile-up effect), with a maximum value that increases with the distance as the jet propagates. The predicted  relationship between the  magnetic field intensity and the distance as described in Eqs. 7, 8 and 11, was confirmed by the simulations. In particular, we have found that the increase in the magnetic field amplification, though initially similar for both collimated and wide jets,  saturates  earlier for increasing jet opening angles. This effect is smaller as the jet Lorentz factor increases.
These results have been found to be nearly insensitive to the density ratio $\eta$, but heavier jets present larger magnetic field coherence lengths than lighter ones. Also smaller coherence lengths have been found for larger  jet opening angles.
In summary, heavy, collimated jets tend to maximize the piling-up and the coherence length of the magnetic field lines.

The results above have been also found to be  nearly independent on the adiabatic index ($\gamma$), although the maximum intensities of the compressed magnetic fields are a little larger in the non-adiabatic cases, as one should expected from the jump conditions and the larger density amplification behind the shocks in these cases. This general behaviour can be explained by the fact that, after a maximum compression behind the double shock structure at the jet head, the magnetized shocked material is forced to expand sideways, along the cocoon that surrounds the jet. Apparently, all the cases reach similar saturation ratios for the density and magnetic field at the contact discontinuity, regardless of the differences in the jet upstream  conditions. Nevertheless, these differences obviously affect the final state of the shocked material that deposits into the cocoon which is clearly distinct in each of the simulated systems as discussed in Section 4.1 (see Figs.\ref{fig:density} and \ref{fig:emag}).

We notice here that in more realistic calculations, with a more consistent treatment of the radiative cooling, the effective value of $\gamma$ would not be homogeneous over the whole computational domain. The `non-adiabatic' models above actually represent extreme examples. In more realistic models, with an adiabatic jet beam (with $\gamma=4/3$) interacting with a radiative cooling cocoon, we would expect the beam structure to be less affected by the  shocked cooled gas of the cocoon than in Fig.\ref{fig:density} and the propagation velocity of the jet head slightly smaller.

Since we have considered a very broad parametric space, the results above can be in principle applicable to all classes of relativistic jets, including microquasars, AGNs and GRBs, but below  we will discuss the implications for the afterglow emission of GRB jets.

\subsection{Implications for GRB afterglows}

\subsubsection{Magnetic field amplification}

Observations
of the afterglow phase of GRBs are explained by synchrotron emission of electrons interacting with nearly equipartition magnetic field intensities of
$B_{\rm equip} \sim 1$G, at distances of $\sim 10^{15}$cm away of the central source \citep[see review of][]{piran2005}. 
As explained in Section 2
the equipartition radius for the magnetic field amplification
depends on the shock width $\lambda$, which can be roughly estimated from Equation 10 (see also Eq. 11).
In this equation, as stressed in Section 2, we need the jet radius at the  breakout from the stellar progenitor envelope. 
This can be estimated from previous analytical and numerical studies of GRB jets 
(e.g., Zhang, Woosley \& MacFadyen 2006; Mizuta \& Aloy 2009, Bromberg et al. 2014, Levinson \& Begelman 2013,  Mizuta \& Ioka 2013). 
For a Poynting flux dominated jet propagating inside the envelope of a Wolf-Rayet progenitor, analytical predictions suggest that   
$r_j \sim r_L \sim 10^7$ cm, where $r_L$ is the radius of the light cylinder near the source (Levinson \& Begelman 2013, Bromberg et al. 2014), 
while for matter dominated jets $r_j$ can be larger. Numerical simulations indicate $r_j \sim  10^9$ cm 
(Zhang, Woosley \& MacFadyen 2006; Mizuta \& Aloy 2009, Mizuta \& Ioka 2013).
Thus, if we assume $r_j$ at the breakout to be  $r_j \sim 10^7 - 10^9$cm, and
$\eta = 10^{-4} - 10^{2}$, then we obtain $\lambda \sim 10^3 - 10^{11}$cm. Despite the simplified geometry, and
absence of magnetic field, assumed on the estimation of Eq.10, these values are in rough agreement
with the $\lambda/r_j$ ratio observed in the simulations. In this case, equipartition should occur
at $x_{bs} \sim 10^{9} - 10^{17}$cm.
These values are compatible with the observed afterglow
distances ($\sim 10^{15}$cm).

It is worth mentioning that the calculation above considers the magnetic field estimated
assuming equipartition between the magnetic and relativistic particles component of the
synchrotron emitting plasma. In principle, the equipartition magnetic field at the
emitting region may strongly deviate from the actual saturation magnetic field, which is related
to the dynamical equilibrium between the jet kinetic pressure and the downstream
magnetic field. It is difficult to estimate the later from a physical background since the
dynamical evolution of the jet as it propagates through the medium is hardly known {\it a priori}.
For this reason, the numerical simulations may provide a good insight.
The saturation on the amplification of the magnetic field  can be estimated from the
conservation of momentum equation, at the shock reference frame at the jet axis, as:

\begin{equation}
\rho_j \Gamma_j^2 (\beta_j - \beta_{sh})^2 \approx \rho_a \beta_ {sh}^2 + \frac{B_s^2}{8\pi \Gamma_s^2}
\end{equation}

\noindent
where $\beta=v/c$, and indices $j$, $a$ and $sh$ stand for jet, ambient and shock, respectively. For instance,
in model AD3, with $\Gamma_j =100$ and $\rho_{a} = 100 \rho_{j} = 1.67 \times 10^{-24}$, we obtained
$\Gamma_s \sim 2.3$, i.e. $\beta_{sh} \sim 0.9$. Therefore, the saturation in the simulation would occur for
$B_s \sim 1.4$G, in agreement with the observations. Naturally, this condition is even more confortable for heavy jets ($\rho_{a} \ll \rho_{j}$), for which one obtains a much larger limit $B_{\rm sh} \gg 1$G.

For jets with opening angles, i.e. $\theta_{\rm j}>0^\circ$, the maximum amplification is reduced.
Our results indicate that wide jets present similar behaviour as their cylindrical collimated counterparts at small distances, for which the conclusions made above would be sustained. This is not true though at larger distances. While the well collimated jets result in a quasi indefinitely increase of magnetic pressure (until equipartition is reached), jets with large opening angles saturate at earlier stages. Here the main cause for the saturation os not equipartition but the widening of the shock region width. In conical jets the ratio between the fluxes out and inwards the shocked region becomes smaller with time, resulting in the width growth. 
Models with $\theta = 20^{\circ}$ saturate with $B_{\rm max}$ approximately
 1 order of magnitude smaller than those with $\theta = 10^{\circ}$. As the width of the shock region increases, the  amplification of the magnetic field is smaller for a lower saturation value in agreement with Eq.\ref{eq:pileup}. For instance, for a $\tan \theta \sim 0.1$ and $\Gamma \sim 10^2$ jet, the saturation radius is expected to be at $\simeq 10^3 r_{\rm j} \sim 10^{12}$cm and, as shown above, this length scale would be large enough to amplify the magnetic fields to the observed intensities. 

However, despite the apparent sufficient amplification factors obtained from 
the pile-up process, the large magnetization of the observed afterglow emission 
cannot be fully explained yet. In the current GRB paradigm, adopted in this 
work, the afterglow emission is assumed to be radiated from the freshly injected 
plasma at vicinity of the shock. It should be further noticed that the freshly 
injected plasma, just downstream of the shock, is weakly magnetized\footnote{as 
given by the standard Rankine-Hugoniot conditions.}. The main effect of pile-up 
only occurs as the matter flows further downstream whereas the frozen field 
slowly grows. Though the field strength could be large it peaks at the contact 
discontinuity region, which is at a distance $\sim \lambda$ of the freshly 
injected shocked plasma, where particles are supposed to accelerate. 
Therefore the pile-up effect, even if it is strong enough, may not directly 
affect the afterglow emission in such a scenario, and would  not  provide 
``the'' solution to the magnetization problem in GRBs.

\subsubsection{Correlation lengths}

Also, by means of second order structure functions (SFs), we
obtained the correlation lengths of the amplified magnetic fields at the jet head. We find
$l_{\rm corr}\sim 3 - 5 \lambda \sim 10^8 - 10^{12}$cm for  light jets, and
$\sim 10^9 - 10^{14}$cm for the heavy jet models. There is no obvious trend between
the correlation length and the jet opening angle.
 For the Weibel instability the correlation lengths obtained are of the order of the plasma skin depth, i.e. $\delta=(c/\omega_{p})\sim 10^{6}$cm, while
observations point towards much larger correlation lengths, of $10^{16}$cm \citep[e.g.][]{waxman2006}.

Still the values obtained in this work are 2 orders of magnitude smaller than those obtained from
observations. Since our models revealed that the correlation length depends on the
jet-to-ambient gas density ratio, heavier jets, compared to those simulated here, would result in
larger $l_{\rm corr}$, closer to observations.
Another possible solution to this problem is that, due to the strong downstream
turbulence as seen in part of our models,
magnetic reconnection could be induced resulting in more uniform fields.
Considering that the equipartition occurs at short timescales (specially for jets),
any magnetic energy loss due to reconnection would be shortly replenished
by further piled-up field lines. It is possible then that the field lines would have
larger observed correlation lengths once the systems reaches the afterglow phase.

In any case, the fact that our present  study of unmagnetized jets impinging into a magnetized ISM results  
magnetic correlation lengths in the shock front  2 orders of magnitude smaller than the expected for GRBs points to the necessity of 
exploring the magnetic field amplification and  pile-up  in magnetized relativistic jets. The recent polarization observations by \citet{Wiersema2014} 
indicate that this may be the correct way to solve this question.

\subsection{Final Remarks}

Two further important remarks are in order. First,
we have  assumed a two-dimensional (2D) jet geometry.
A more realistic 3D geometry can
reduce the pile-up efficiency since this geometry allows another degree of freedom for the
magnetic field lines (and gas) to leave the shock region. However, since in this case
the degree of freedom of field lines is still smaller than that of the gas, the pile-up
must still occur, though not as efficient as in the two-dimensional well-collimated jet case.
We stress that the main goal of this work was not to reproduce the actual emission properties of the afterglow, but to verify if the magnetic field could be amplified in the jet-envelope shock context. In order to determine exactly the emission properties, on top of the 
 amplification process, three-dimensional simulations are mandatory. 
 
A third dimension will naturally introduce another degree of freedom for the downstream flow, which implies an extra dimension to which magnetic field lines may be carried away from the shock region thus decreasing the pile-up effect. We may therefore expect that in a 3D jet model the amplification of the magnetic field intensity will slow down, but the final picture of the pile-up in this case is still unknown.
Another possible effect that was not taken into account in our study regards the fact
that, at the time that the jet breaks out from the stellar surface into the ambient medium,
the GRB central engine has probably turned off already. This implies that the continuous
injection should stop, giving place to a propagating jet parcel with a forward bow shock
at the head slowly detaching from the reverse shock. This effect will also weaken
the piling-up of the magnetic field in the bow shock. Both effects will be investigated in
depth in a forthcoming work.

However, even in the case of efficient pile-up we must be careful in 
attributing to this mechanism the solution for the magnetization problem in the 
afterglow emission. The magnetic energy in our models has been found to be 
concentrated at the contact discontinuity while the emitting particles are 
expected to be located the the downstream side of the shock surface. A large 
distance between these two different regions result in an effective low 
magnetization where the emitting particles actually are. This issue must be 
pursued in forthcoming works.

Finally, as stressed before, the afterglow emission is generally believed to be due to relativistic particles accelerated by a first-order Fermi process occurring mostly at the shock region, at the jet head.
Examining Fig.\ref{fig:emag}, we note that other regions in the  beam and the cocoon than the shock head itself have also reached a magnetized turbulent structure with high intensity magnetic fields. These  magnetic fields, in part also amplified by the instabilities developed in the cocoon and by turbulent shear, can equally  help to accelerate particles to relativistic velocities.  In these regions, first-order Fermi acceleration by magnetic reconnection, as first proposed by \citep{dalpinolazarian2005}, can be also very efficient, as well as second order Fermi to pre-accelerate the particles, as indicated by recent numerical MHD studies of particle acceleration in different domains of magnetic reconnection \citep{kowal12} \citep[see also][for a review]{dgdpkowal2013}. This issue will be further  explored by means of  "in situ" particle  acceleration simulations in relativistic jets as in \citet{dgdpkowal2013} where  preliminary tests have been presented (see also applications to GRBs in \citet{giannios2010,cerutti2013}).

\section*{Acknowledgments}

G.R.S. thanks CNPQ for financial support.
D.F.G. thanks the European Research Council (ADG-2011 ECOGAL) and the Brazilian agencies CNPq (No. 300382/2008-1), CAPES (3400-13-1) and FAPESP (No. 2011/12909-8) for financial support.
G.K. thanks FAPESP (No. 2009/50053-8, 2011/51275-4, 2013/04073-2, 2013/18815-0)  for financial support.
E.M.G.D.P. thanks FAPESP (No. 2006/50654-3) and CNPq (306598/2009-4) for financial support.
The authors also acknowledge very fruitful discussions  with T. Piran, J. Stone and G. Lugones. This work has made use of the computing facilities of the Laboratory of Astroinformatics (IAG/USP, NAT/Unicsul), purchased by FAPESP (grant 2009/54006-4),  and of the Hydra cluster at EACH-USP.

\end{document}